\documentclass[preprint,amsmath,amssymb,aps,a4]{revtex4-2}
\usepackage{graphicx}
\usepackage{lineno}
\usepackage{bm}
\usepackage{txfonts}
\usepackage{hyperref}
\usepackage{caption}
\usepackage{amsmath}
\usepackage{amssymb}
\usepackage[section]{placeins}
\usepackage{tabularx}
\usepackage{braket}
\usepackage{multirow}
\usepackage{subcaption}
\usepackage{rotating}
\usepackage{xcolor}
\date{\today}
\newcommand{\be}{\begin{eqnarray}}
\newcommand{\ee}{\end{eqnarray}}

\newcommand{\bfk}{{\bf k}_{\perp}}

\begin{document}
\title{Valence quark distribution of light $\rho$ and heavy $J/\psi$ vector mesons in light-cone quark model}
\author{Tanisha$^{1}$}
\email{tanisha220902@gmail.com}
\author{Satyajit Puhan$^{1}$}
\email{puhansatyajit@gmail.com}
\author{Anurag Yadav$^{1}$}
\email{anuragyadav1431905@gmail.com}
  \author{Harleen Dahiya$^{1}$}
\email{dahiyah@nitj.ac.in}
\affiliation{$^1$ Computational High Energy Physics Lab, Department of Physics, Dr. B.R. Ambedkar National
	Institute of Technology, Jalandhar, 144008, India}

\date{\today}%
\begin{abstract}
	In this work, we have investigated the valence quark structure of light $\rho$ and heavy $J/\psi$ vector mesons using the light-cone quark model through unpolarized quark generalized parton distributions (GPDs). By solving the quark-quark correlator, we have represented the quark GPDs in the form of light-front wave function (LFWFs). The charge, magnetic, and quadrupole form factors of these particles have been derived from the unpolarized quark GPDs and compared with the available theoretical predictions and lattice simulation data. The structure functions corresponding to the Rosenbluth scattering cross section of these mesons have also been calculated and compared with the available NJL model predictions.  The results of our calculations are found to follow a similar trend as the other model results. We have also calculated the parton distribution functions (PDFs) of these particles in the forward limit of GPDs. The calculated PDFs have also been evolved to $5$ GeV$^2$ through next to next leading order Dokshitzer–Gribov–Lipatov–Altarelli–Parisi (DGLAP) evolutions.

	\vspace{0.1cm}
	\noindent{\it Keywords}: light and heavy vector mesons, generalized parton distributions (GPDs), parton distribution functions (PDFs), form factors (form factors), structure functions, light-cone quark model.
\end{abstract}
%
\maketitle
%
%

\section{Introduction\label{secintro}}
Vector mesons such as $\rho$ and $J/\psi$ play an important role in advancing our understanding of the Standard Model by offering a versatile testing ground for both the theoretical and experimental techniques. Owing to their differing mass scales and spin-1 character, they enable the exploration of polarization-sensitive effects and tensor structures that are not present in scalar or spin-1/2 systems. As such, these mesons provide a valuable context for assessing the applicability of effective models in describing hadronic phenomena across a wide range of energy scales. At the low energy scale $\mu^2\le 1$ GeV$^2$, the valence quark contribution is more compared to gluons than at the high energy scale $\mu^2\ge 1$ GeV$^2$ which is primarily due to confinement. Therefore, it becomes interesting to explore the valence quark behavior within the low energy quark models such as Nambu-Jona-Lasinio model (NJL) \cite{Volkov:2005kw,Arbuzov:2006ia,Ninomiya:2014kja}, light-front quark models  (LFQM) \cite{Belyaev:1997iu,Choi:2017zxn,Yadav:2025txk}, light-front holographic model (LFHM) \cite{Brodsky:2014yha,deTeramond:2008ht}, spectator model \cite{Bacchetta:2021oht,Lorce:2016ugb}, quark-diquark model \cite{Strodthoff:2011tz,Maji:2016yqo}, MIT-Bag model \cite{Signal:2021aum,Johnson:1975zp}, contact interaction \cite{Zhang:2024adr,Zhang:2020ecj}, lattice simulations \cite{Lin:2017snn,Savage:2011xk,Glozman:2009rn}, etc. 

Using these models, one can understand the internal structure of a hadron through the five-dimensional generalized transverse momentum distributions (GTMDs) \cite{Puhan:2025kzz,Echevarria:2016mrc,Meissner:2009ww,Meissner:2008ay}, three-dimensional generalized parton distribution functions (GPDs) \cite{Diehl:2003ny,Chavez:2021llq,Broniowski:2022iip,Guidal:2004nd}, transverse momentum parton distribution functions (TMDs) \cite{Diehl:2015uka,Angeles-Martinez:2015sea,Pasquini:2008ax,Puhan:2023hio}, one-dimensional parton distribution functions (PDFs) \cite{Collins:1981uw,Martin:1998sq,Gluck:1994uf}, etc. The transverse and spatial structure of the valence quark inside a hadron can be studied using the TMDs and GPDs. TMDs carry information about the transverse structure of the hadron in the form of transverse ($\bfk$) and longitudinal ($x$) momentum of the constituents with no information about the spatial structure, which can be accessed through GPDs as a functions of momentum transferred between final and initial state ($\Delta_\perp$), longitudinal momentum $(x)$ and skewness parameter ($\xi$). Elastic form factors (Eform factors), orbital angular momentum, charge radii, pressure and shear distributions, gravitational form factors, single spin asymmetry, etc of hadrons can be accessed through the GPDs \cite{Freese:2020mcx,Diehl:2003ny,Guidal:2013rya}. GPDs can be extracted directly from deeply virtual Compton scattering (DVCS) \cite{Ji:1996nm,Xie:2023xkz} and deeply virtual meson
production (DVMP) \cite{Favart:2015umi}. TMDs and GPDs are the extended form of one-dimensional PDFs, which carry information about the longitudinal momentum fraction $x$ carried by the constituent from the hadrons. 
PDFs can be extracted experimentally from the deep inelastic scattering process (DIS) \cite{Alekhin:2002fv}. One major issue in hadron physics has been and still is the lack of accurate information about the shapes of the above distribution functions generated from the first principle of quantum chromodynamics (QCD).

In this work, we have calculated the quark GPDs of spin-$1$ vector mesons within the light-cone quark model (LCQM) \cite{Puhan2023,Qian:2008px,Brodsky:2000xy} as they  have higher spin degrees of freedom and more polarizations as compared to spin-$1/2$ nucleons and spin-$0$ mesons. These polarizations occur due to the transverse and longitudinal polarizations of spin projections, resulting in the formation of tensor structures of spin-$1$ mesons. There are a total of nine GPDs present at the leading twist compared to eight of spin-$1/2$ nucleons and two of spin-$0$ mesons \cite{Cano:2003ju}. Out of nine quark GPDs, five are unpolarized and four are polarized quark GPDs for spin-$1$ vector mesons. In the present work, we have limited our work to compute the unpolarized quark GPDs only. However, polarized quark GPDs have been studied in NJL model \cite{Zhang:2022zim} and LFQM \cite{Sun:2018ldr} previously. These quark GPDs have been calculated at zero skewness from the quark-quark correlator and have been presented in the overlap form of light-front wave functions (LFWFs). It is established that the LFWF is Lorentz-invariant and can be represented in terms of the constituents' momentum fractions, which are unaffected by the overall hadron momentum.  Thus, it is quite advantageous to learn how to derive LFWFs from a fundamental standpoint and are essential for our knowledge of the hadron structure and the light-front quantization approach to QCD. 

For the numerical calculations, we have considered both the light $\rho$ and heavy $J/\psi$ vector mesons. Out of five $H_i (x,0,-\Delta_\perp^2)$ ($i=1,..5$) GPDs, $H_4 (x,0,-\Delta_\perp^2)$ comes out to be zero, $H_1$ gives rise to the $F_1(x)$ structure function of the DIS process and $H_5$ carries information about the polarized $b_1(x)$ structure function of the spin-1 mesons. The quark GPDs of vector mesons have been studied in different theoretical models \cite{Kumar:2019eck, Adhikari:2018umb, Zhang:2022zim, Shi:2023oll,Sun:2017gtz,Puhan:2025uat} whereas till now there are no lattice simulations and experimental results available. A total of three form factors are  present for the leading twist unpolarized GPDs case: charge ($G_c$), magnetic ($G_M$), and quadrupole ($G_Q$) form factors \cite{Choi:2004ww}. These form factors encode information about the charge radii, magnetic moment, and quadrupole moment of the mesons. These form factors can be used to calculate the structure functions corresponding to the Rosenbluth cross section \cite{,Hofstadter:1957wk}. There has been a wide range of theoretical studies reported for these form factors \cite{Karmanov:1996qc,Sun:2017gtz,Chung:1988mu,Brodsky:1992px,Frankfurt:1993ut,DeMelo:2018bim,Allahverdiyeva:2023fhn,Kumar:2019eck,Adhikari:2018umb,Zhang:2022zim,Shi:2023oll,Sun:2017gtz,Puhan:2025uat} and few lattice simulation studies \cite{QCDSF:2008tjq,Shultz:2015pfa}.  However, there has been a complete lack of experimental studies for both the particles. Further, for the sake of completeness, we have also predicted the quark PDFs for both the particles. There are a total of four PDFs present at the leading twist for spin-$1$ mesons compared to one for spin-$0$ mesons and three for spin-$1/2$ nucleons \cite{Bacchetta:2000jk}. These are unpolarized $f_1(x)$, transversity $h_1(x)$, helicity $g_1(x)$, and tensor $f_{1LL}$ quark PDFs. The unpolarized $f_1(x)$ and tensor polarized $f_{1LL}(x)$ quark PDFs can be derived from $H_1$ and $H_5$ quark GPDs at the forward limit $\Delta_\perp=0$. However, the $f_{1LL}$ PDF is found to be zero for our case. For numerical calculations, we have adopted the LCQM with Brodsky-Huang-Leapage (BHL) prescription for the momentum space wave function \cite{Lepage:1980fj} and spin wave functions through Melosh-Wigner rotation \cite{Choi:1996mq,Qian:2008px}. 

The paper is arranged as follows. In Sec. \ref{satya}, we have discussed the LCQM along with the spin and momentum wave function in light-front (LF) formalism. The derivation and discussion of unpolarized quark GPDs from quark-quark correlator has been presented in Sec. \ref{gpds}. In this section, we have also derived and discussed the different form factors along with structure functions. In Sec. \ref{pdfs}, we have presented the quark parton distribution functions for spin-$1$ vector mesons. We have finally summarized our work in Sec. \ref{con}.

\section{Methodology}\label{satya}
\subsection{Light-cone quark model}
Understanding hadrons as relativistic bound states of partons is a central goal of modern particle physics. The LF formalism provides an elegant approach to this problem, expressing the hadron’s wave function as a sum over Fock-states defined by the momentum and spin of its constituents. The hadron wave function based on LF quantization QCD using the multi-particle Fock-state expansion is expressed as \cite{Puhan2023, Qian:2008px, Brodsky:2000xy}
\begin{eqnarray*}\label{fockstate}
	|M_{h}(P, S_z) \rangle
	&=&\sum_{n,\lambda_j}\int\prod_{j=1}^n \frac{\mathrm{d} x_j \mathrm{d}^2
		\mathbf{k}_{\perp j}}{\sqrt{x_j}~16\pi^3}
	16\pi^3 ~ \delta\Big(1-\sum_{j=1}^n x_j\Big)\delta^{(2)}\Big(\sum_{j=1}^n \mathbf{k}_{\perp j}\Big) ~\psi_{n/M}(x_j,\mathbf{k}_{\perp j},\lambda_j)   | n ; \mathbf{k}^+_j, \mathbf{k}_{\perp j},
	\lambda_j \rangle.
\end{eqnarray*}
Here, $|M_h(P, S_z) \rangle$ is the hadron eigenstate with $P=(P^+,P^-,P_{\perp})$ and $S_z$ being the hadron's total average momentum and longitudinal spin projection, respectively. $n$ is the number of constituents inside the hadron, $\lambda_j$ is the helicity of the $j$-th constituent of the hadron, $\mathbf{k_j}=(\mathbf{k}^+_j,\mathbf{k}^-_j,\mathbf{k}_{j \perp})$ is the momentum of the $j$-th constituent of the hadron and $x_j=P^+/\mathbf{k}_j^+$ is the longitudinal momentum fraction carried by the $j$-th constituent from the hadron. Both longitudinal momentum fraction and transverse momenta of the constituents of the hadron obey the momentum sum rule
\begin{eqnarray}
	\sum_{j=1}^n \textbf{k}_{j}=0, \ \ \ \ \sum_{j=1}^n x_{j}=1.
\end{eqnarray}
As we are dealing with mesons in this work, the meson Fock-state can be written in terms of quarks, gluons, and sea-quark degrees of freedom as \cite{Lepage:1980fj, Brodsky:1997de, Pasquini:2023aaf}
\begin{eqnarray}
	|M\rangle &=& \sum |q\bar{q}\rangle \psi_{q\bar{q}}
	+ \sum
	|q\bar{q}g\rangle \psi_{q\bar{q}g} + \sum
	|q\bar{q}gg\rangle \psi_{q\bar{q}g g} + |q\bar{q}(q \bar q)_{sea}\rangle \psi_{q\bar{q}(q \bar q)_{sea}} + \cdots  \, ,
\end{eqnarray}
where $|M\rangle$ denotes the meson eigenstate. Since the higher Fock-state contribution is very less compared to $|q\bar{q}\rangle$ leading state for the case of heavy mesons \cite{Shi:2022erw}, we have not considered the higher Fock-state contribution in this work. The two-particle meson minimal Fock-state can be written in terms of the intrinsic momenta of the quark and antiquark from Eq. (\ref{fockstate}) as 
\begin{eqnarray}
	|M(P, S_Z)\rangle &=& \sum_{\lambda_1,\lambda_2}\int
	\frac{\mathrm{d} x \mathrm{d}^2
		\mathbf{k}_{\perp}}{\sqrt{x(1-x)}2 (2 \pi)^3}
	\Psi_{S_Z}(x,\mathbf{k}_{\perp},\lambda_1,\lambda_2)|x,\mathbf{k}_{\perp},
	\lambda_1,\lambda_2 \rangle
	.
	\label{meson}
\end{eqnarray}
Here, $\lambda_{1(2)}$ are the helicities of quark (antiquark), $\Psi_{S_Z}(x,\mathbf{k}_{\perp},\lambda_1,\lambda_2)$ is the LF meson wave function with different spin ($S_z$) and quark (antiquark) helicity ($\lambda$) projections. $x$ and $1-x$ are the momentum fractions carried by the active quark and antiquark, respectively. The LF meson wave function can be expressed as 
\begin{eqnarray}
	\Psi_{S_z}(x,\textbf{k}_\perp, \lambda_1, \lambda_2)= \mathcal{S}_{S_z}(x,\textbf{k}_\perp, \lambda_1, \lambda_2) \psi_{q \bar q}(x, \textbf{k}_\perp).\
	\label{space}
\end{eqnarray}
Here, $\mathcal{S}_{S_z}(x,\textbf{k}_\perp, \lambda_1, \lambda_2)$ and $\psi_{q \bar q}(x, \textbf{k}_\perp)$ are the spin and momentum space wave functions of the mesons respectively. By adopting the Brodsky-Huang-Lepage (BHL) prescription, the momentum space wave function in Eq. (\ref{space}) can be expressed as \cite{Qian:2008px,Xiao:2002iv} 
\begin{eqnarray}
	\psi_{q \bar q}(x,\textbf{k}_\perp)= A \ {\rm exp} \Bigg[-\frac{\frac{\textbf{k}^2_\perp+m_q^2}{x}+\frac{\textbf{k}^2_\perp+m^2_{\bar q}}{1-x}}{8 \beta^2}\Bigg]\, ,
	\label{bhl-k}
\end{eqnarray}
where $m_q$ and $m_{\bar q}$ are the quark and antiquark masses respectively. $\beta$ and $A$ are the harmonic scale parameter and normalization constant of the meson. The quark masses $(m_{q(\bar q)})$ and $\beta$ are calculated by fitting with the mass spectra and decay constant of the respective mesons. For this work, we have adopted the quark mass $m_{u(\bar d)}=0.2$ GeV and $m_{c \bar c}=1.68$ GeV for light $\rho (u \bar d)$ \cite{Kaur:2020emh} and heavy $J/\psi(c \bar c)$ vector mesons \cite{Puhan:2023hio}, respectively. The $\beta$ parameter is taken to be $0.410$ and $0.699$ for $\rho (u \bar d)$ and  $J/\psi(c \bar c)$ vector mesons, respectively \cite{Kaur:2020emh,Puhan:2023hio}. The normalization constant is calculated by normalizing the momentum space wave function as  
\begin{eqnarray}
	\int \frac{dx d^2\bfk}{2 (2\pi)^3}|\psi_{q \bar q}(x,\bfk)|^2=1.
\end{eqnarray}
The four-vector momenta of the  constituent quark ($k_1$) and antiquark ($k_2$) in the LF frame are respectively defined as 
\begin{eqnarray}
	k_1&\equiv&\bigg(x P^+, \frac{\textbf{k}_\perp^2+m_q^2}{x P^+},\textbf{k}_\perp \bigg),\label{n1}\\
	k_2&\equiv&\bigg((1-x) P^+, \frac{\textbf{k}_\perp^2+m_{\bar q}^2}{(1-x) P^+},-\textbf{k}_\perp \bigg),
	\label{n3}
\end{eqnarray}
with $M_{q \bar q}$ being the invariant mass of the composite meson system defined in terms of its quark mass $m_q$ and antiquark mass $m_{\bar q}$ as
\begin{eqnarray}
	M_{q \bar q}^2=\frac{\bfk^2+m^2_q}{x} +\frac{\bfk^2+m^2_{\bar q}}{1-x}\, .
\end{eqnarray}
$\mathcal{S}_{S_z}(x,\bfk,\lambda_1,\lambda_2)$ in Eq. (\ref{bhl-k}) is front-form spin wave function derived from the instant form by Melosh-Wigner rotation \cite{Qian:2008px,Xiao:2002iv,Kaur:2020vkq}.
This transformation of instant form state $\Phi(T)$ and front form state $\Phi(F)$ is expressed as 
\begin{eqnarray}
	\Phi_i^\uparrow(T)&=&-\frac{[\textbf{k}_i^R \Phi_i^\downarrow(F)-(\textbf{k}_i^+ +m_{q(\bar q)})\Phi_i^\uparrow(F)]}{\omega_{i(q \bar q)}},\label{instant-front1}\\
	\Phi_i^\downarrow(T)&=&\frac{[\textbf{k}_i^L\Phi_i^\uparrow(F)+(\textbf{k}_i^+ +m_{q(\bar q)})\Phi_i^\downarrow(F)]}{\omega_{i (q \bar q)}}.
	\label{instant-front}
\end{eqnarray}
Here, $\Phi(F)$ is a two-component Dirac spinor and $\textbf{k}_i^{R(L)}=\textbf{k}_i^1 \pm \iota \textbf{k}_i^2$. $\omega_{i(q \bar q)}$ is defined as $\omega_{i(q \bar q)}=1/ \sqrt{2 \textbf{k}^+_i (\textbf{k}^0+m_{q(\bar q)})}$. Now applying different momenta forms from Eqs. (\ref{n1})-(\ref{n3}) in  Melosh-Wigner rotation, the spin wave function is obtained in the form of $\kappa_{S_z}^F(x,\textbf{k}_\perp, \lambda_1, \lambda_2)$ coefficient as 
\begin{equation}
	\mathcal{S}_{S_z}(x,\textbf{k}_\perp, \lambda_1, \lambda_2)=\sum_{\lambda_1, \lambda_2}\kappa_{S_z}^F(x,\textbf{k}_\perp, \lambda_1, \lambda_2) \Phi_1^{\lambda_1}(F) \Phi_2^{\lambda_2}(F).
\end{equation}
These spin-wave function coefficients satisfy the following normalization relation
\begin{eqnarray}
	\sum_{\lambda_1,\lambda_2} \kappa_{S_z}^{F*}(x, \textbf{k}_\perp, \lambda_1, \lambda_2)\kappa_{S_z}^F(x, \textbf{k}_\perp, \lambda_1, \lambda_2)=1.
\end{eqnarray}
Similarly, the same spin-wave function can be calculated using the proper vertex chosen for the vector meson \cite{Choi:1996mq,Qian:2008px} as
\begin{equation}
	\mathcal{S}_{S_z}(x,\textbf{k}_\perp, \lambda_1, \lambda_2) = \bar u (k_1,\lambda_1) \Bigg[-\frac{1}{\sqrt{2}
		\sqrt{M_{q \bar q}^2-(m_q-m_{\bar q})^2}}(\gamma^\mu-\frac{k_1^\mu-k_2^\mu}{M_{q\bar q}+m_q+m_{\bar q}})\epsilon(P,S_z) \Bigg]v(k_2,\lambda_2) \, .
\end{equation}
Here, $u$ and $v$ are the Dirac spinors. In this work, we have only considered the light $\rho(u \bar d)$ vector and heavy $J/\psi(c \bar c)$ vector mesons, so $m_q=m_{\bar q}$ for this case. $\epsilon(P,S_z)$ is the polarization vector of the vector mesons, which can be expressed for both transverse ($S_z=\pm 1$)and longitudinal polarization ($S_z=0$) as 
\begin{eqnarray}
	\epsilon(P,S_z=\pm 1) & =& ( \epsilon_{\pm 1}^+,  \epsilon_{\pm 1}^-, \epsilon_{\pm 1}^{\perp})  =  \Bigg(0, \frac{\mp \sqrt{2}P^{R,L}}{P^+}, \mp \frac{1}{\sqrt{2}},-\frac{\iota}{\sqrt{2}}\Bigg), 
	\nonumber\\
	\epsilon(P,S_z=0)&=&( \epsilon_{0}^+, \epsilon_{0}^-, \epsilon_{0}^{\perp}) = \frac{1}{M_{q \bar q}}\Bigg(P^+, \frac{P_\perp^2-M^2_{q \bar q}}{P^+}, P_\perp \Bigg).
	\label{polar}
\end{eqnarray}
Here, $P^{R,L}=P_{1}\pm i P_2$. Both the above methods give rise to the same form of spin wave function. The spin wave functions for vector mesons for different spin projections/polarization and different helicities of quark antiquark ($\lambda=(\uparrow,\downarrow)$) are expressed for $S_z=+1$ as \cite{Qian:2008px} 
\begin{equation}
	\begin{array}{lll}
		\mathcal{S}_{(S_z=+1)}
		(x,\mathbf{k}_\perp,\uparrow,\uparrow)&=&\frac{1}{\omega_{q \bar q}}[\mathbf{k}_\perp^2+(M_{q \bar q}+m_q+m_{\bar q})((1-x)m_q+x m_{\bar q})],\\
		\mathcal{S}_{(S_z=+1)}(x,\mathbf{k}_\perp,\uparrow,\downarrow)&=&\frac{1}{\omega_{q \bar q}}[(\textbf{k}_1 + \iota \textbf{k}_2)(xM_{q \bar q}+m_q)],\\
		\mathcal{S}_{(S_z=+1)}(x,\mathbf{k}_\perp,\downarrow,\uparrow)&=&\frac{1}{\omega_{q \bar q}}[-(\textbf{k}_1 + \iota \textbf{k}_2)((1-x)M_{q \bar q}+m_{\bar q})],\\
		\mathcal{S}_{(S_z=+1)}(x,\mathbf{k}_\perp,\downarrow,\downarrow)&=&\frac{1}{\omega_{q \bar q}}[-(\textbf{k}_1 + \iota \textbf{k}_2)^2].
	\end{array}
\end{equation}
For longitudinal spin projection $S_z$=0, the spin wave functions are
\begin{equation}
	\begin{array}{lll}
		\mathcal{S}_{(S_z=0)}(x,\mathbf{k}_\perp,\uparrow,\uparrow)&=&\frac{1}{\sqrt{2}}\frac{1}{\omega_{q \bar q}}[(\textbf{k}_1-\iota \textbf{k}_2)((1-2x)M_{q \bar q}+(m_{\bar q}-m_q))],\\
		\mathcal{S}_{(S_z=0)}(x,\mathbf{k}_\perp,\uparrow,\downarrow)
		&=& \frac{1}{\sqrt{2}}\frac{1}{\omega_{q \bar q}}[2 \mathbf{k}_\perp^2 + (M_{q \bar q}+m_q+m_{\bar q})((1-x)m_q+x m_{\bar q})],\\
		\mathcal{S}_{(S_z=0)}(x,\mathbf{k}_\perp,\downarrow,\uparrow)
		&=& \frac{1}{\sqrt{2}}\frac{1}{\omega_{q \bar q}}[2 \mathbf{k}_\perp^2 + (M_{q \bar q}+m_q+m_{\bar q})((1-x)m_q+x m_{\bar q})],\\
		\mathcal{S}_{(S_z=0)}(x,\mathbf{k}_\perp,\downarrow,\downarrow)&=&\frac{1}{\sqrt{2}}\frac{1}{\omega_{q \bar q}}[(-\textbf{k}_1 - \iota \textbf{k}_2)((1-2x)M_{q \bar q}+(m_{\bar q}-m_q))].
	\end{array}
\end{equation}
Finally, the spin wave functions for transverse spin projection $S_z=-1$ are
\begin{equation}
	\begin{array}{lll}
		\mathcal{S}_{(S_z=-1)}(x,\mathbf{k}_\perp,\uparrow,\uparrow)&=&\frac{1}{\omega_{q \bar q}}[-(\textbf{k}_1-\iota \textbf{k}_2)^2],\\
		\mathcal{S}_{(S_z=-1)}(x,\mathbf{k}_\perp,\uparrow,\downarrow)&=&\frac{1}{\omega_{q \bar q}}[(\textbf{k}_1-\iota \textbf{k}_2)((1-x)M_{q \bar q}+m_{\bar q})],\\
		\mathcal{S}_{(S_z=-1)}(x,\mathbf{k}_\perp,\downarrow,\uparrow)&=&\frac{1}{\omega_{q \bar q}}[-(\textbf{k}_1-\iota \textbf{k}_2)(xM_{q \bar q}+m_q)],\\
		\mathcal{S}_{(S_z=-1)}(x,\mathbf{k}_\perp,\downarrow,\downarrow)&=&\frac{1}{\omega_{q \bar q}}[\mathbf{k}_\perp^2+(M_{q \bar q}+m_q+m_{\bar q})((1-x)m_q+x
		m_{\bar q})].
	\end{array}
\end{equation}
Here, $\omega_{q \bar q}=(M_{q \bar q}+m_q+m_{\bar q})\sqrt{x(1-x)[M_{q \bar q}^2-(m_q-m_{\bar q})^2]}$. The two-particle Fock-state in Eq. (\ref{meson}) can be written in the form of LFWFs with all possible helicities of its constituent quark and antiquark for vector mesons as
\begin{eqnarray}
	\ket{M (P^+,\textbf{P}_\perp,S_z)}&=&\int \frac{{ {\rm d}x  \rm d}^2\textbf{k}_\perp}{2 (2 \pi)^3 \sqrt{x(1-x)}}\big[\Psi_{S_z}(x,\textbf{k}_\perp, \uparrow, \uparrow)\ket{x P^+, \textbf{k}_\perp, \uparrow, \uparrow} \nonumber\\
	&&+\Psi_{S_Z}(x,\textbf{k}_\perp, \downarrow, \downarrow)\ket{x P^+, \textbf{k}_\perp, \downarrow, \downarrow}+\Psi_{S_Z}(x,\textbf{k}_\perp, \downarrow, \uparrow)\nonumber\\
	&&\ket{x P^+, \textbf{k}_\perp, \downarrow, \uparrow}+\Psi_{S_Z}(x,\textbf{k}_\perp, \uparrow, \downarrow)\ket{x P^+, \textbf{k}_\perp, \uparrow, \downarrow}\big].
	\label{overlap}
\end{eqnarray}

\begin{figure*}
	\centering
	\begin{minipage}[c]{0.98\textwidth}
		(a)\includegraphics[width=7.5cm]{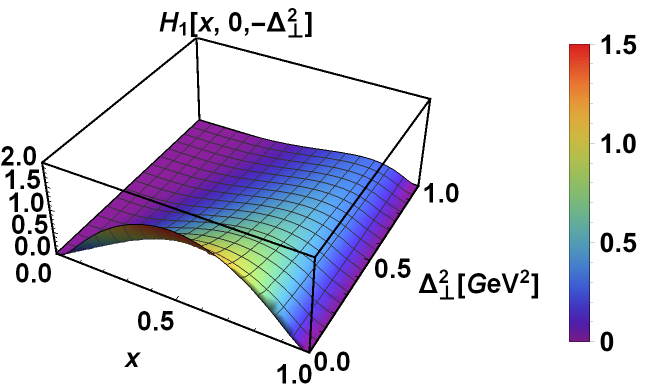}
		\hspace{0.03cm}	
		(b)\includegraphics[width=7.5cm]{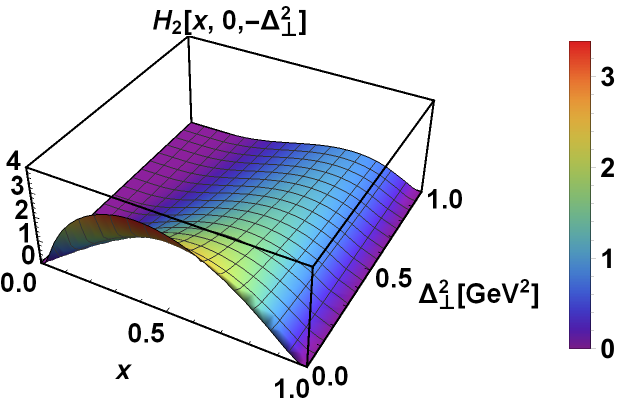} 
		\hspace{0.03cm}
	\end{minipage}
	\centering
	\begin{minipage}[c]{0.98\textwidth}
		(c)\includegraphics[width=7.5cm]{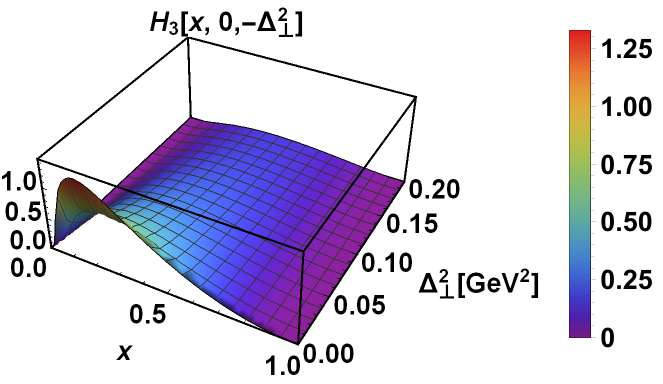}
		\hspace{0.03cm}	
	\end{minipage}
	\caption{\label{fig1} (Color online) The three-dimensional quark GPDs have been plotted with respect to longitudinal momentum fraction ($x$) and transverse momentum transferred $\Delta_\perp^2$ (GeV$^2$) for $\rho$-meson in (a) for $H_1(x,0,-\Delta_\perp^2)$, (b) for $H_2(x,0,-\Delta_\perp^2)$, and (c) for $H_3(x,0,-\Delta_\perp^2)$.}
\end{figure*}
\begin{figure*}
	\centering
	\begin{minipage}[c]{0.98\textwidth}
		(a)\includegraphics[width=7.5cm]{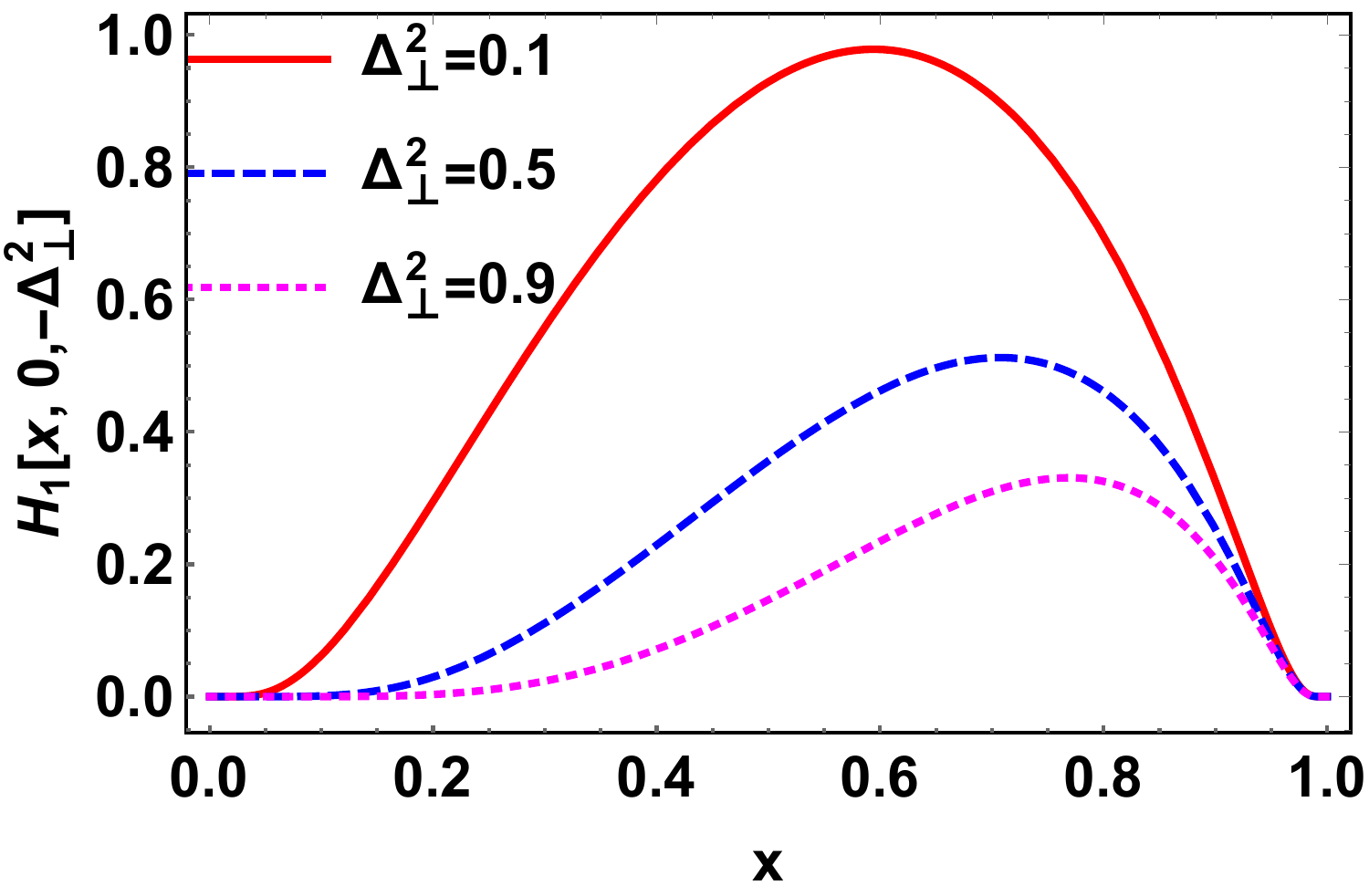}
		\hspace{0.03cm}	
		(b)\includegraphics[width=7.5cm]{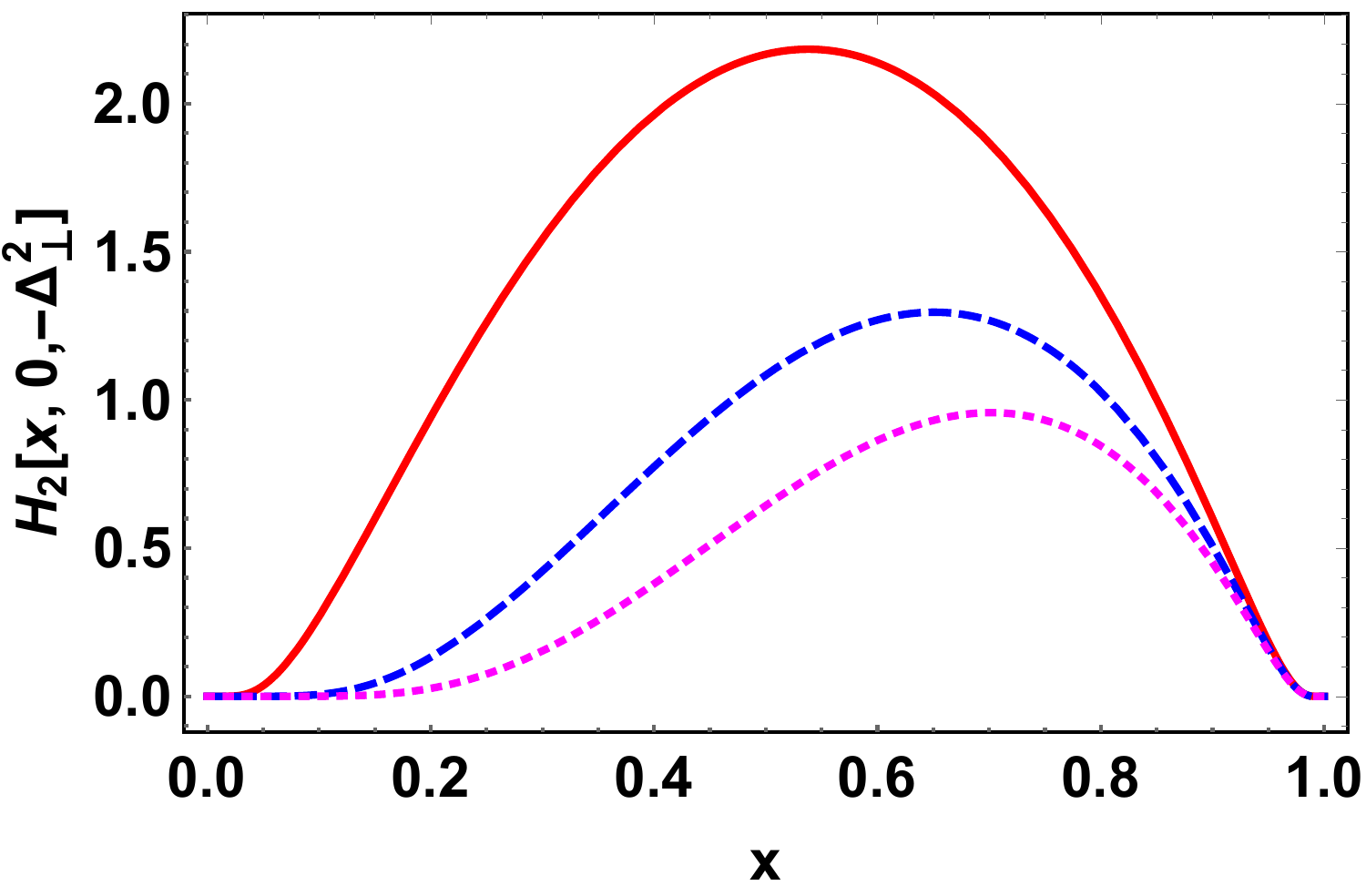} 
		\hspace{0.03cm}
	\end{minipage}
	\centering
	\begin{minipage}[c]{0.98\textwidth}
		(c)\includegraphics[width=7.5cm]{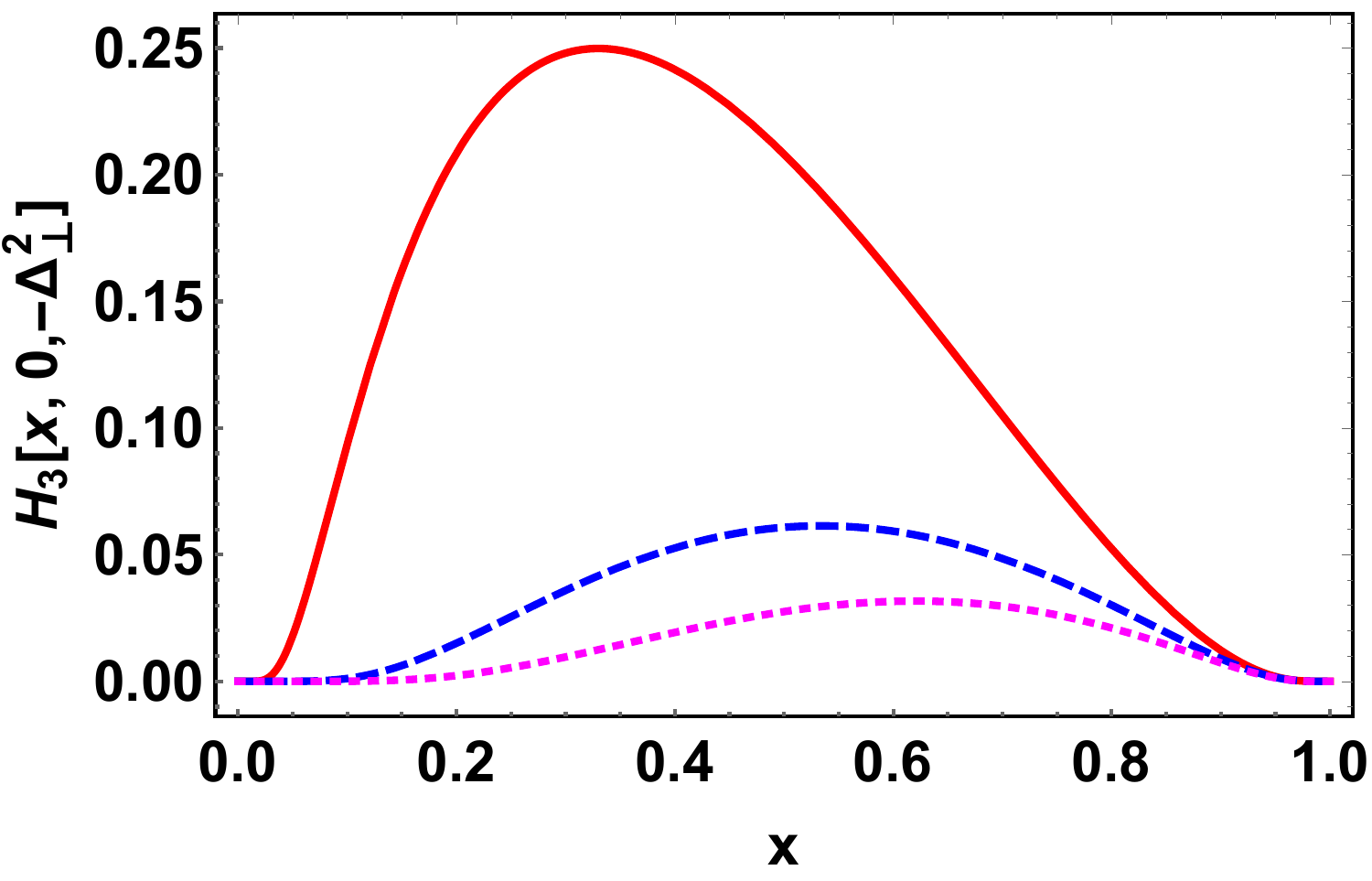}
		\hspace{0.03cm}	
		(d)\includegraphics[width=7.5cm]{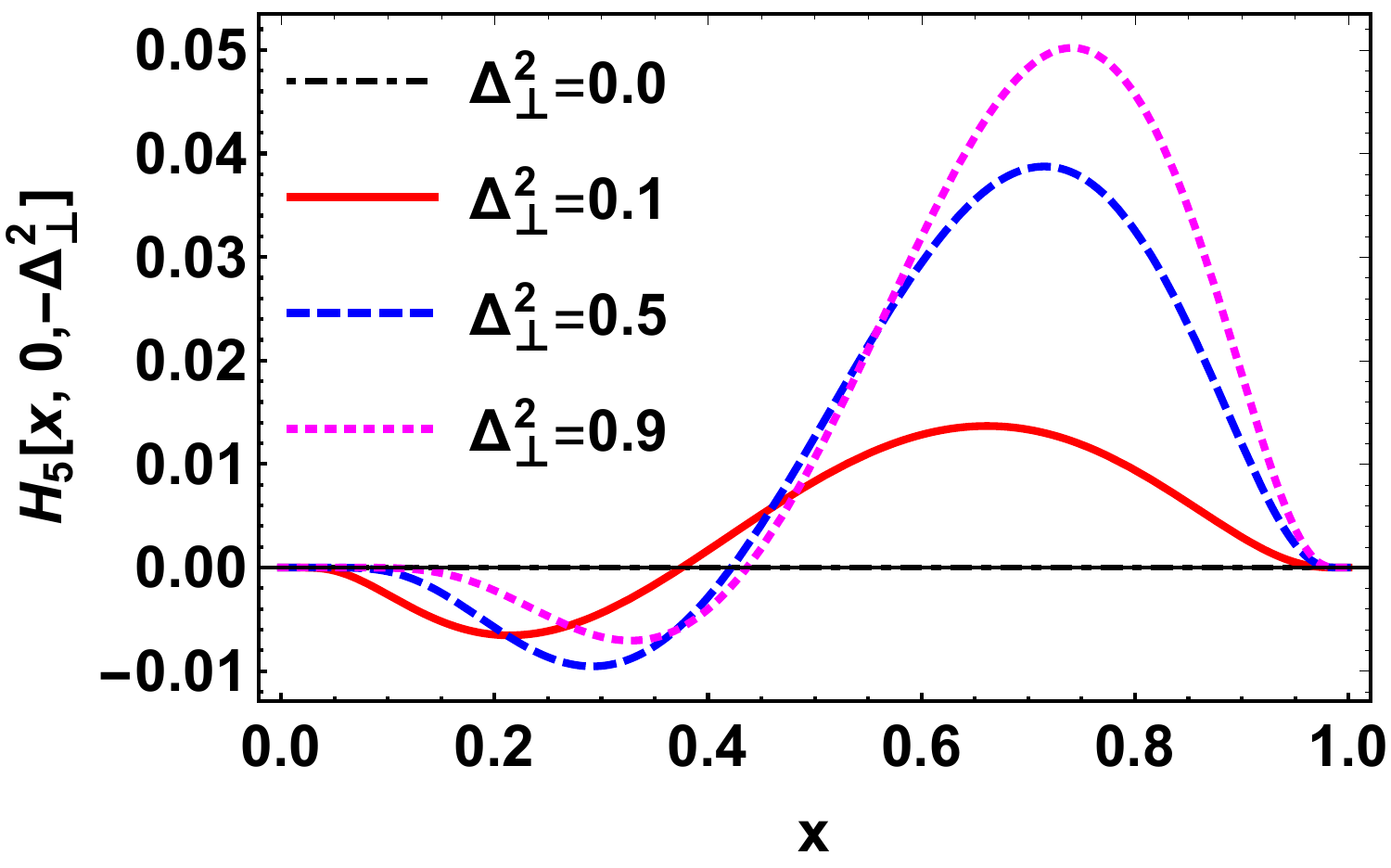} 
		\hspace{0.03cm}
	\end{minipage}
	\caption{\label{fig2} (Color online) The quark GPDs of $\rho$-meson (a) $H_1(x,0,-\Delta^2_\perp)$, (b) $H_2(x,0,-\Delta^2_\perp)$, (c) $H_3(x,0,-\Delta^2_\perp)$, and (d) $H_5(x,0,-\Delta^2_\perp)$ have been plotted with respect to longitudinal momentum fraction ($x$) at fixed value of $\Delta_\perp^2=0$, $0.1$, $0.2$ and $0.9$ GeV$^2$.}
\end{figure*}
\begin{figure*}
	\centering
	\begin{minipage}[c]{0.98\textwidth}
		(a)\includegraphics[width=7.5cm]{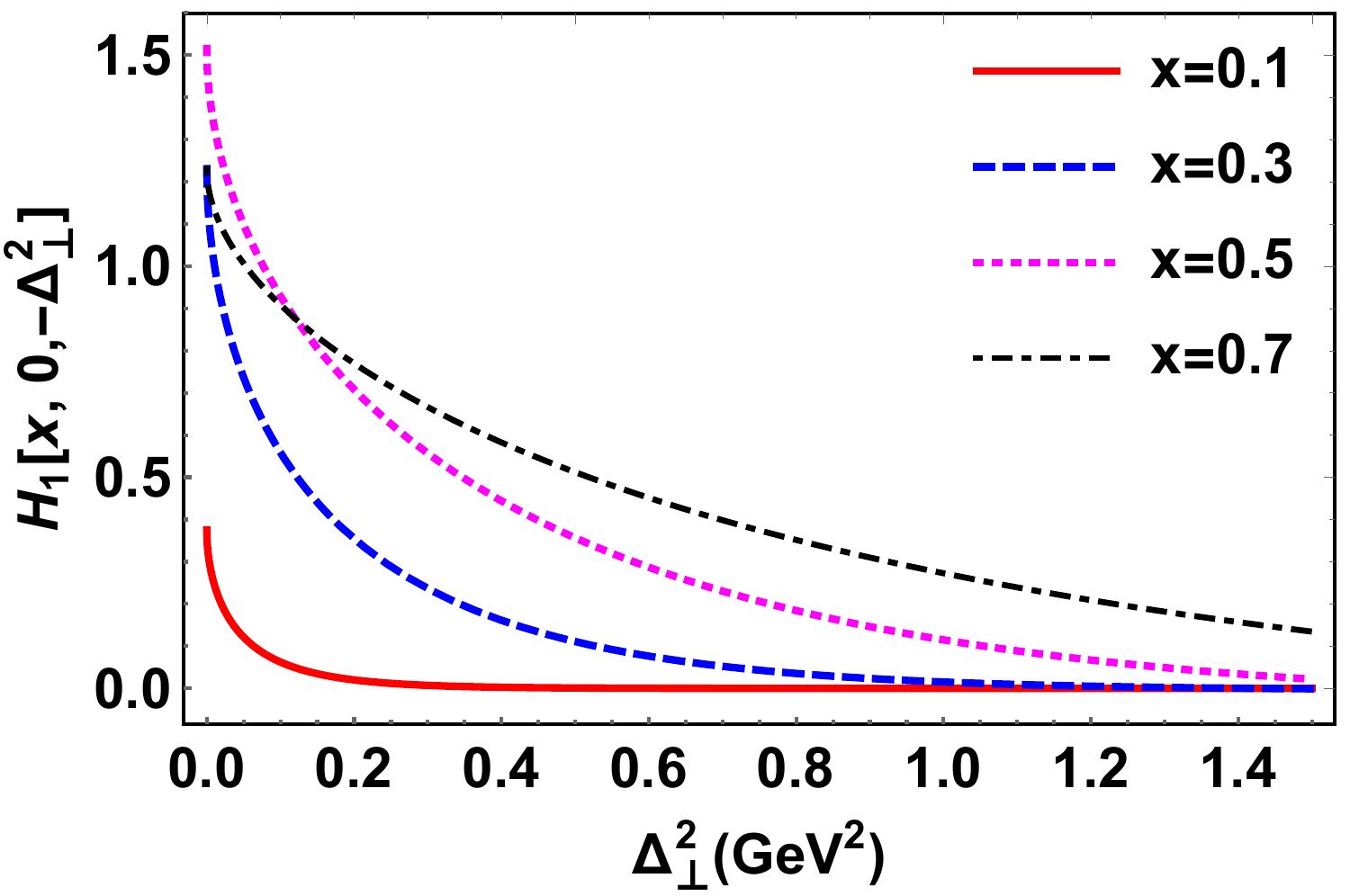}
		\hspace{0.03cm}	
		(b)\includegraphics[width=7.5cm]{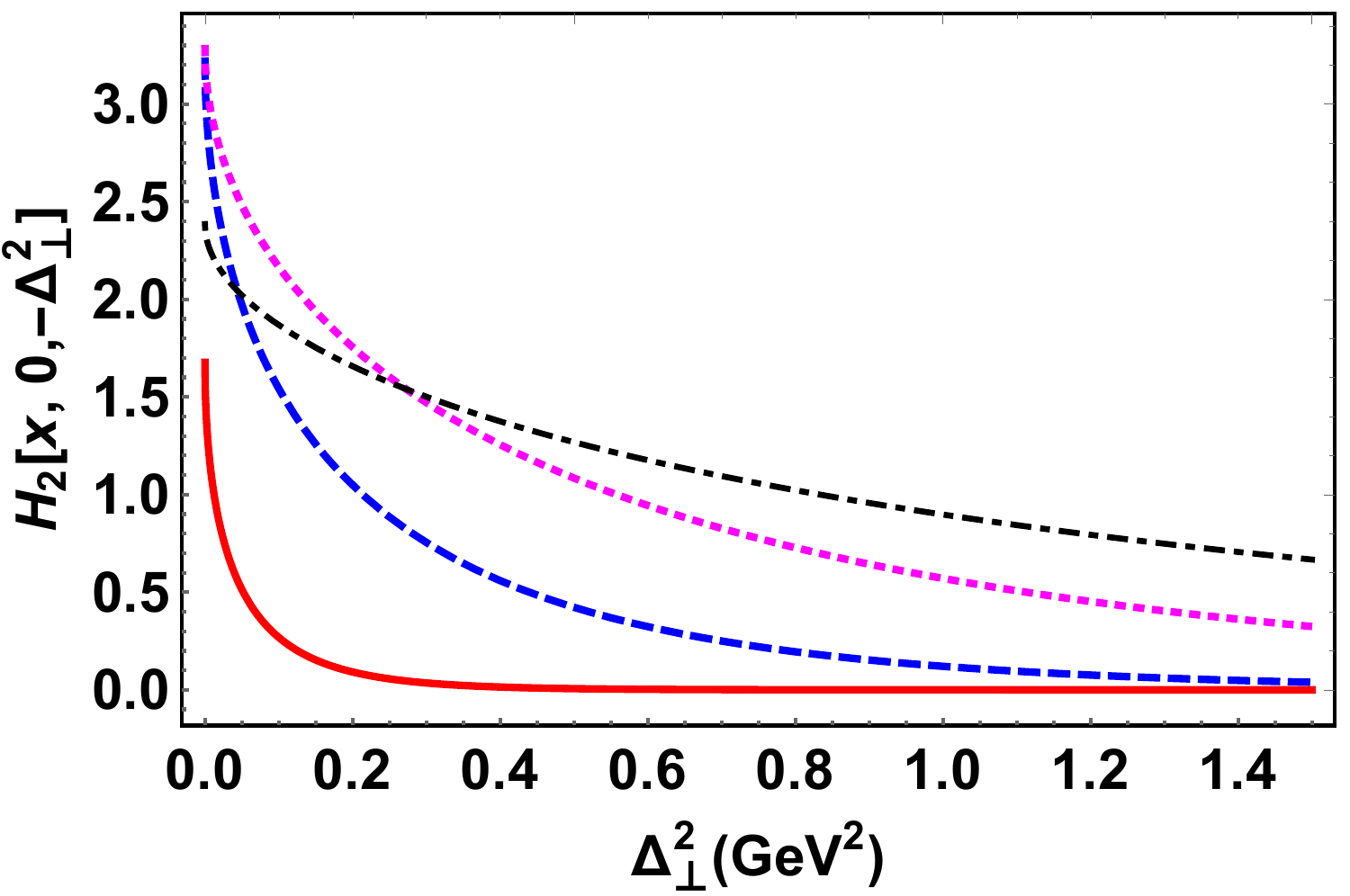} 
		\hspace{0.03cm}
	\end{minipage}
	\centering
	\begin{minipage}[c]{0.98\textwidth}
		(c)\includegraphics[width=7.5cm]{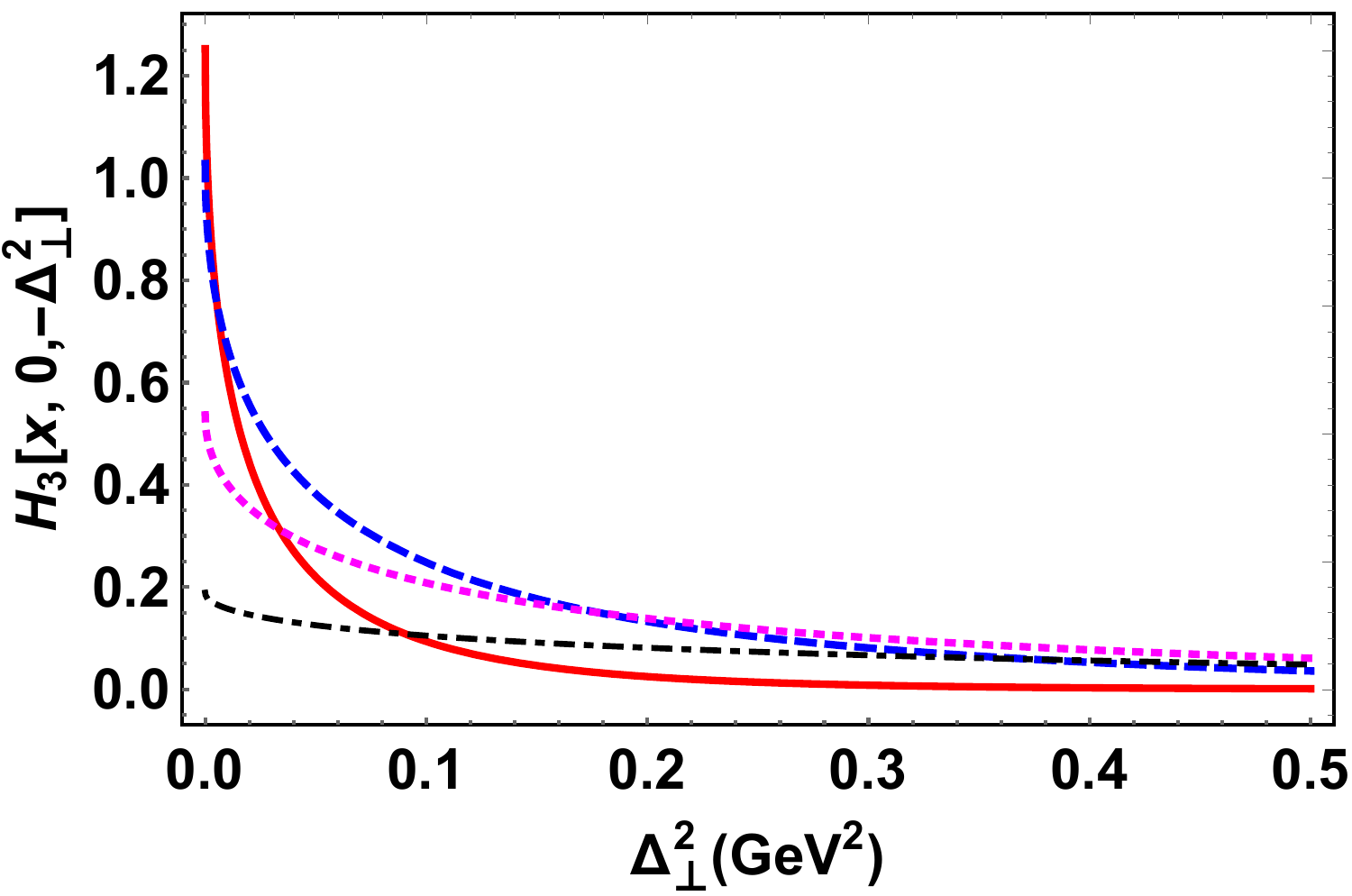}
		\hspace{0.03cm}	
		(d)\includegraphics[width=7.5cm]{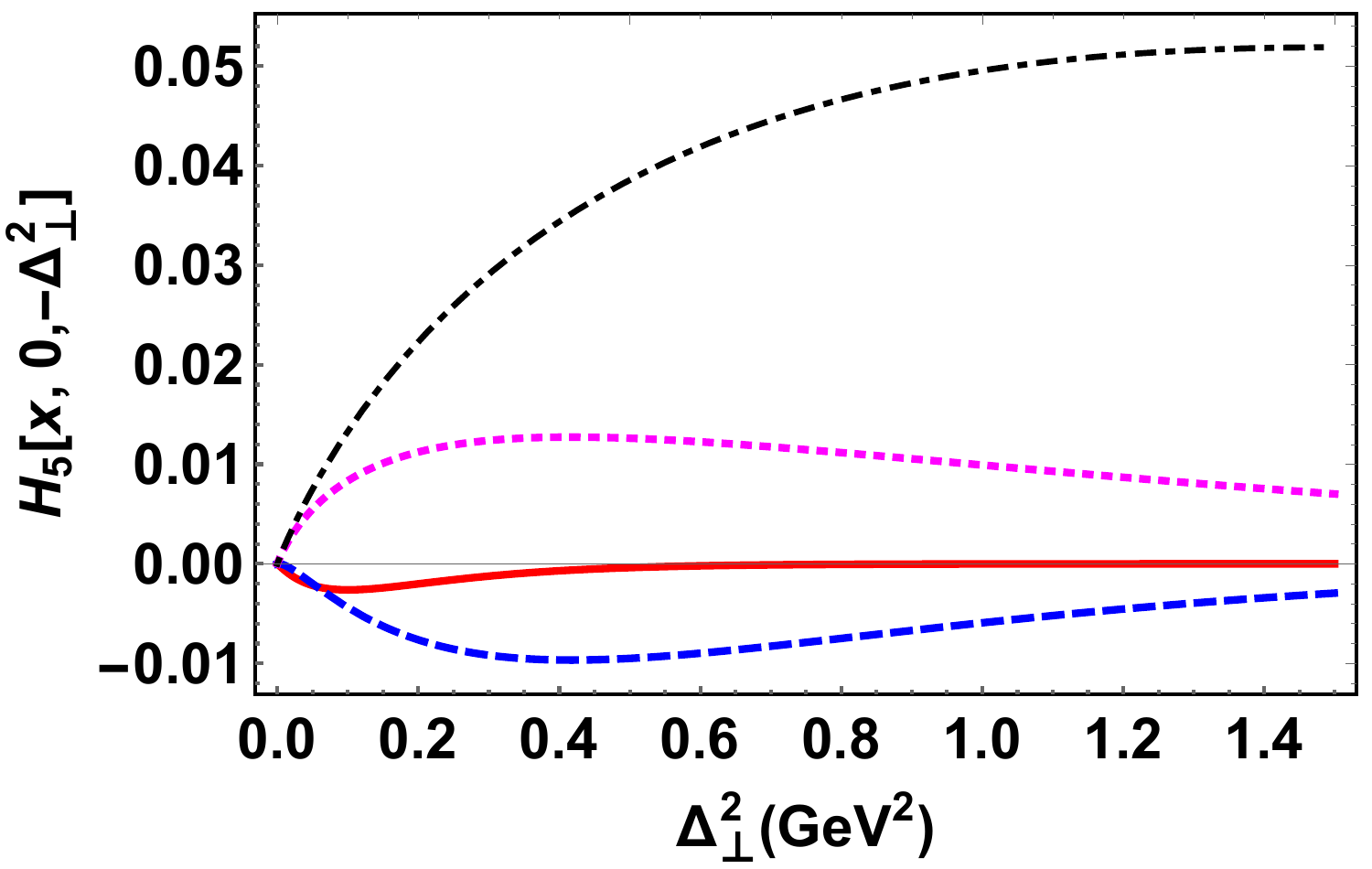} 
		\hspace{0.03cm}
	\end{minipage}
	\caption{\label{fig3} (Color online) The quark GPDs of $\rho$-meson (a) $H_1(x,0,-\Delta^2_\perp)$, (b) $H_2(x,0,-\Delta^2_\perp)$, (c) $H_3(x,0,-\Delta^2_\perp)$, and (d) $H_5(x,0,-\Delta^2_\perp)$  have been plotted with respect to $\Delta_\perp^2$ at fixed value of longitudinal momentum fraction $x=0.1$, $0.3$, $0.5$ and $0.7$.}
\end{figure*}
\begin{figure*}
	\centering
	\begin{minipage}[c]{0.98\textwidth}
		(a)\includegraphics[width=7.5cm]{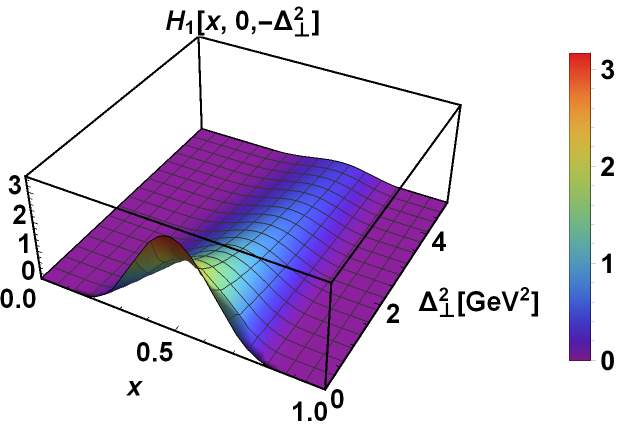}
		\hspace{0.03cm}	
		(b)\includegraphics[width=7.5cm]{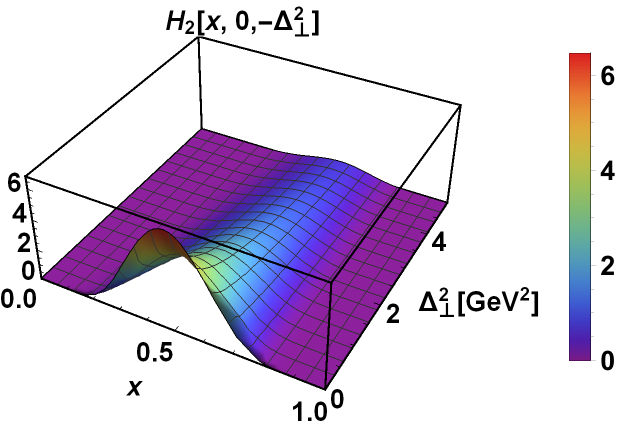} 
		\hspace{0.03cm}
	\end{minipage}
	\centering
	\begin{minipage}[c]{0.98\textwidth}
		(c)\includegraphics[width=7.5cm]{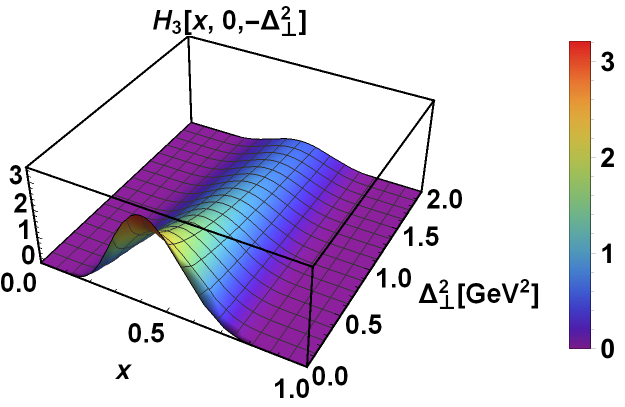}
		\hspace{0.03cm}	
	\end{minipage}
	\caption{\label{fig4} (Color online) The three-dimensional quark GPDs have been plotted with respect to longitudinal momentum fraction ($x$) and transverse momentum transferred $\Delta_\perp^2$ (GeV$^2$) for $J/\psi$-meson in (a) for $H_1(x,0,-\Delta_\perp^2)$, (b) for $H_2(x,0,-\Delta_\perp^2)$, and (c) for $H_3(x,0,-\Delta_\perp^2)$.}
\end{figure*} 
\begin{figure*}
	\centering
	\begin{minipage}[c]{0.98\textwidth}
		(a)\includegraphics[width=7.5cm]{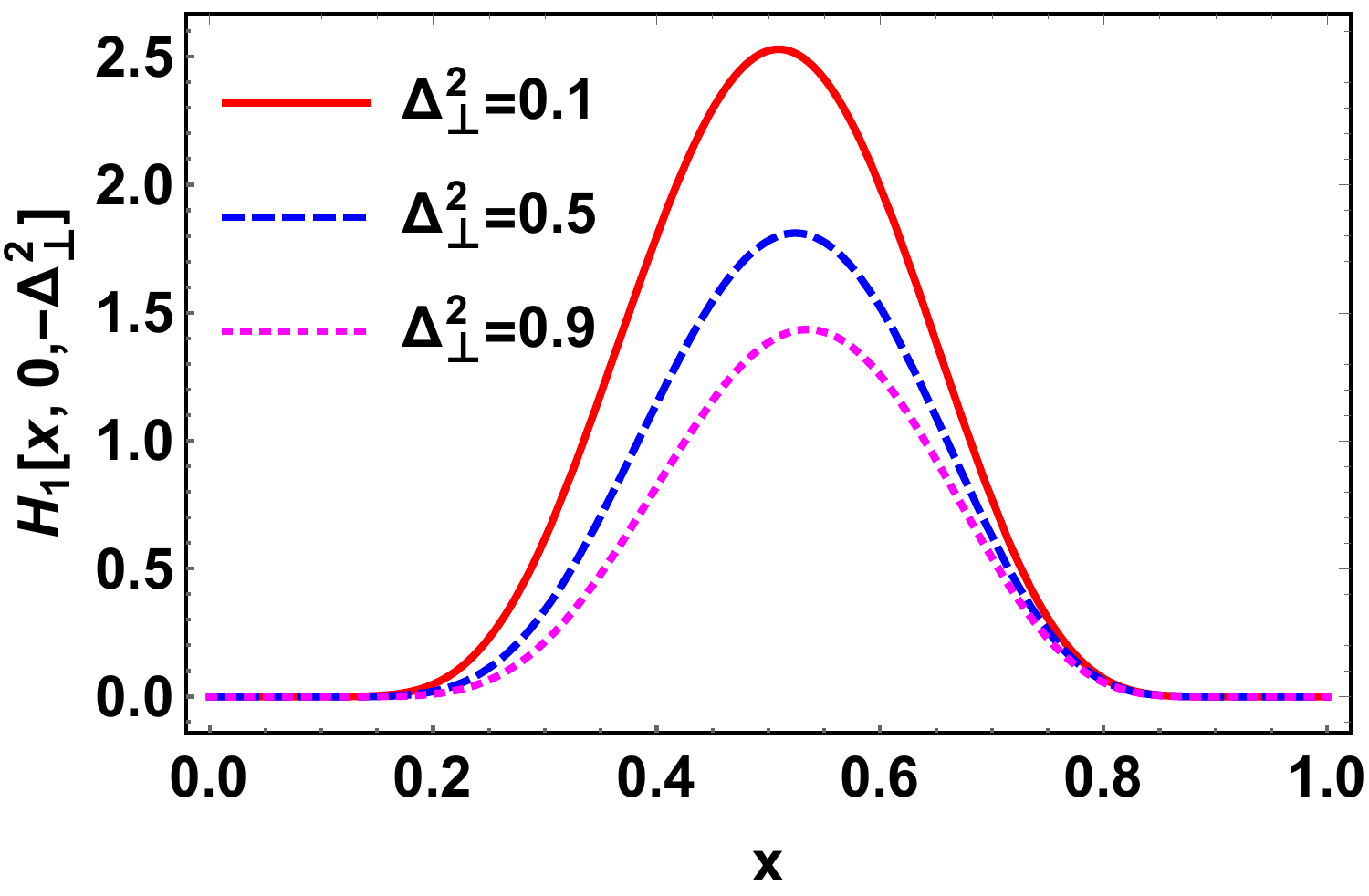}
		\hspace{0.03cm}	
		(b)\includegraphics[width=7.5cm]{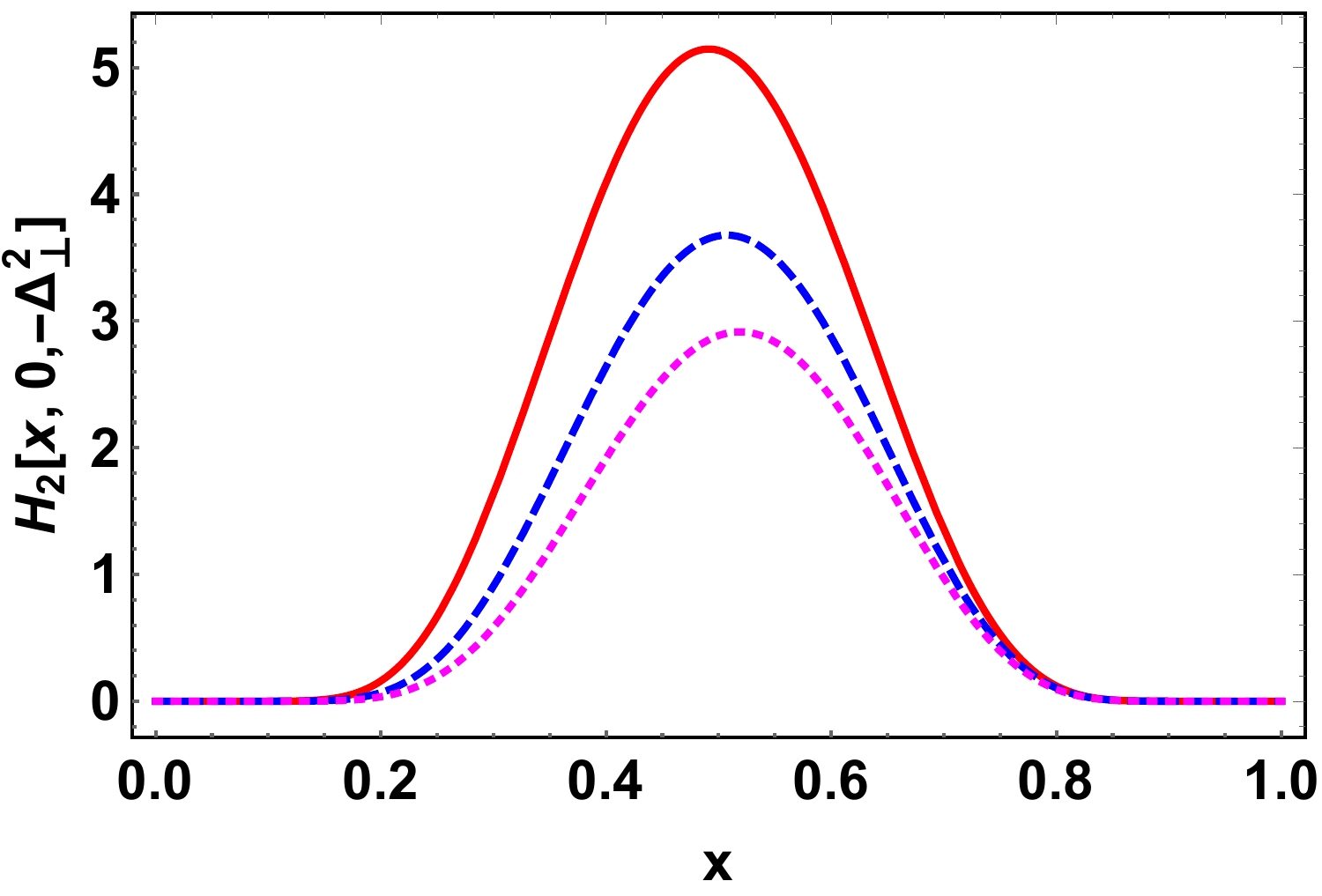} 
		\hspace{0.03cm}
	\end{minipage}
	\centering
	\begin{minipage}[c]{0.98\textwidth}
		(c)\includegraphics[width=7.5cm]{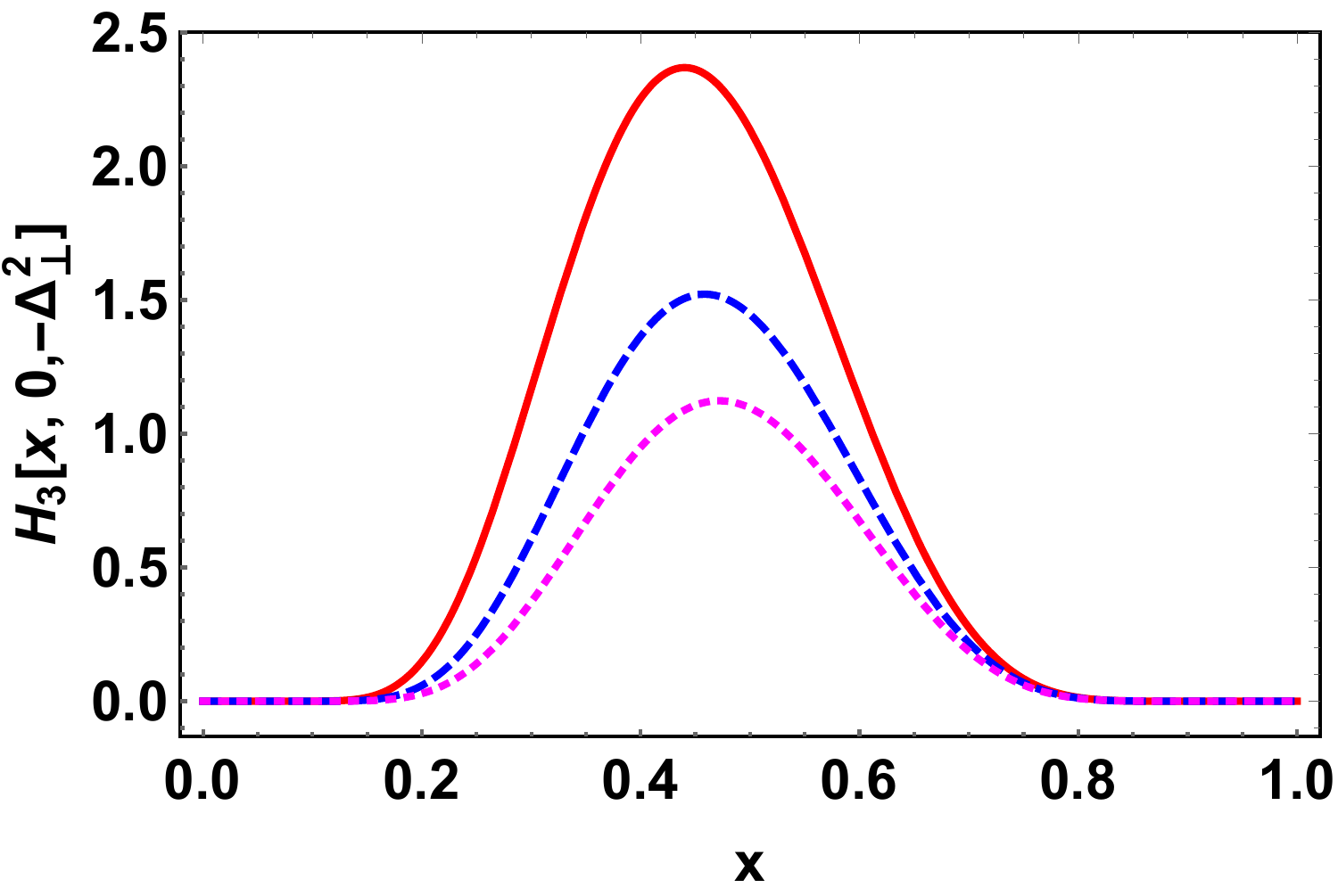}
		\hspace{0.03cm}	
		(d)\includegraphics[width=7.5cm]{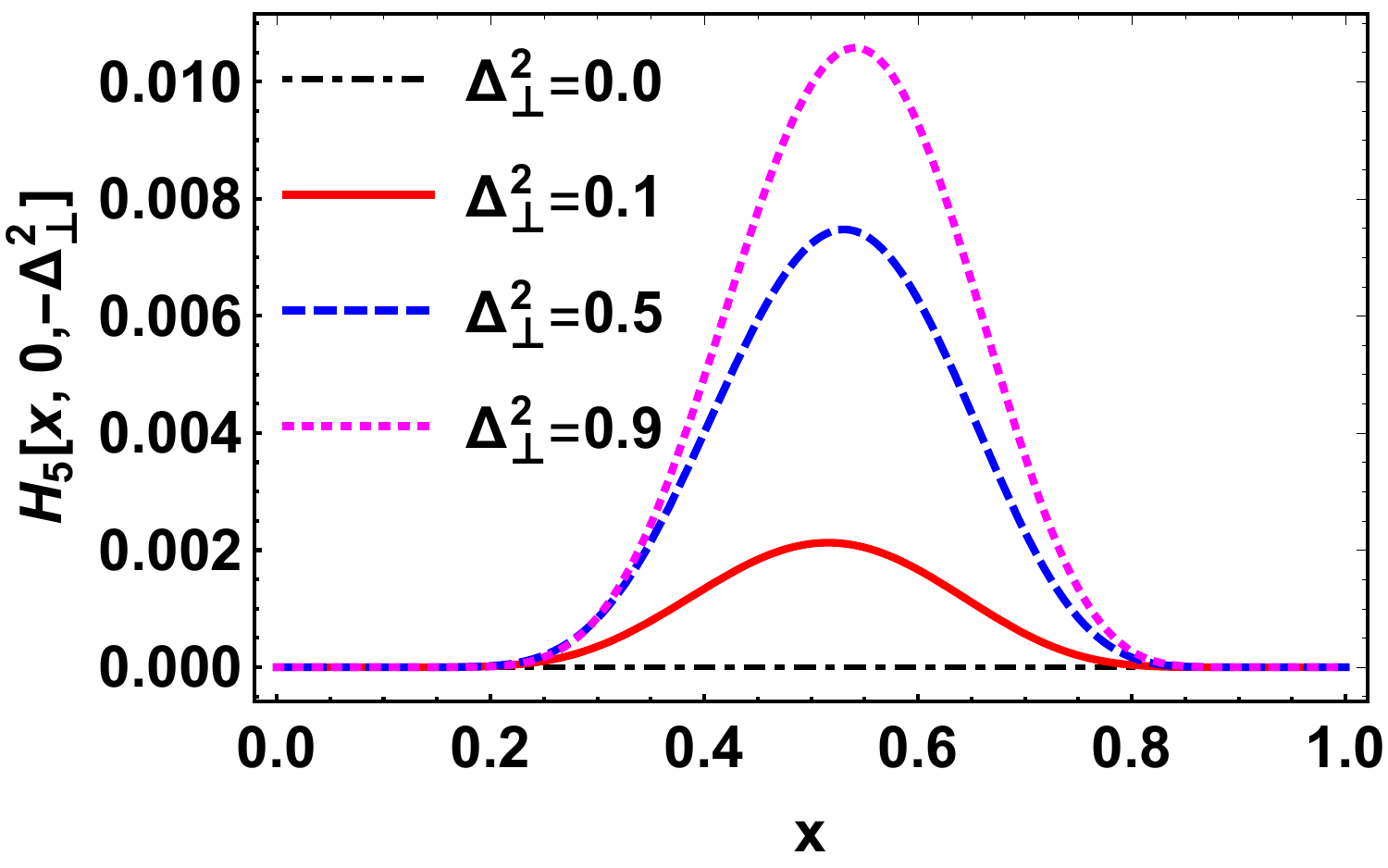} 
		\hspace{0.03cm}
	\end{minipage}
	\caption{\label{fig5} (Color online) The quark GPDs of $J/\psi$-meson (a) $H_1(x,0,-\Delta^2_\perp)$, (b) $H_2(x,0,-\Delta^2_\perp)$, (c) $H_3(x,0,-\Delta^2_\perp)$, and (d) $H_5(x,0,-\Delta^2_\perp)$  have been plotted with respect to longitudinal momentum fraction ($x$) at fixed value of $\Delta_\perp^2=0.1$, $0.2$ and $0.9$ GeV$^2$.}
\end{figure*}
\begin{figure*}
	\centering
	\begin{minipage}[c]{0.98\textwidth}
		(a)\includegraphics[width=7.5cm]{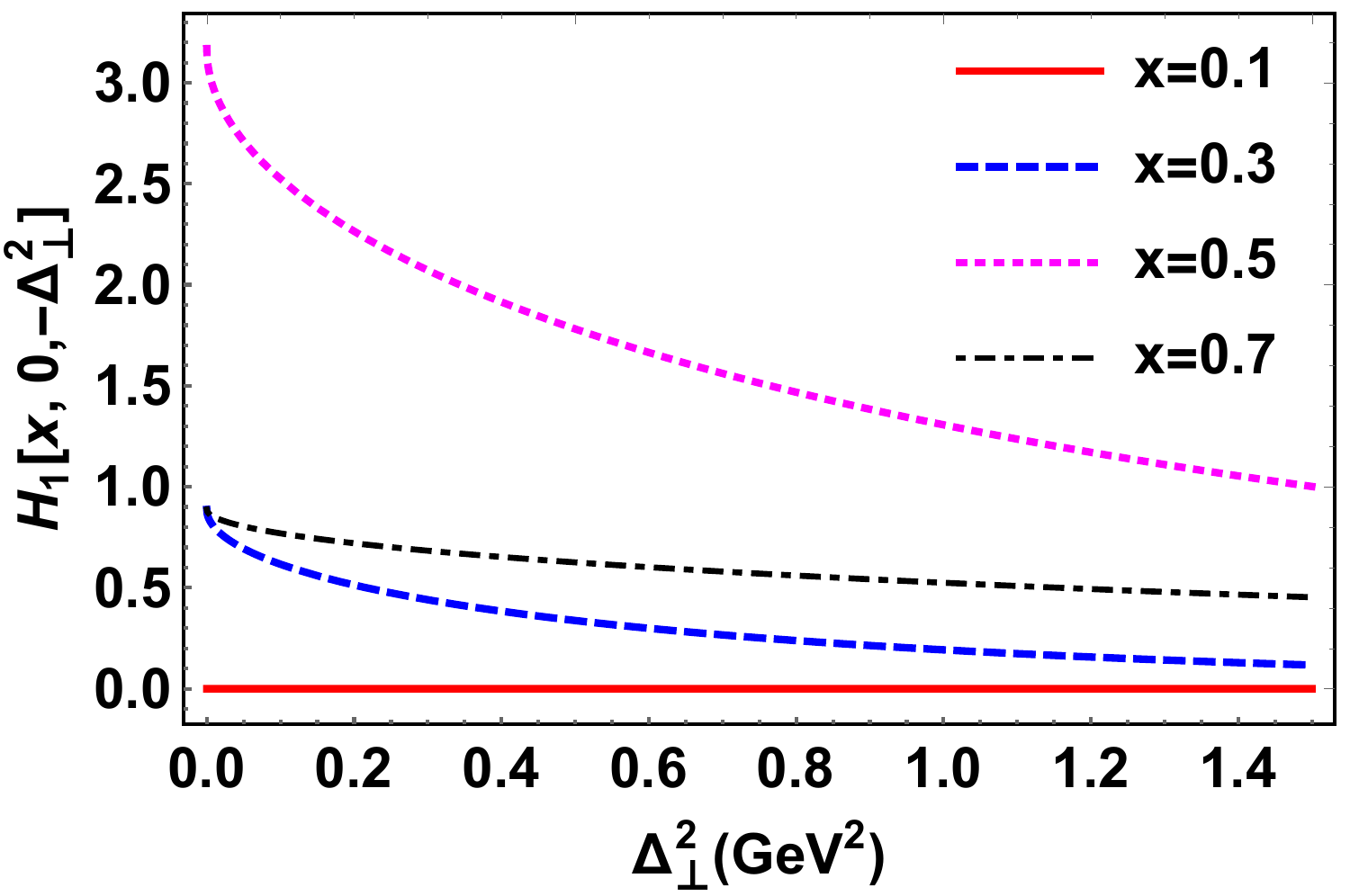}
		\hspace{0.03cm}	
		(b)\includegraphics[width=7.5cm]{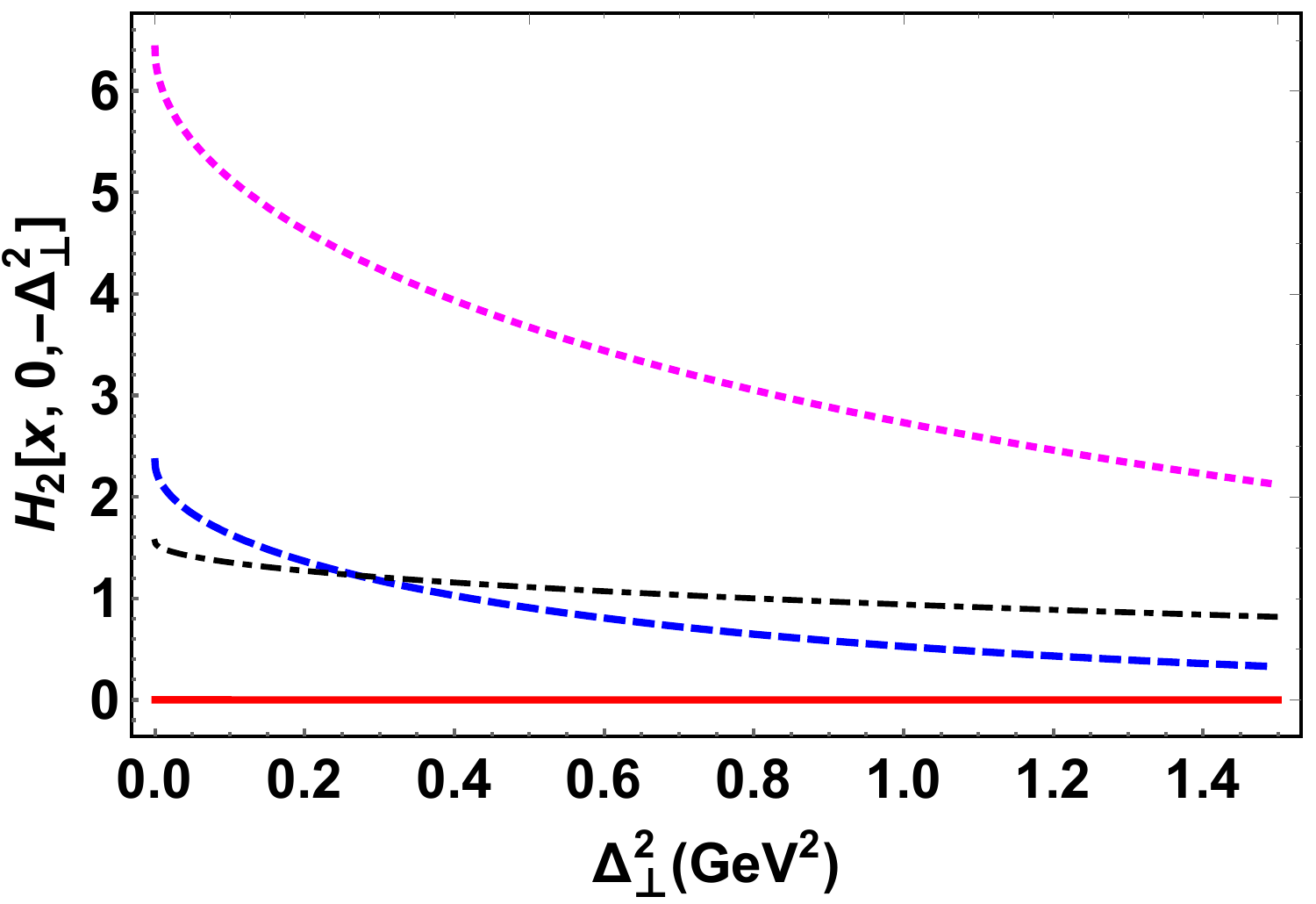} 
		\hspace{0.03cm}
	\end{minipage}
	\centering
	\begin{minipage}[c]{0.98\textwidth}
		(c)\includegraphics[width=7.5cm]{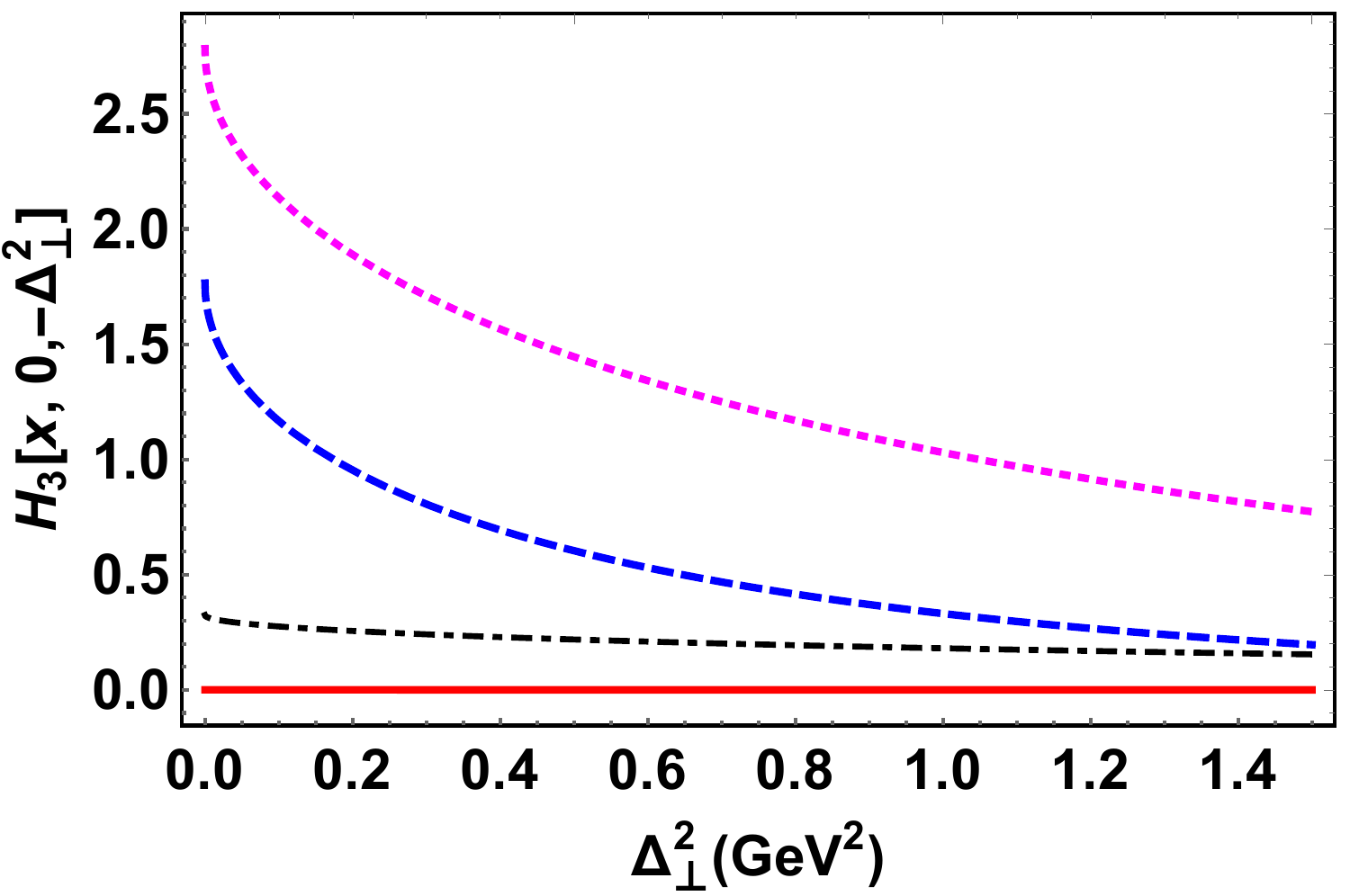}
		\hspace{0.03cm}	
		(d)\includegraphics[width=7.5cm]{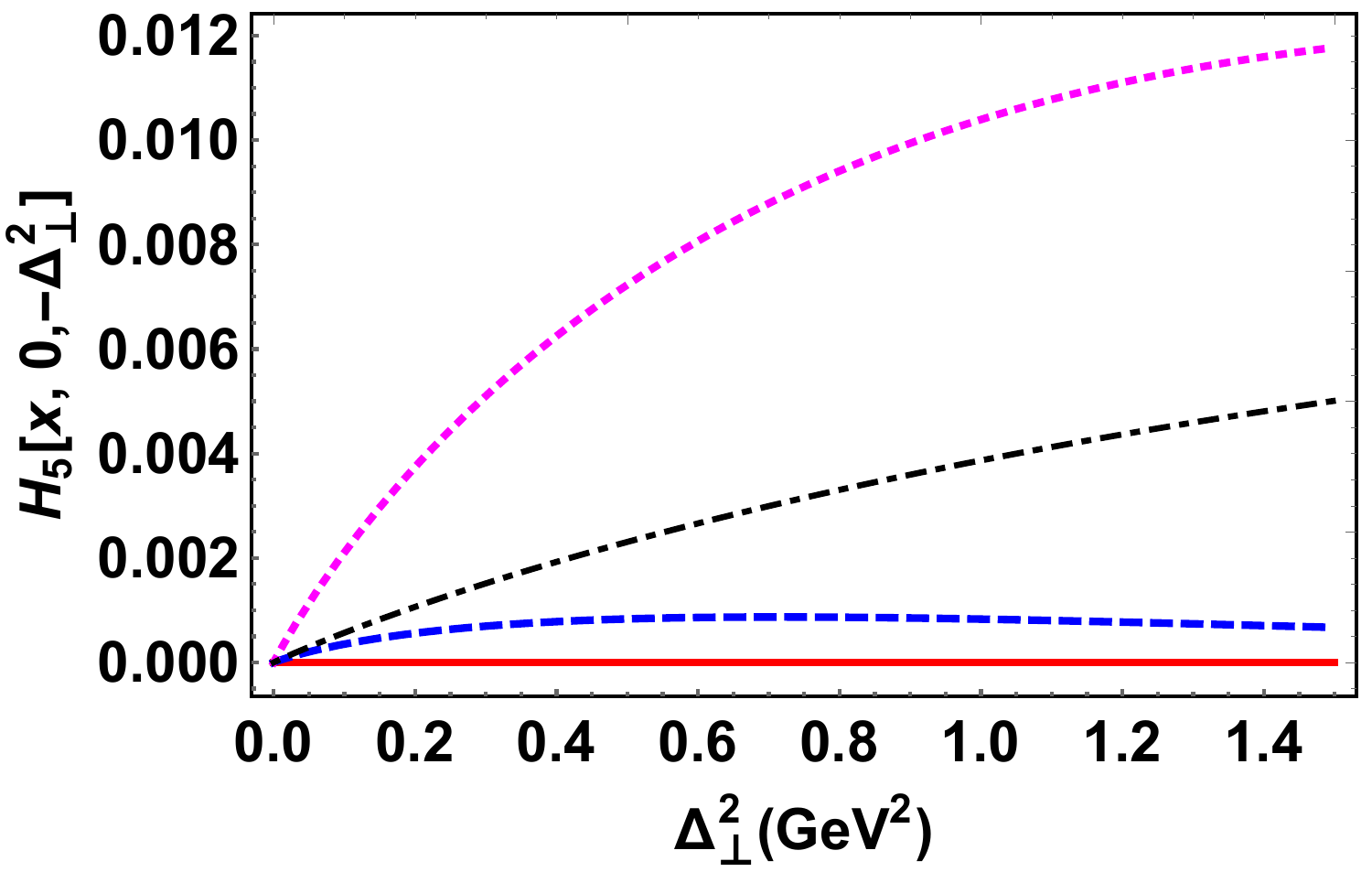} 
		\hspace{0.03cm}
	\end{minipage}
	\caption{\label{fig6} (Color online) The quark GPDs of $J/\psi$-meson (a) $H_1(x,0,-\Delta^2_\perp)$, (b) $H_2(x,0,-\Delta^2_\perp)$, (c) $H_3(x,0,-\Delta^2_\perp)$, and (d) $H_5(x,0,-\Delta^2_\perp)$  have been plotted with respect to $\Delta_\perp^2$ at fixed value of longitudinal momentum fraction $x=0.1$, $0.3$, $0.5$ and $0.7$.}
\end{figure*}

\begin{figure*}
	(a)\includegraphics[width=7.5cm]{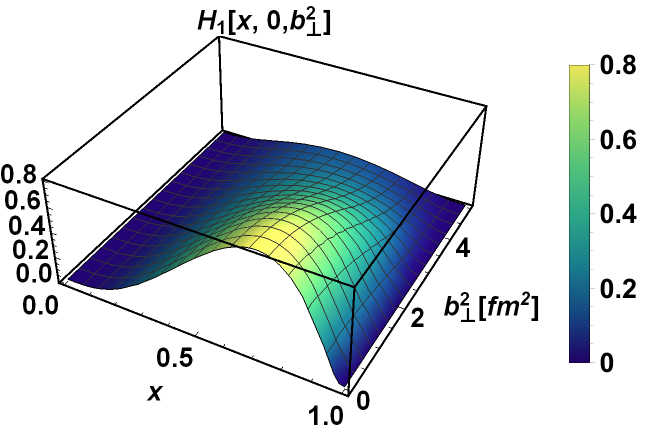}
	(b)\includegraphics[width=7.5cm]{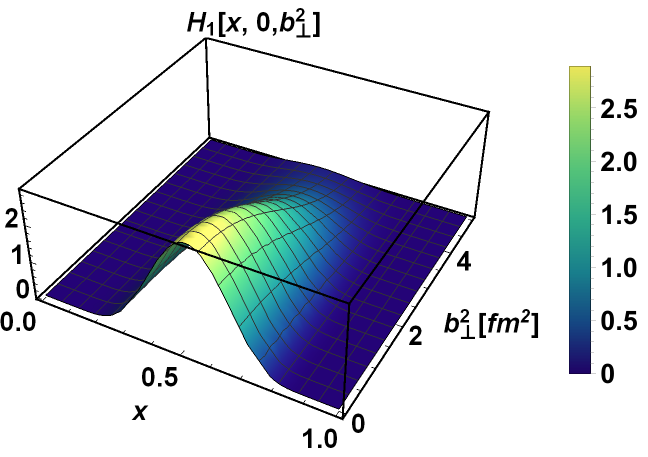} 
	(c)\includegraphics[width=7.5cm]{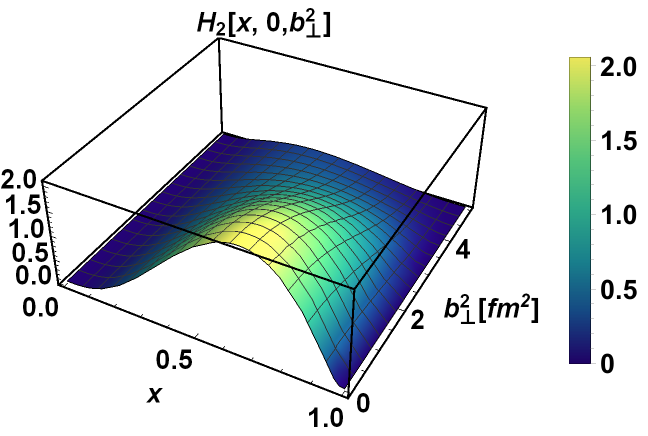} 
	(d)\includegraphics[width=7.5cm]{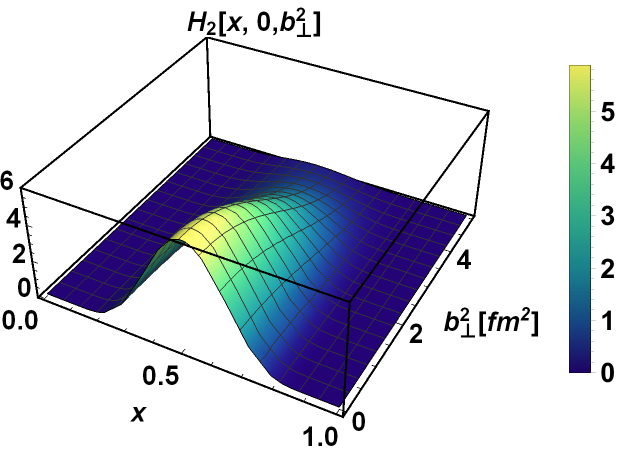} 
	(e)\includegraphics[width=7.5cm]{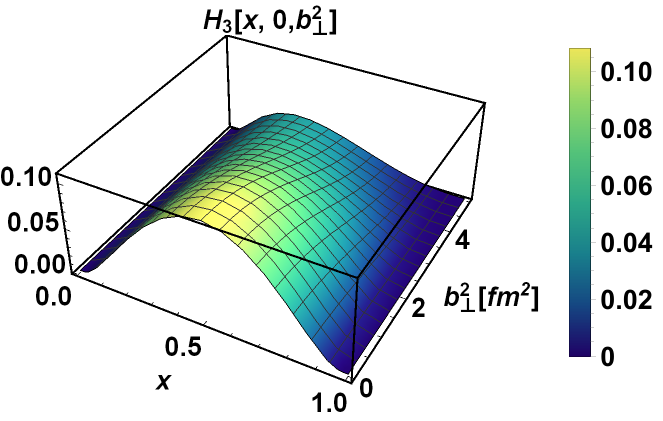}
	(f)\includegraphics[width=7.5cm]{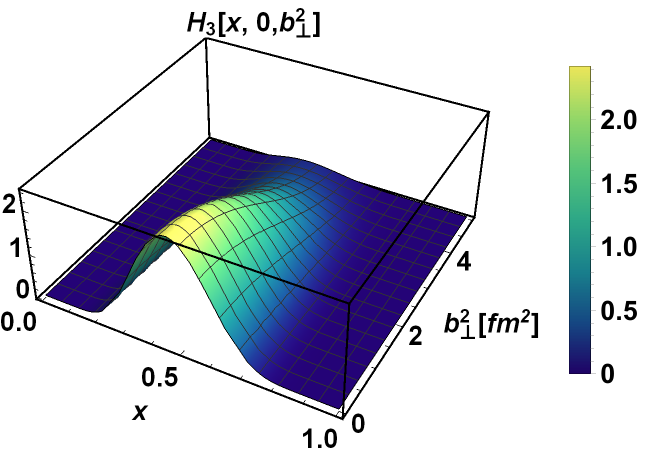}
	\caption{\label{fig7} (Color online) The $H_1(x,0,b^2_\perp)$,  $H_2(x,0,b^2_\perp)$, and $H_3(x,0,b^2_\perp)$ quark GPDs for $\rho$-meson (left panel (a), (c), and (e)) and $J/\psi$-meson (right panel (b), (d), and (f)) have been plotted with respect to $b_\perp^2$ (fm$^2$) and longitudinal momentum fraction $x$.}
\end{figure*}

\section{Unpolarized quark GPDs}\label{gpds}
The amplitudes in DVCS and deep exclusive meson electroproduction (DEMP) on a spin-one target are determined by non-perturbative matrix components that are parametrized in terms of nine GPDs for the quark sector out of which five are unpolarized GPDs. The correlation function for the unpolarized quark GPDs of spin-1 mesons is expressed through the quark-quark correlator as \cite{Cano:2003ju,Kumar:2019eck,Shi:2023oll}
\begin{eqnarray}
	V_{S_z^{\prime},S_z} (x,\xi,- \Delta_{\perp}^{2})=\int  \frac{dz^-}{2\pi}  e^{ixP_a^+z^-}
	\bra{M (P^{+\prime},\textbf{P}^{\prime}_\perp,S_z^\prime)} | \bar{\psi}\left(-\frac{z^-}{2}\right) \gamma^+ \psi\left(\frac{z^-}{2}\right)  \ket{M (P^+,\textbf{P}_\perp,S_z)}.
\end{eqnarray}
Here, $ V_{S_z^{\prime},S_z} (x,\xi,- \Delta_{\perp}^{2})$ is the light-front (LF) overlap correlation function between the initial and final state meson. $P$ and $P'$ are the four-momenta of the initial and final state of the meson having spin projections $S_z$ and $S_z^{\prime}$, respectively. \(P_a=(P'+P)/2\) is the average momentum of the meson. \(\Delta(\Delta^+,\Delta^-,\Delta_\perp)=P'-P\) is the momentum transferred between the final and initial meson states. The skewness variable is denoted by \(\xi=\frac{-\Delta^{+}}{2P_a^{+}}\). However, for this work, we have not considered the longitudinal momentum transfer between the initial and final state mesons, therefore, $\xi=0$ as $\Delta^+=0$. Consequently, the correlator can be decomposed to five spin-1 unpolarized GPDs having the form \(V_{S_z',S_z} (x,\xi=0,- \Delta_{\perp}^{2})\) as \cite{Shi:2023oll,Adhikari:2018umb}
\begin{eqnarray}
	V_{S_z^\prime,S_z}(x,0,- \Delta_{\perp}^{2}) =\ & -(\epsilon^{\prime *} \cdot \epsilon) H_1 (x,0,- \Delta_{\perp}^{2})
	+ \frac{(\epsilon \cdot n)(\epsilon' \cdot P) + (\epsilon' \cdot n)(\epsilon \cdot P)}{P \cdot n} H_2 (x,0,- \Delta_{\perp}^{2})\\ \notag 
	&- 2 \frac{(\epsilon \cdot P)(\epsilon^{\prime *} \cdot P)}{M_{q \bar q}^2} H_3 (x,0,- \Delta_{\perp}^{2})+ \frac{(\epsilon \cdot n)(\epsilon' \cdot P) - (\epsilon^{\prime *} \cdot n)(\epsilon \cdot P)}{P \cdot n} H_4 (x,0,- \Delta_{\perp}^{2})\\ \notag 
	&+ \left\{ M_{q \bar q}^2 \frac{(\epsilon \cdot n)(\epsilon' \cdot n)}{(P \cdot n)^2} + \frac{1}{3} (\epsilon^{\prime *} \cdot \epsilon) \right\} H_5  (x,0,- \Delta_{\perp}^{2}),
	\label{haan}
\end{eqnarray}
with $\epsilon \equiv \epsilon(P,S_z)$ and $\epsilon' \equiv \epsilon(P',S_z')$ being the polarization vector of the initial and final state meson (discussed in Eq. (\ref{polar})). The four-vector notations used in the above equation are expressed in the Breit frame as \cite{Choi:1996mq,Shi:2023oll}
\begin{eqnarray}
	n&=&(0,\sqrt2,0,0), \\
	\Delta&=& (0,0,\Delta_\perp,0), \\
	P&=&\frac{M_{q \bar q}}{\sqrt{2}}(\sqrt{1+\eta},\sqrt{1+\eta},-\frac{\Delta_\perp}{M_{q \bar q}\sqrt{2}},0),\\
	P^{\prime}&=&\frac{M_{q \bar q}}{\sqrt{2}}(\sqrt{1+\eta},\sqrt{1+\eta},\frac{\Delta_\perp}{M_{q \bar q}\sqrt{2}},0).
	\label{notation}
\end{eqnarray}
Here, $\eta=\frac{\Delta_\perp^2}{4 M^2_{q \bar q}}$. Consequently, the LF overlap correlation function $V_{S_z', S_z}(x, \xi, - \Delta_{\perp}^{2})$ obey the parity and time reversal invariance as \cite{Berger:2001zb,Shi:2023oll,Cano:2003ju}
\begin{eqnarray}
	V_{S_z', S_z}(x, 0, - \Delta_{\perp}^{2}) &= (-1)^{S_z' - S_z} V_{-S_z', -S_z}(x, 0, - \Delta_{\perp}^{2}), \\
	V_{S_z', S_z}(x, \xi, - \Delta_{\perp}^{2}) &= (-1)^{S_z'- S_z} V_{S_z', S_z}(x, -\xi, -\Delta_{\perp}^{2}).
\end{eqnarray}
Now, applying the above parity and time-reversal condition to calculate the unpolarized quark GPDs at zero skewness, we are left with four independent $V_{S_z', S_z}(x, \xi, - \Delta_{\perp}^{2})$ functions in the form of different spin projections, which are $V_{0,0}(x,0, - \Delta_{\perp}^{2})$, $V_{0,1}(x,0, - \Delta_{\perp}^{2})$, $V_{1,1}(x,0, - \Delta_{\perp}^{2})$, and $V_{1,-1}(x,0, - \Delta_{\perp}^{2})$. Reversing Eq. (\ref{haan}), by using the meson momentum and polarization vector, the unpolarized GPDs are found to be
\begin{eqnarray}
	H_1 (x,0,-\Delta_\perp^2)&=&\frac{1}{3}[V_{0,0}(x,0,-\Delta_\perp^2)-2(\eta-1)V_{1,1}(x,0,-\Delta_\perp^2) \nonumber\\ 
	&& +2\sqrt{2\eta}V_{1,0}(x,0,-\Delta_\perp^2)+2V_{1,-1}(x,0,-\Delta_\perp^2)],\label{eq:H1}\\
	H_2 (x,0,-\Delta_\perp^2) &=& 2V_{1,1}(x,0,-\Delta_\perp^2) -\frac{2}{\sqrt{2\eta}} V_{1,0}(x,0,-\Delta_\perp^2),\label{eq:H2}\\
	H_3(x,0,-\Delta_\perp^2)&=&-\frac{V_{1,-1}(x,0,-\Delta_\perp^2)}{\eta},\label{eq:H3}\\
	H_4(x,0,-\Delta_\perp^2)&=&0,\\
	H_5(x,0,-\Delta_\perp^2)&=& V_{0,0}(x,0,-\Delta_\perp^2)-(1+2\eta)V_{1,1}+2\sqrt{2\tau}V_{1,0}(x,0,-\Delta_\perp^2)-V_{1,-1}(x,0,-\Delta_\perp^2). \label{eq:H5}
\end{eqnarray}
The LF overlap correlation
function $V_{S_z', S_z}(x, 0, - \Delta_{\perp}^{2})$ can be written in the overlap form of LFWFs of the total meson wave function as
\begin{eqnarray}
	V_{S_z', S_z}(x, 0, - \Delta_{\perp}^{2})=\sum_{\lambda_1,\lambda_2}\int \frac{d^2\bfk}{2 (2\pi)^3} \Psi^{*}_{S_z}(x,\bfk^{\prime},\lambda_1,\lambda_2)\Psi_{S_z}(x,\bfk,\lambda_1,\lambda_2).
\end{eqnarray}
Here, $\bfk^{\prime}=\bfk+(1-x)\Delta_\perp$ and $\bfk$ are the transverse momentum of the quark of the final and initial meson. $H_4$ quark GPD for our case is found to be zero similar to NJL model, DS-BSE approach, BLFQ, and LFQM. 

\par The behavior of these unpolarized quark GPDs for \(\rho\)-meson have been provided through three-dimensional plots in Fig.~\ref{fig1} as a function of longitudinal momentum fraction \(x\) and squared momentum transfered \(\Delta_{\perp}^{2}\). \(H_{1} (x,0,-\Delta_\perp^2)\) quark GPD shows a smooth, positive distributions all over, and symmetric behavior at around \(x=0.5\) at $\Delta_\perp=0$ under \(x\leftrightarrow (1-x)\) as shown in Fig.~\ref{fig1} (a), reflecting the equal constituent quark masses of \(\rho\)-meson.  A similar kind of positive distributions of $H_1(x,0,-\Delta_\perp^2)$ can also be seen in LFQM \cite{Kumar:2019eck} and NJL model \cite{Zhang:2022zim}. The symmetric behavior at $\Delta_\perp^2=0$ observed is in line with the NJL model with the distribution nearly vanishing around the end points \(x=0\) and \(x=1\). As \(\Delta_{\perp}^{2}\) increases, the distribution \(H_{1} (x,0,-\Delta_\perp^2)\) peak suppresses and becomes zero after $\Delta_\perp \ge 1.2$ GeV. A higher peak is observed in the \(H_{2} (x,0,-\Delta_\perp^2)\) quark GPD as compared to other quark GPDs as shown in Fig. \ref{fig1}. It has more of a sharp symmetric behavior around \(x \approx 0.6-0.7\) with larger amplitude. The \(H_{2} (x,0,-\Delta_\perp^2)\) exhibits  stronger suppression and a noticeable shift of the peak toward higher \(x\) as \(\Delta_{\perp}^{2}\) increases. The GPDs $H_2(x,0,-\Delta_\perp^2)$ and $H_3(x,0,-\Delta_\perp^2)$  show positive distributions in  the entire range of $x$ and $\Delta_\perp$ similar to NJL model observations  \cite{Zhang:2022zim}. In case of LFQM, $H_2$ shows positive distribution whereas $H_3$ has both positive and negative distribution clearly contradicting our model results. 

The $H_5(x,0,-\Delta_\perp^2)$ quark GPD has a lower peak amplitude distribution compared to others and have zero distribution at $\Delta_\perp=0$. The terms \(V_{0,0}\) and \(V_{1,1}\) show the contribution of S-wave as there is no increment in orbital angular momentum (\(L_{z}=0\)), whereas the contribution of the P-wave is shown by the presence of \(V_{1,0}\) term indicating the increment in the angular momentum by unity (\(L_{z}=\pm 1\)). Additionally, the term \(V_{1,-1}\) contributes to the presence of D-wave as the orbital angular momentum is raised by two units (\(L_{z}=\pm 2\)). Consequently, the quark GPDs \(H_{1(5)} (x,0,-\Delta_\perp^2)\) include the superposition of S, P, and D-wave, whereas only \(H_{2} (x,0,-\Delta_\perp^2)\) contributes S and P-wave, and \(H_{5} (x,0,-\Delta_\perp^2)\) only contains D-wave. In Figs. \ref{fig2} and \ref{fig3}, we have plotted the quark GPDs of $\rho$-meson with respect to $x$ at fixed values of $\Delta_\perp$ and with respect to $\Delta_\perp$ at fixed values of  $x$, respectively. We observed that the distribution decreases with $x$ with an increase in momentum transferred between the states for all the quark GPDs except $H_5$. For the case $H_5$, we found that the distribution shows both negative and positive behavior, and with an increase in $\Delta_\perp$, the positive distribution increases. This kind of behavior shows up to $\Delta_\perp^2=20$ GeV$^2$, after which the distribution becomes zero. A similar kind of behavior is also seen with an increase in $x$ also. 

To have an idea about the heavy quark structure, we have also studied the quark GPDs of the heavy charmonia $J/\psi$ vector meson. The quark GPDs of $J/\psi$-meson have been plotted with respect to $x$ and $\Delta_\perp^2$ through three-dimensional and two-dimensional plots in Figs. \ref{fig4}-\ref{fig6}. All the quark GPDs are found to have maximum distributions around $x=0.5$ and show $x\leftrightarrow (1-x)$ symmetry. The distributions of quark GPDs decrease with an increase in $\Delta_\perp$, as like the $\rho$-meson case. The $H_5$ quark GPD shows only positive behavior for the case of $J/\psi$-meson. We observed that $H_{1,2,3}$ quark GPDs of our results have a similar kind of trend with BLFQ results \cite{Adhikari:2018umb}. Overall, we found that the momentum fraction carried by the quark decreases with an increase in momentum transferred to the final state meson.
\par All these quark GPDs have also been plotted in impact parameter space by taking the Fourier transform of $\Delta_\perp$ as
\begin{eqnarray}
	H_i\left(x,0,b^2_{\perp}\right)&=\int \frac{\mathrm{d}^2\bm{\Delta}_{\perp}}{(2 \pi )^2}e^{-i\bm{b}_{\perp}\cdot \bm{\Delta}_{\perp}}H_i\left(x,0,-\bm{\Delta}_{\perp}^2\right). 
\end{eqnarray} 
These quark GPDs at impact parameter space have been studied using the three-dimensional plots in Fig. \ref{fig7} for both light $\rho$ and heavy $J/\psi$ vector mesons. All the quark GPDs are found to have positive distributions with a sharply decreasing behavior with maximum distributions at $b_\perp=0$.

\subsection{Vector meson Form Factors}
\begin{figure*}
	\centering
	\begin{minipage}[c]{0.98\textwidth}
		(a)\includegraphics[width=7.5cm]{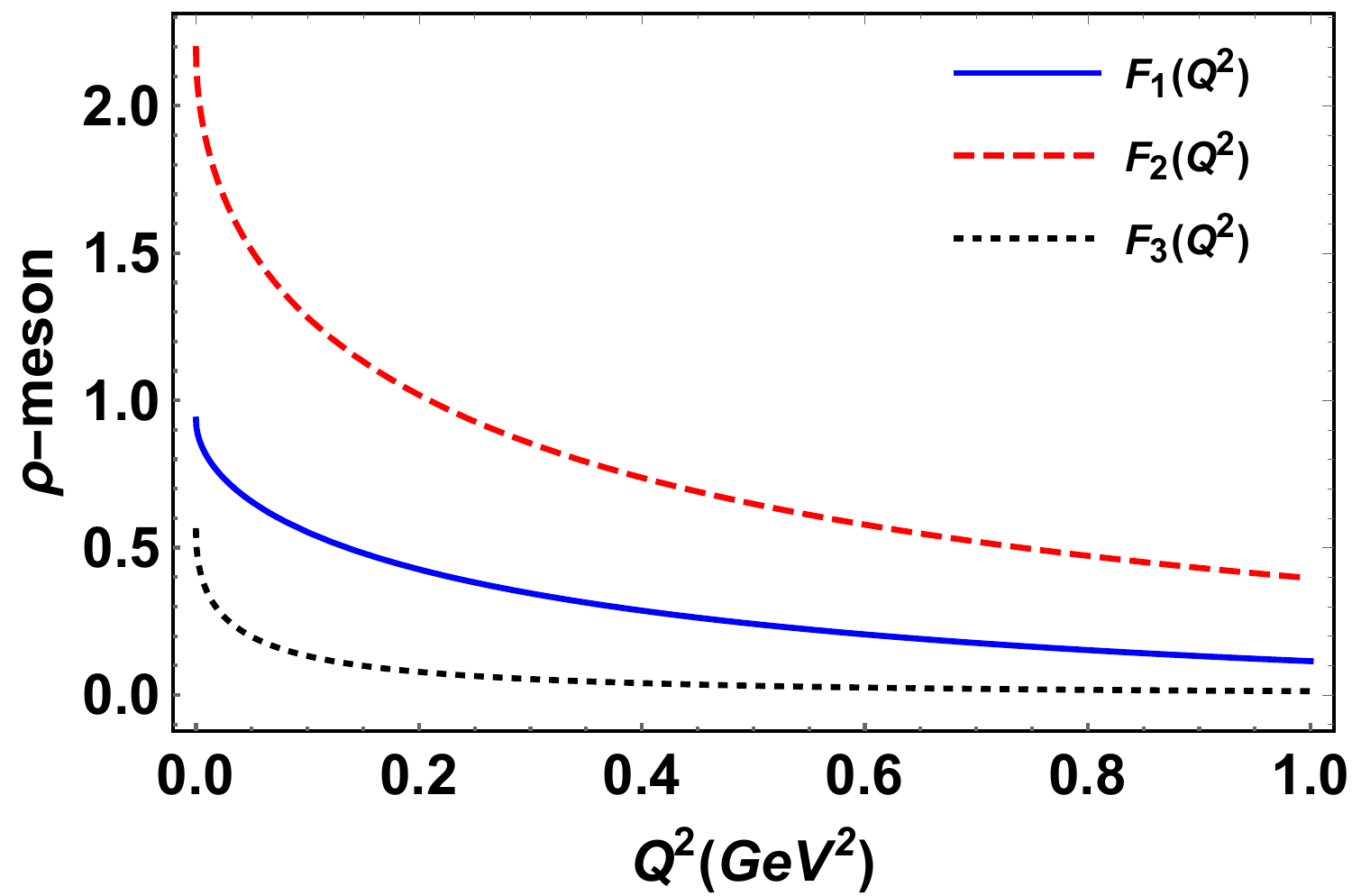}
		\hspace{0.03cm}	
		(b)\includegraphics[width=7.5cm]{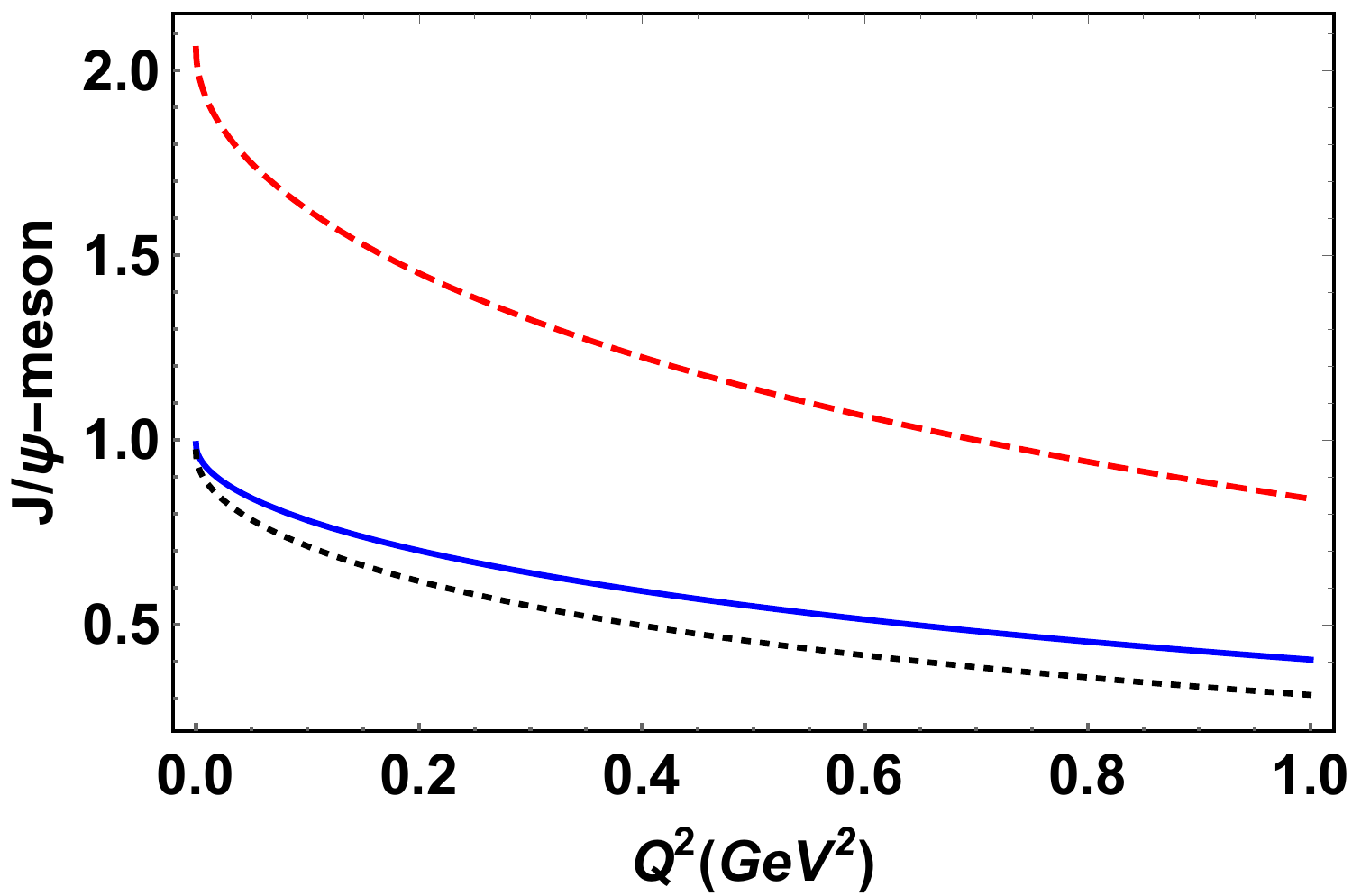} 
		\hspace{0.03cm}
	\end{minipage}
	\caption{\label{fig8} (Color online) The elastic form factors $F_1(Q^2)$, $F_2(Q^2)$, and $F_3(Q^2)$ have been plotted with respect to $Q^2$ for (a) 
		light $\rho$ and (b) heavy $J/\psi$ vector meson.}
\end{figure*}

\begin{figure*}
	\centering
	\begin{minipage}[c]{0.98\textwidth}
		(a)\includegraphics[width=7.5cm]{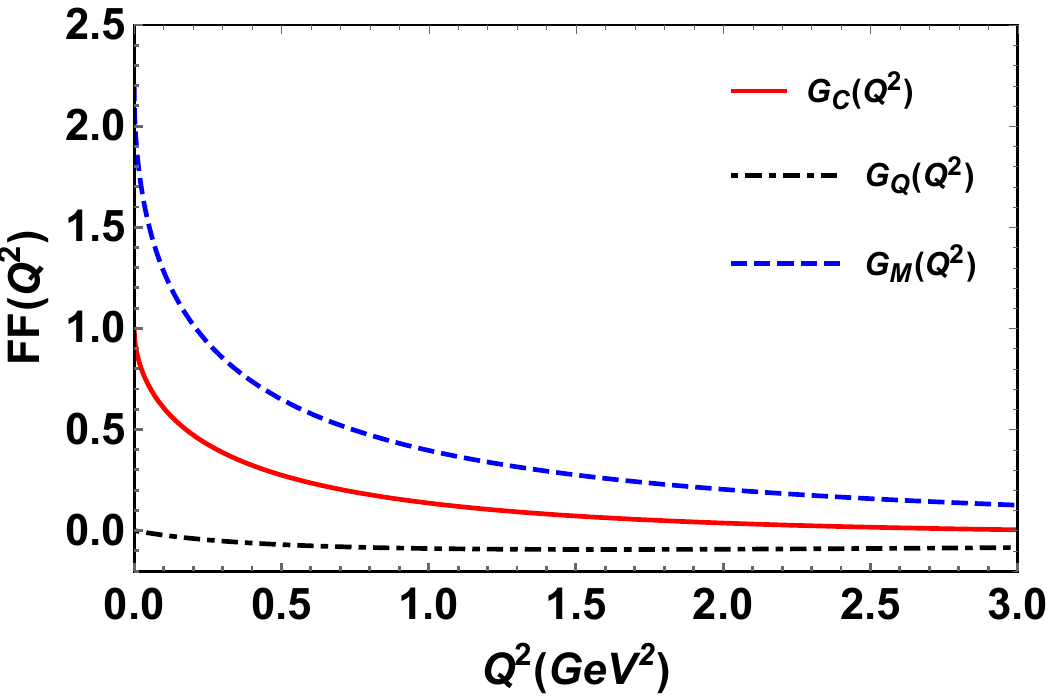}
		\hspace{0.03cm}	
		(b)\includegraphics[width=7.5cm]{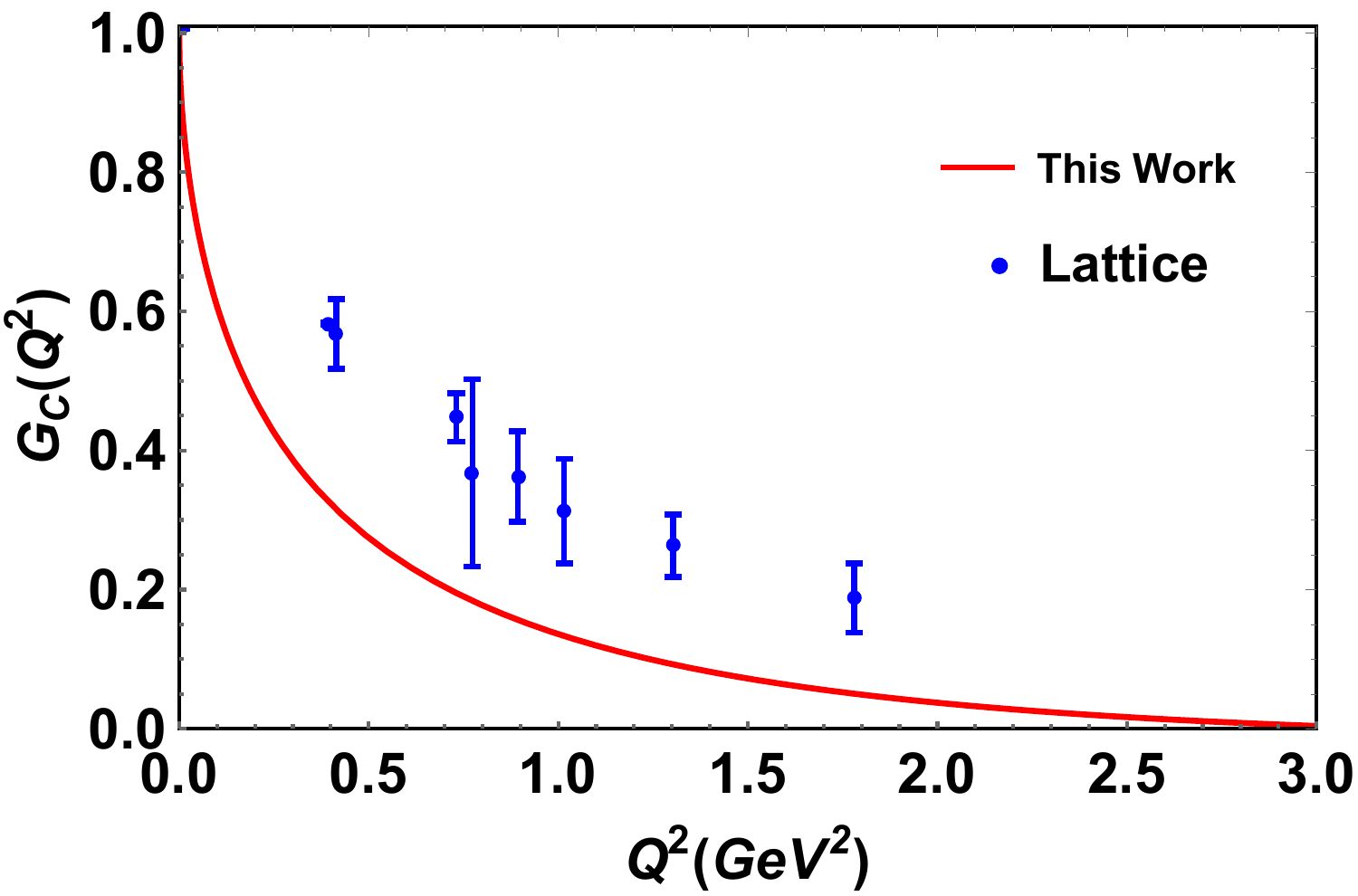} 
		\hspace{0.03cm}
	\end{minipage}
	\centering
	\begin{minipage}[c]{0.98\textwidth}
		(c)\includegraphics[width=7.5cm]{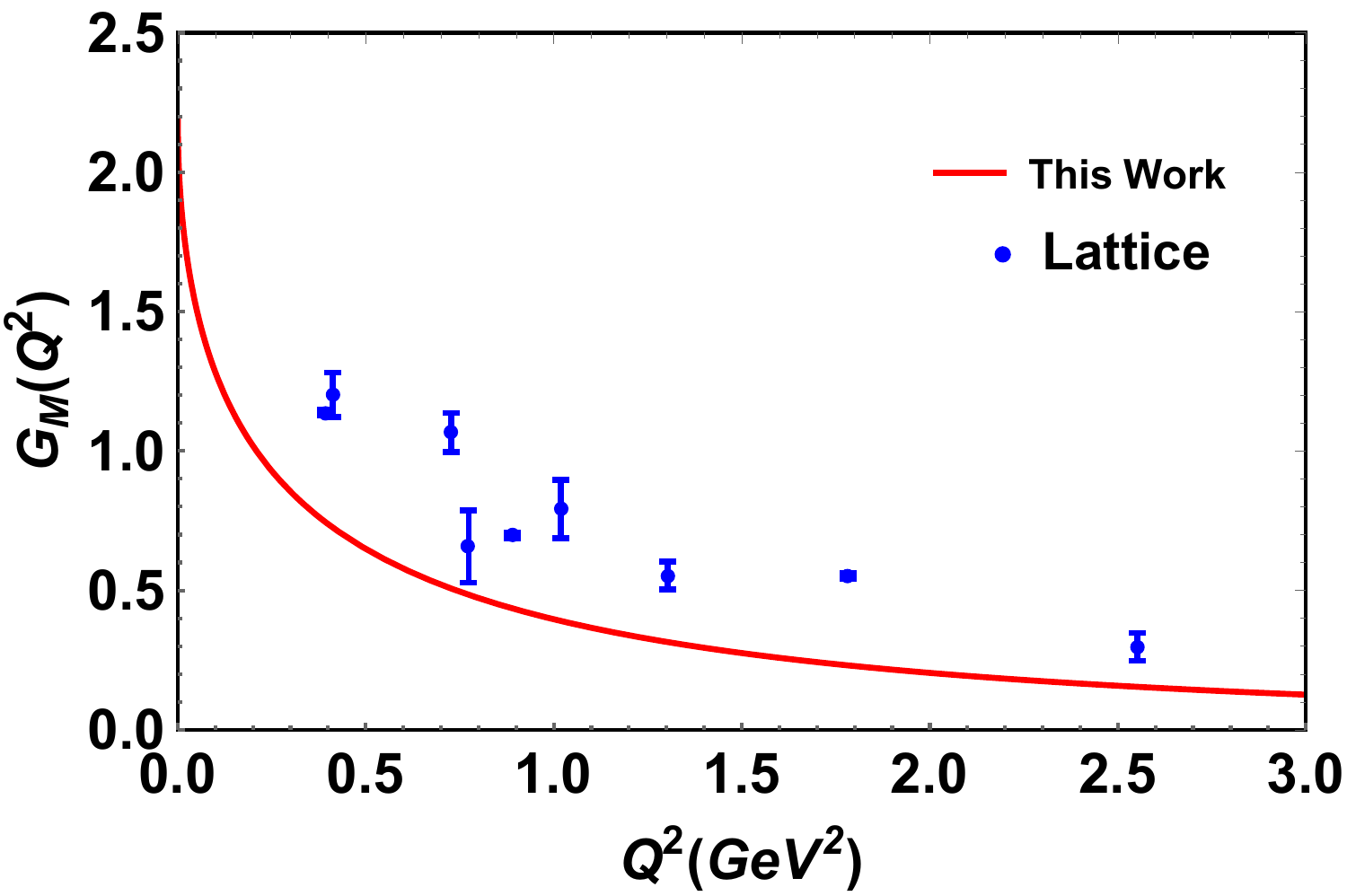}
		\hspace{0.03cm}	
		(d)\includegraphics[width=7.5cm]{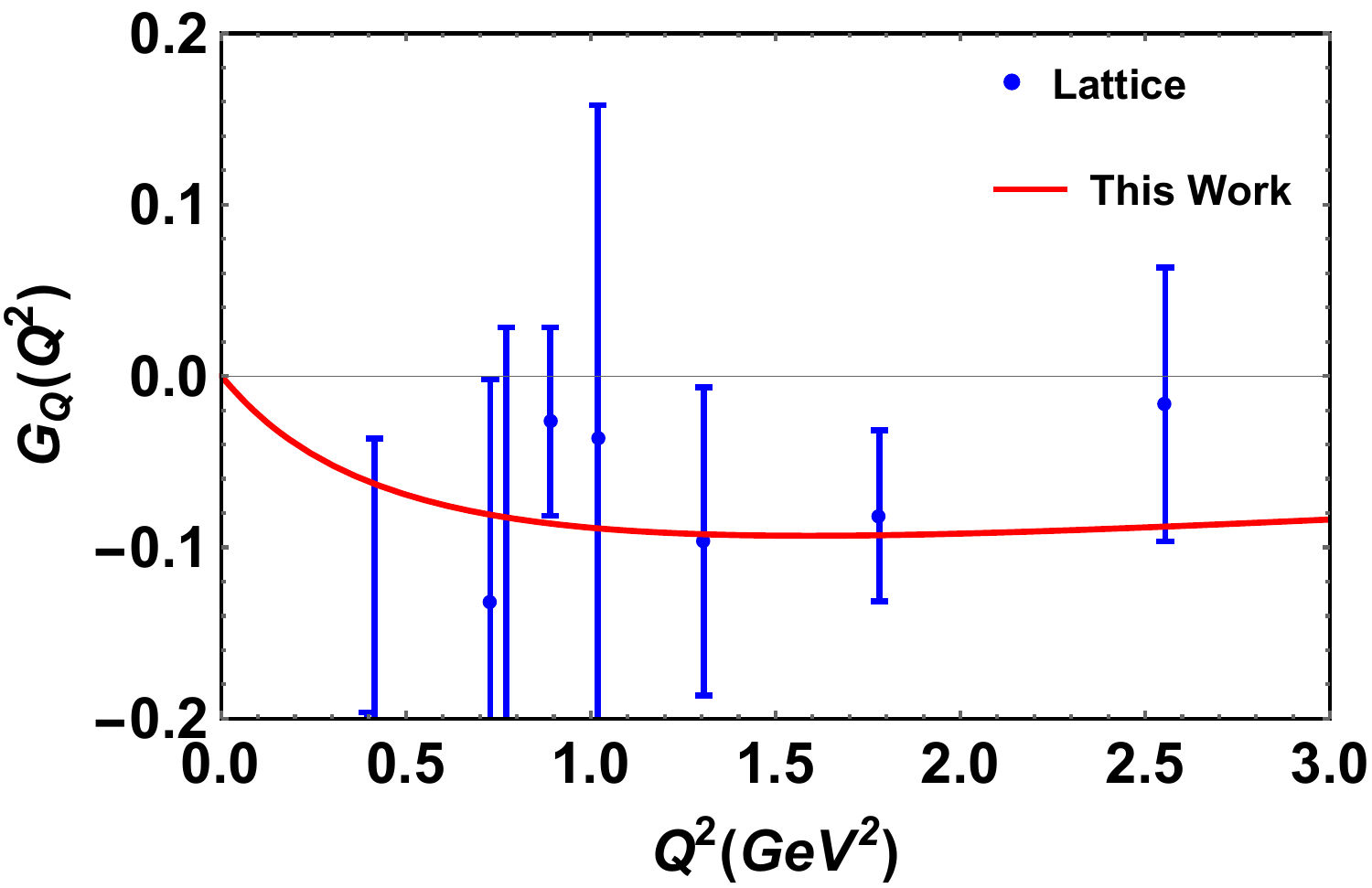} 
		\hspace{0.03cm}
	\end{minipage}
	\centering
	\begin{minipage}[c]{0.98\textwidth}
		(e)\includegraphics[width=7.5cm]{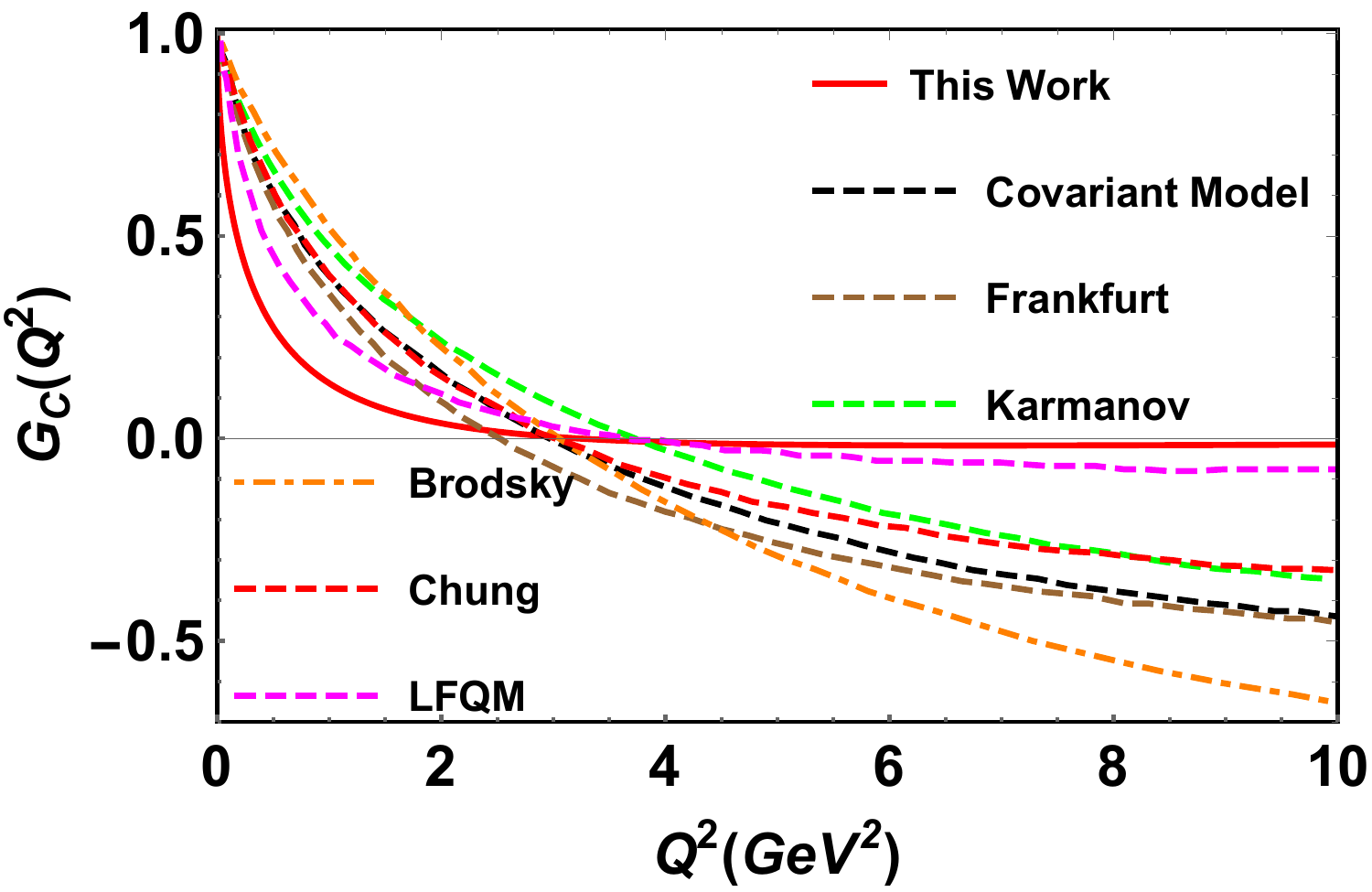}
		\hspace{0.03cm}	
		(f)\includegraphics[width=7.5cm]{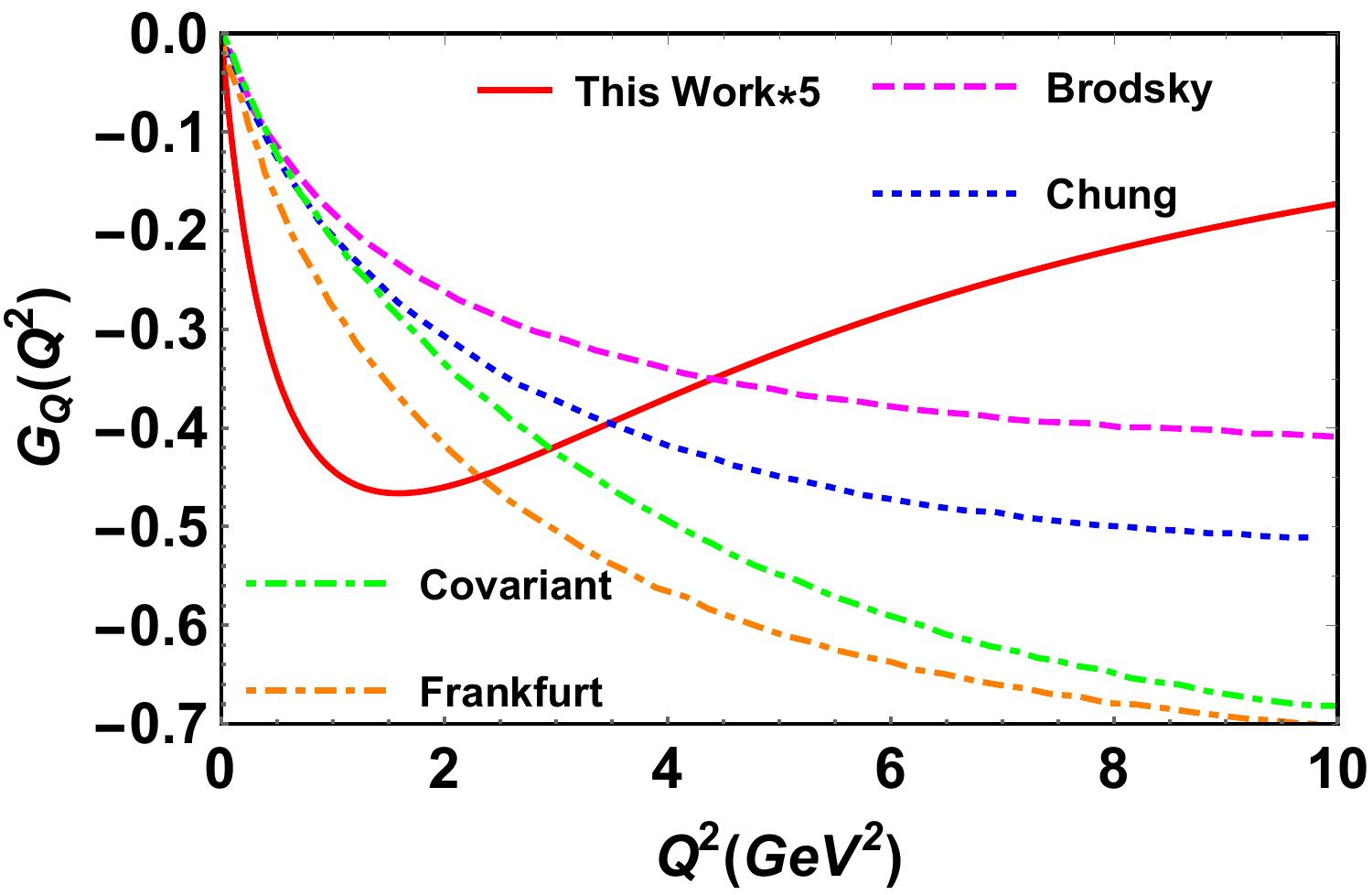} 
		\hspace{0.03cm}
	\end{minipage}
	\caption{\label{fig9} (Color online) (a) The charge ($G_C(Q^2)$), magnetic ($G_M(Q^2)$), and quadrupole ($G_Q(Q^2)$) form factors have been plotted with respect to $Q^2$ for $\rho$-meson. The charge ($G_C(Q^2)$), magnetic ($G_M(Q^2)$), and quadrupole ($G_Q(Q^2)$) form factors have been compared with available lattice simulations data \cite{Lasscock:2006nh} in (b), (c), and (d), respectively. The charge $G_C(Q^2)$ and quadrupole $G_M(Q^2)$ form factors have been compared with available theoretical models \cite{Karmanov:1996qc,Sun:2017gtz,Chung:1988mu,Brodsky:1992px,Frankfurt:1993ut,DeMelo:2018bim}.}
\end{figure*}

\begin{figure*}
	\centering
	\begin{minipage}[c]{0.98\textwidth}
		(a)\includegraphics[width=7.5cm]{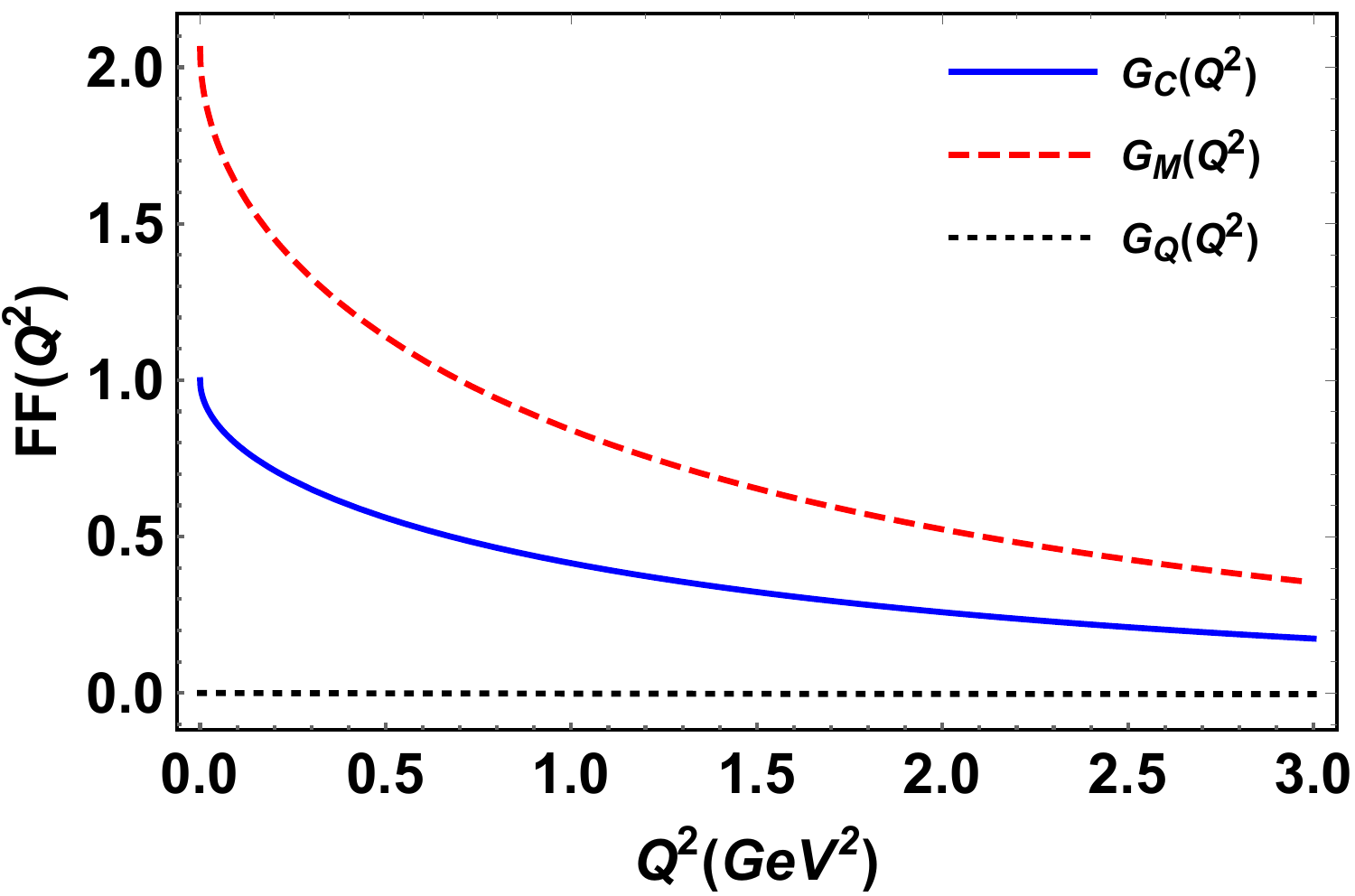}
		\hspace{0.03cm}	
		(b)\includegraphics[width=7.5cm]{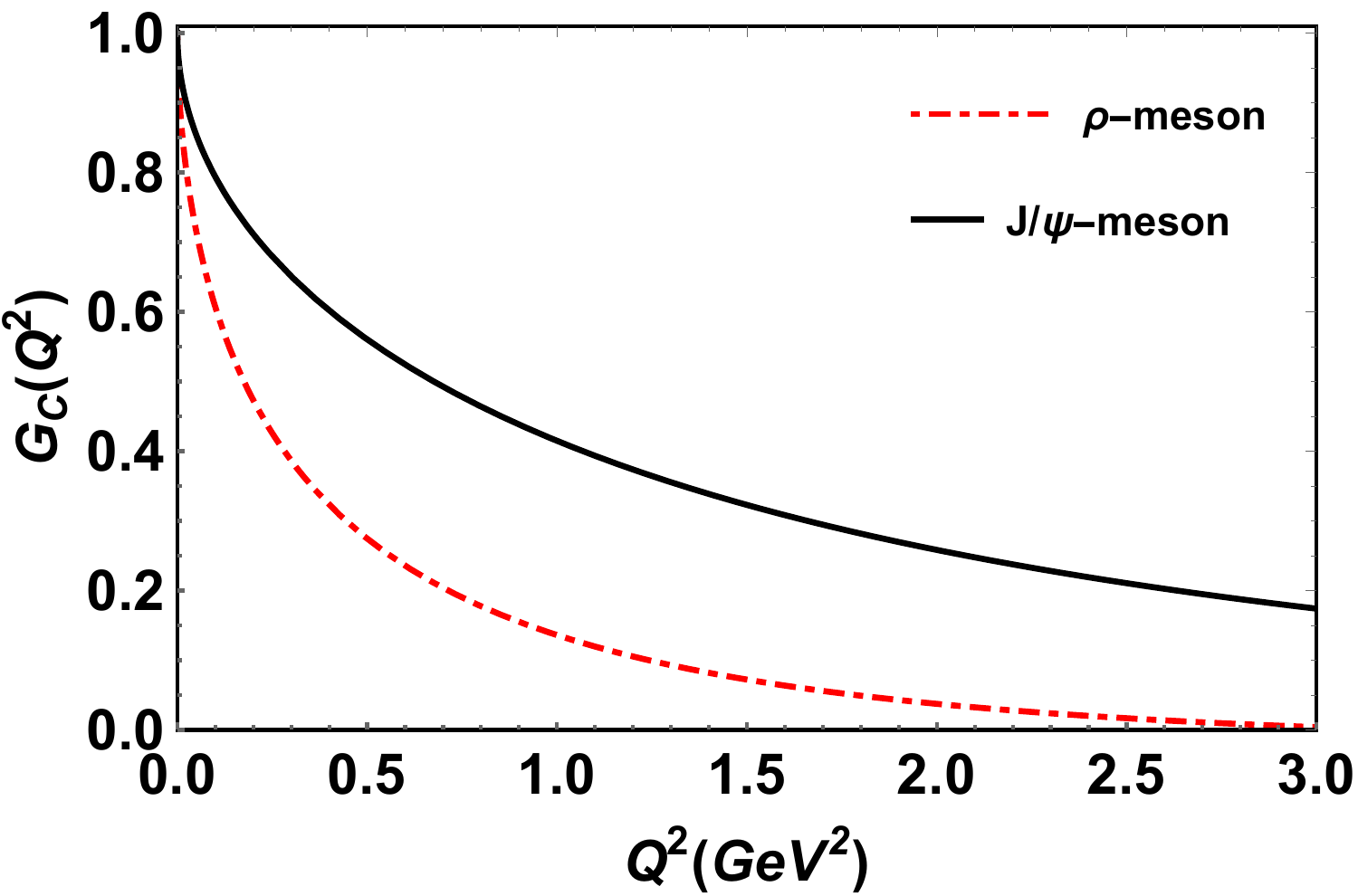} 
		\hspace{0.03cm}
	\end{minipage}
	\centering
	\begin{minipage}[c]{0.98\textwidth}
		(c)\includegraphics[width=7.5cm]{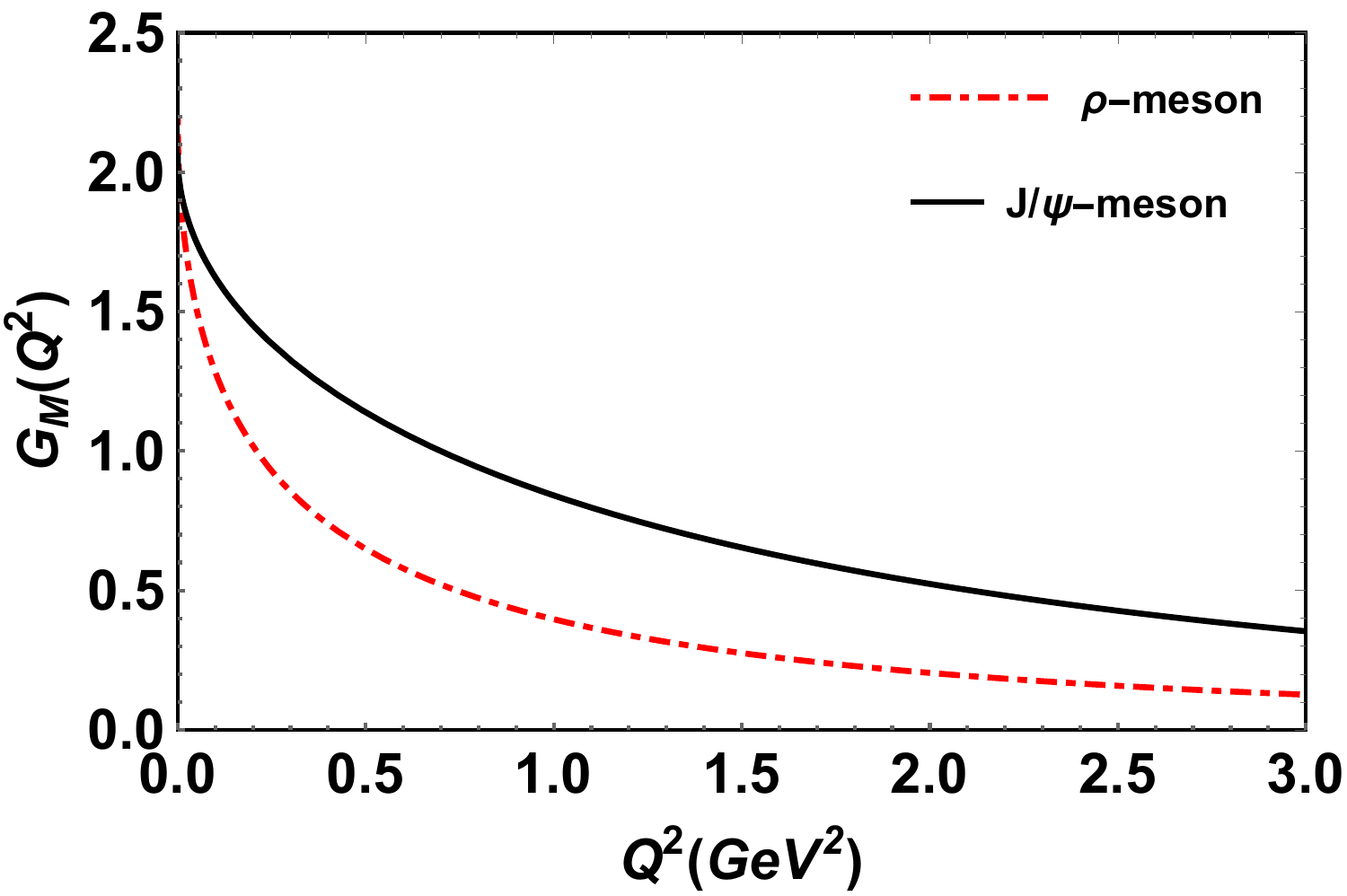}
		\hspace{0.03cm}	
		(d)\includegraphics[width=7.5cm]{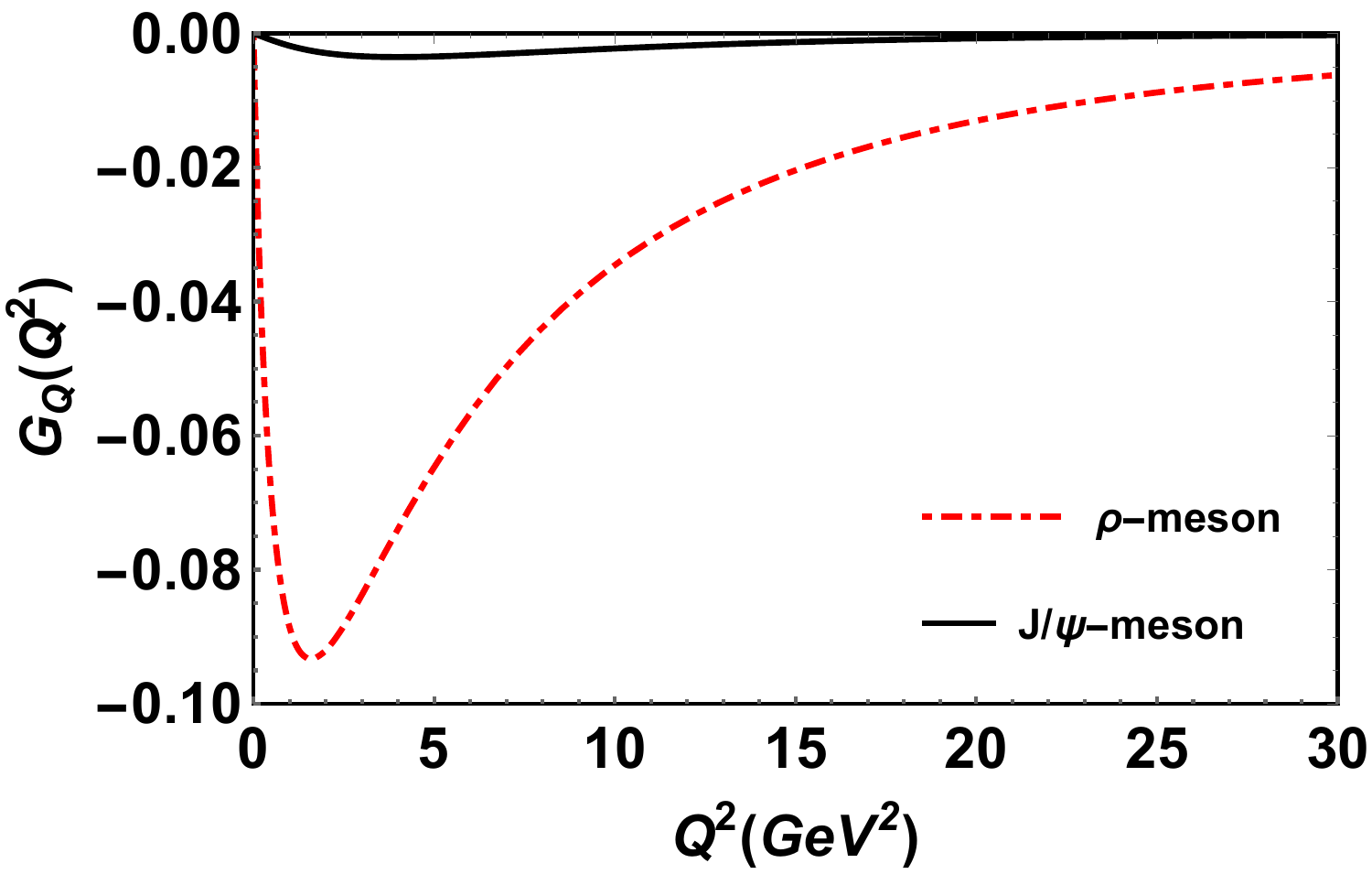} 
		\hspace{0.03cm}
	\end{minipage}
	\caption{\label{fig10} (Color online) (a) The charge ($G_C(Q^2)$), magnetic ($G_M(Q^2)$), and quadrupole ($G_Q(Q^2)$) form factors have been plotted with respect to $Q^2$ for $J/\psi$-meson. The charge ($G_C(Q^2)$), magnetic ($G_M(Q^2)$), and quadrupole ($G_Q(Q^2)$) form factors of $J/\psi$-meson have been compared with $\rho$-meson form factors in (b), (c), and (d), respectively. }
\end{figure*}

There are a total of three electromagnetic form factors $(F_i(Q^2) (i=1,2,3))$ present at the leading twist for the spin-$1$ vector mesons case. These form factors can be calculated either by solving the quark-quark correlation function or from unpolarized quark GPDs. These Lorentz invariant form factors are defined by the matrix elements of the $J^\mu$ current sandwiched between initial $|M(P^+,\textbf{P}_\perp, S_z \rangle$ and final $\langle M(P^{+\prime},\textbf{P}^{\prime}_\perp, S_z'|$ states as follows \cite{Choi:2004ww,Kumar:2019eck}
\begin{eqnarray}
	\langle M(P^{+\prime}, S_z' | J^\mu | M(P^+, \textbf{P}_\perp, S_z \rangle &= & - \epsilon^{\prime} \cdot \epsilon (P+P')^\mu F_1(Q^2) + ( \epsilon^{\prime} n \cdot \epsilon^{\prime} 
	- \epsilon^{\prime} n \cdot \epsilon ) F_2 (Q^2) + \nonumber\\
	&& \frac{(\epsilon^{\prime} \cdot n)(\epsilon \cdot n)}{2 M_v^2} (P+P')^\mu F_3(Q^2),
	\label{electromagnetic}
\end{eqnarray}
where $Q^2=\Delta^2=-\Delta^2_\perp$. On the other hand, these form factors can be calculated from the unpolarized quark GPDs as
\begin{eqnarray}
	\int_{-1}^1\mathrm{d}x H_i(x,0,-\Delta_\perp^2)&=F_i(Q^2), \quad \quad (i=1,2,3) \, \\
	\int_{-1}^1\mathrm{d}x H_i(x,\xi,-\Delta_\perp^2)&=0. \quad \quad (i=4,5)
\end{eqnarray}
Both calculations give rise to equal form factor results. These form factors have been plotted with respect to $Q^2$ GeV$^2$ in Fig. \ref{fig8} for both $\rho$ and $J/\psi$ vector mesons. All the form factors are found to have positive distributions and have a maximum distribution at $Q^2=0$. It is observed that $F_2(Q^2)$ has higher distributions compared to other form factors for both the  particles.
These form factors can be used to calculate the charge, magnetic, and quadrupole form factors of vector mesons. These can be calculated as 
\begin{eqnarray}
	G_C(Q^2)&=&(1+\frac{2}{3}\eta)F_1(Q^2)+ \frac{2}{3}\eta F_2(Q^2)+\frac{2}{3}\eta(1+\eta)F_3(Q^2), \\
	G_M(Q^2)&=& F_2(Q^2),\\
	G_Q(Q^2)&=& F_1(Q^2)+F_2(Q^2)+(1+\eta)F_3(Q^2).
\end{eqnarray}
The light $\rho$-meson and heavy $J/\psi$-meson form factors have been plotted  in Figs. \ref{fig9} and \ref{fig10} with respect to $Q^2$ GeV$^2$, respectively. All the form factors of $\rho$-meson have been compared with lattice simulation results \cite{Lasscock:2006nh} up to $Q^2=3$ GeV$^2$ and are found to have a similar trend but lower distributions except for the case of $G_Q$. At $Q^2=0$, $G_C$, $G_M$, and $G_Q$ are found to be unity, $2.1$, and zero for both the particles, respectively. For a better understanding of these form factors, we have compared our results with theoretical model results \cite{Karmanov:1996qc,Sun:2017gtz,Chung:1988mu,Brodsky:1992px,Frankfurt:1993ut,DeMelo:2018bim} up to $Q^2=10$ GeV$^2$ in Figs. \ref{fig9} (e) and (f). The charge $G_C$ and magnetic $G_M$ form factors are found to be have a similar trend but with lower distributions compared to the NJL model \cite{Zhang:2024nxl,Zhang:2022zim}, LFQM \cite{Kumar:2019eck}, DS-BSE \cite{Shi:2023oll}, contact interaction \cite{Hernandez-Pinto:2024kwg}, AdS/QCD model \cite{Allahverdiyeva:2023fhn}, LFCQM \cite{Sun:2017gtz}, BLFQ \cite{Adhikari:2018umb} and other models \cite{Karmanov:1996qc,Chung:1988mu,Brodsky:1992px,Frankfurt:1993ut,DeMelo:2018bim} results. However, the quadrupole form factor $G_Q$, on the one hand, has a opposite kind of behavior at low $Q^2$ compared to Refs. \cite{Zhang:2024nxl,Zhang:2022zim,Hernandez-Pinto:2024kwg,Kumar:2019eck,Shi:2023oll,DeMelo:2018bim}, whereas, on the other hand,  it has the same kind of trend with Refs. \cite{Adhikari:2018umb,Chung:1988mu,Brodsky:1992px,Frankfurt:1993ut} as shown in Fig. \ref{fig9} (f). The charge form factor $G_C$ shows negative distribution after $Q^2\ge4$ GeV$^2$ as like NJL model \cite{Zhang:2022zim}, LFCQM \cite{Sun:2017gtz} and other models \cite{Karmanov:1996qc,Chung:1988mu,Brodsky:1992px,Frankfurt:1993ut} as shown in Fig. \ref{fig9} (e). We have also observed that the heavier $J/\psi$-meson has higher form factor distributions compared to the light $\rho$-meson as shown in Fig. \ref{fig10}. We have also observed that the asymptotic limit of QCD
reported in Refs. \cite{Brodsky:1992px,Hernandez-Pinto:2024kwg} is strictly followed by our model, which reads as
\begin{eqnarray}
	G_C(Q^2):G_M
	(Q^2):G_Q(Q^2)=^{Q^2\rightarrow\infty}1-\frac{2 \eta}{3}:2:-1.
\end{eqnarray}
At zero momentum transfer $Q^2=0$, one can find the charge, magnetic moment ($\mu_p$), and quadrupole moment $Q_p$ of the vector mesons from these form factors as
\begin{eqnarray}
	eG_C(Q^2=0)=e, \ \ \ \ G_M(Q^2=0)=\mu_p, \ \ \ \, G_Q(Q^2=0)=Q_p.
\end{eqnarray}
Here, $e$ is the static charge. $\mu_p$ and $Q_p$ are in the units of Bohr magneton $e/2 M_{\rho (J/\psi)}$ and $e/ M^2_{\rho (J/\psi)}$ respectively. Here, $M_{\rho (J/\psi)}$ is the physical mass of $\rho$ and $J/\psi$-mesons. The charge radii $\sqrt{\langle r_c^2\rangle}$ of the respective vector mesons can be calculated using the fundamental radius equation as
\begin{eqnarray}
	\langle r_c^2\rangle &= & \frac{-6}{G_C(0)} \frac{\partial G_C(Q^2)}{\partial Q^2}\Big|_{Q^2\rightarrow0}.
\end{eqnarray}
$G_C(0)$ in this case is unity. The value of charge radii $\sqrt{\langle r_c \rangle}$, magnetic moment $\mu_p$, and quadrupole moment $Q_p$ have been presented in Table \ref{table1} for both particles along with the comparison with available predictions \cite{Bhagwat:2006pu,Dudek:2006ej,Hernandez-Pinto:2024kwg,Luan:2015goa,Shultz:2015pfa,Owen:2015gva}. We observe that our charge radii values are slightly higher than other results whereas $\mu_p$ and $Q_p$ are in good agreement with them. 

\begin{figure*}
	\centering
	\begin{minipage}[c]{0.98\textwidth}
		(a)\includegraphics[width=7.5cm]{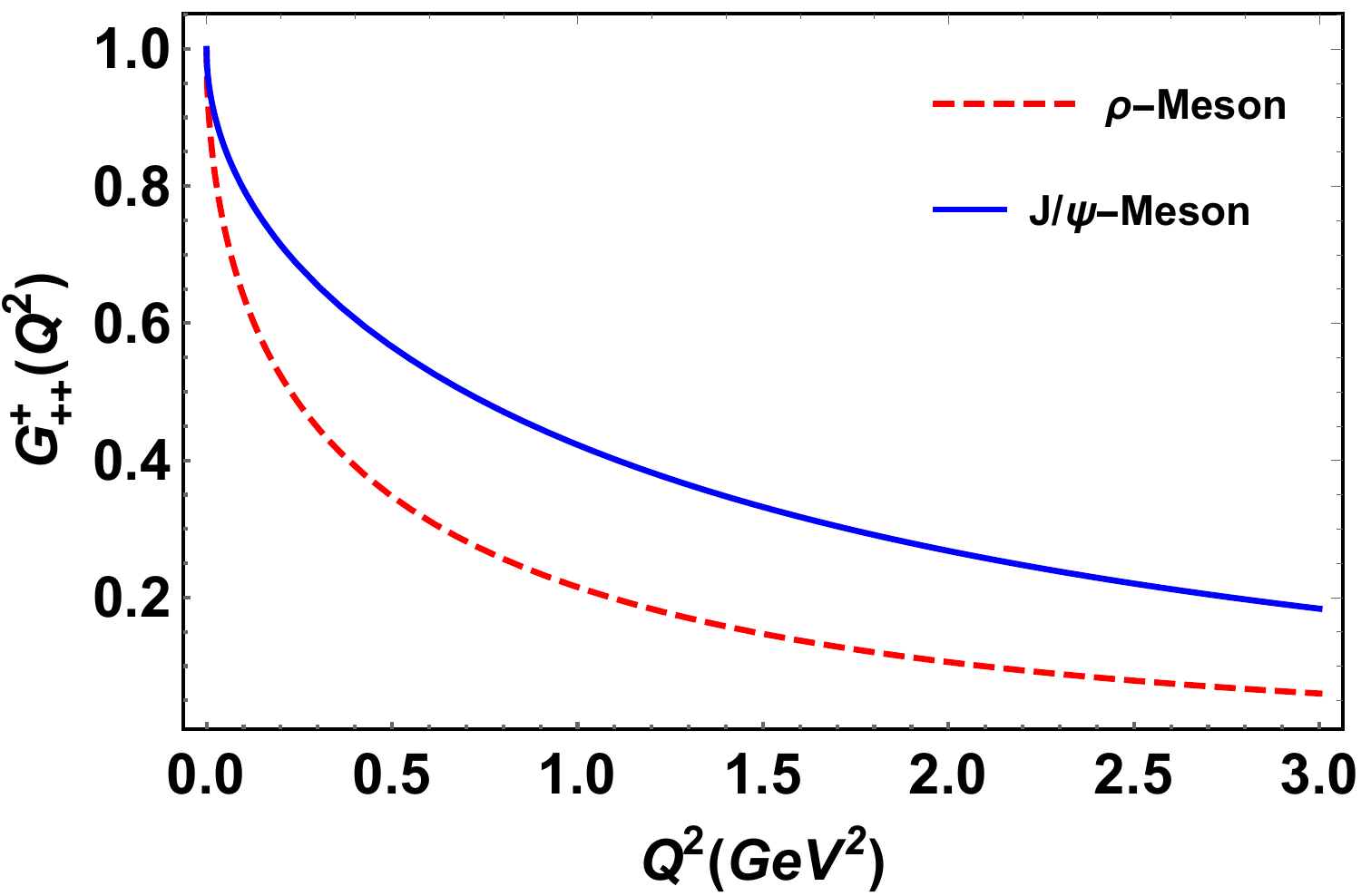}
		\hspace{0.03cm}	
		(b)\includegraphics[width=7.5cm]{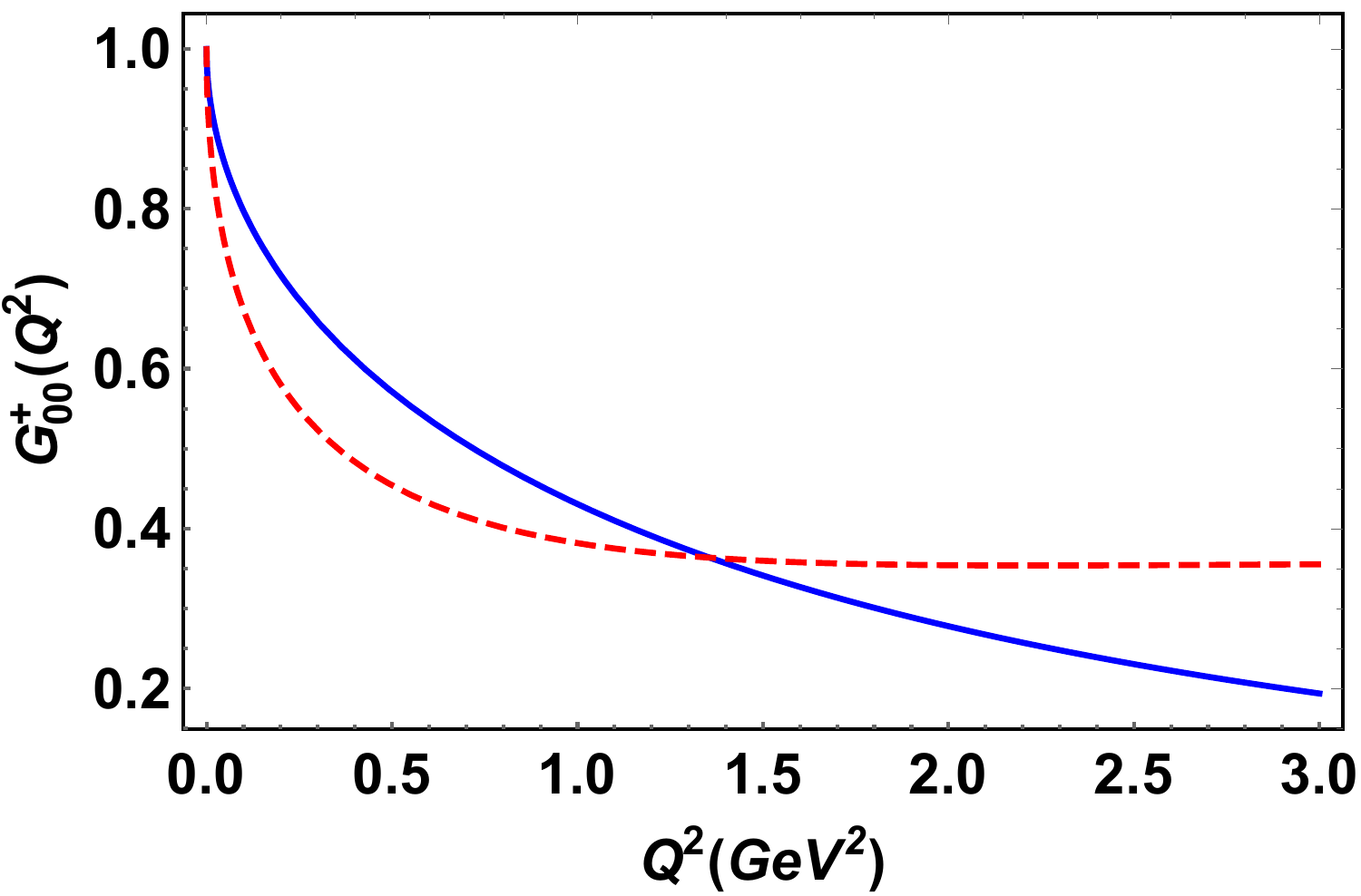} 
		\hspace{0.03cm}
	\end{minipage}
	\centering
	\begin{minipage}[c]{0.98\textwidth}
		(c)\includegraphics[width=7.5cm]{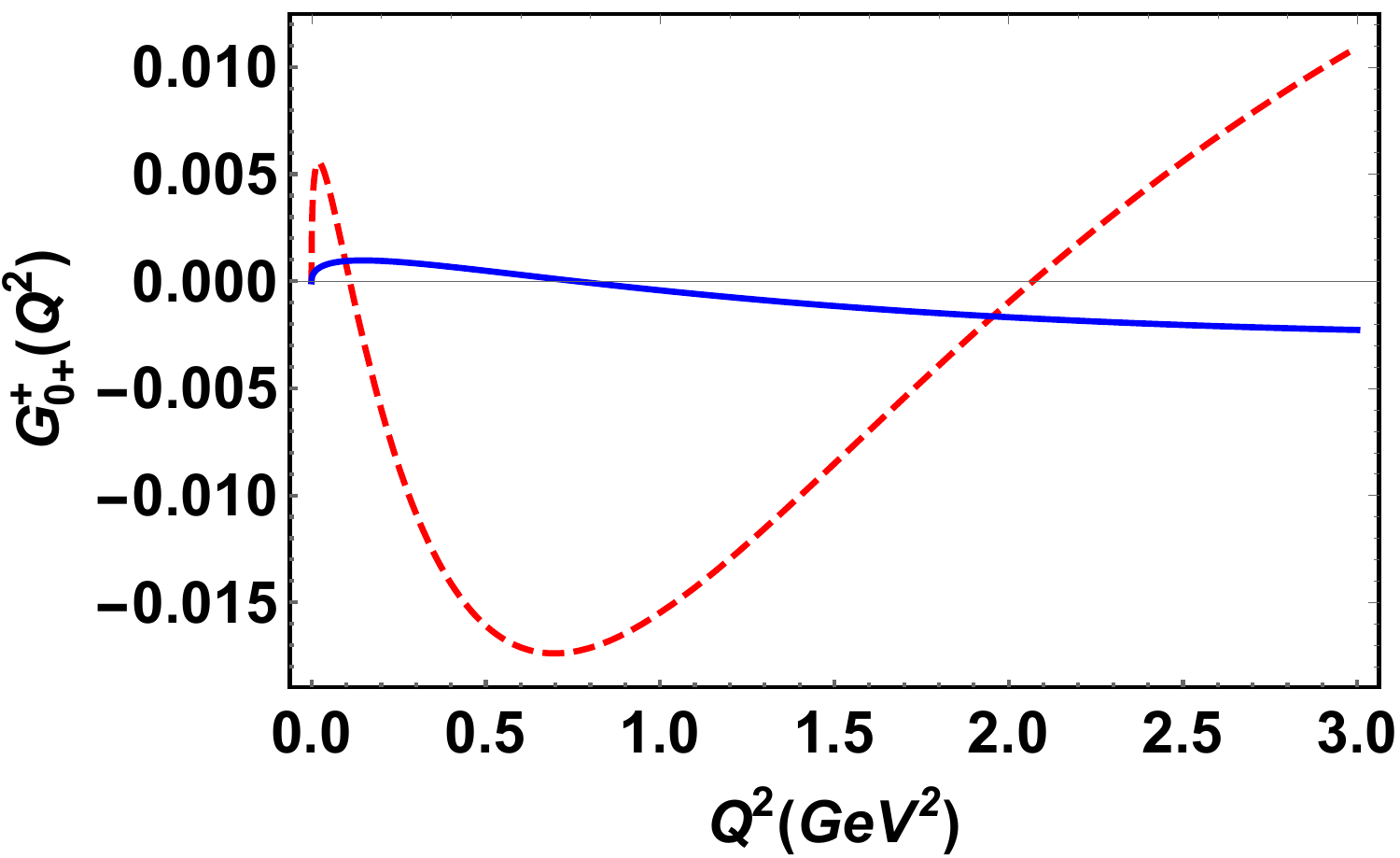}
		\hspace{0.03cm}	
		(d)\includegraphics[width=7.5cm]{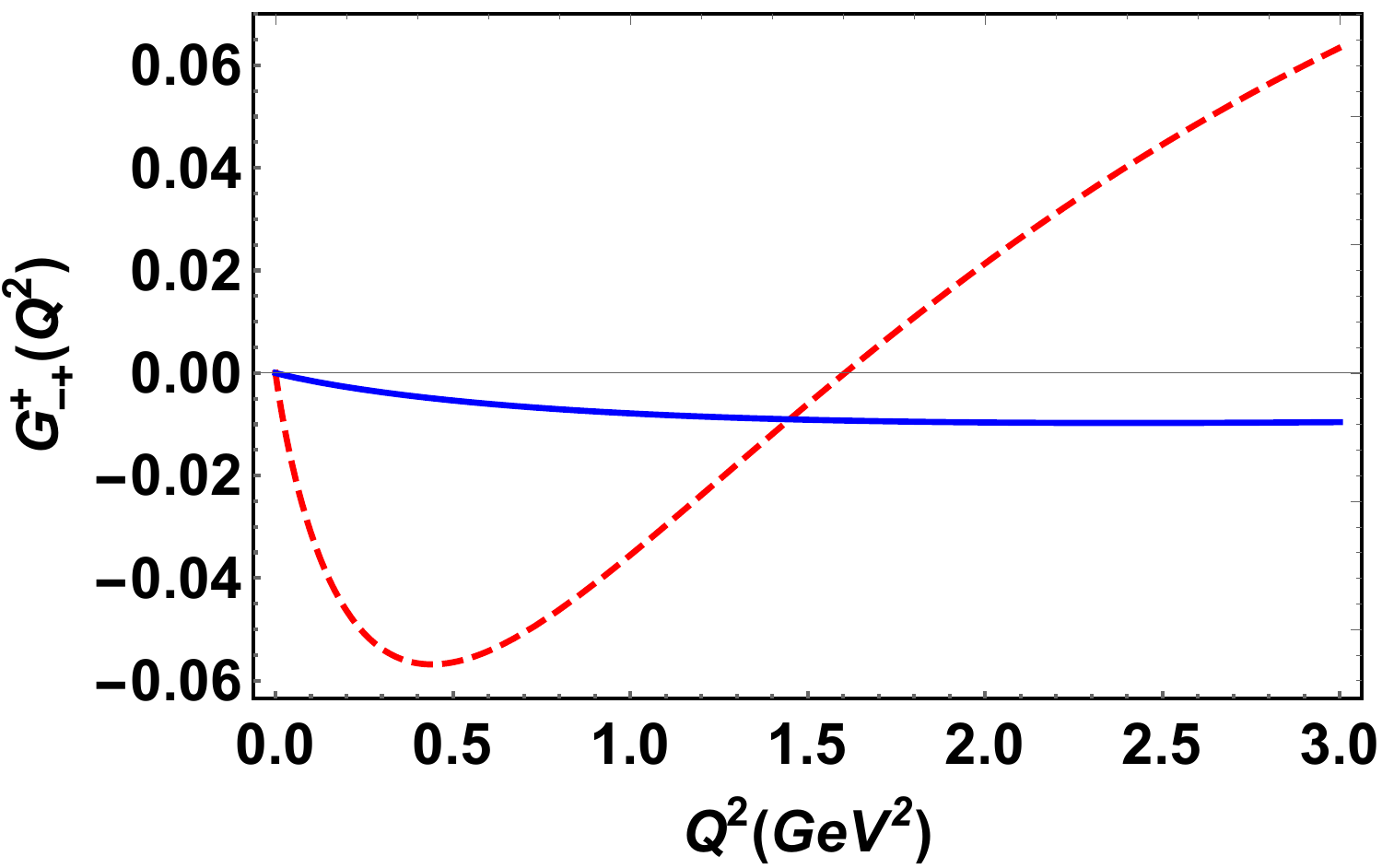} 
		\hspace{0.03cm}
	\end{minipage}
	\caption{\label{fig11} (Color online) The helicity-conserving matrix elements $G^+_{++}(Q^2)$ and $G^+_{00}(Q^2)$ have been plotted with respect to $Q^2$ in (a) and (b), respectively, for both $\rho$ and $J/\psi$-mesons. The helicity-flip matrix element with one unit $G^+_{0+}(Q^2)$ and two units $G^+_{-+}(Q^2)$ of helicity-flip has been plotted in (c) and (d), respectively, for both $\rho$ and $J/\psi$-mesons.}
\end{figure*}

Futher, using $G_C(Q^2)$, $G_M(Q^2)$, and $G_Q(Q^2)$ form factors, one can define the conserving ($G^+_{++}(Q^2)$ and $G^+_{00}(Q^2)$) and non-conserving ($G^+_{-+}(Q^2)$ and $G^+_{0+}(Q^2)$) helicity matrix element as 
\begin{eqnarray}
	G_{++}^+(Q^2)&=&\frac{1}{1+\eta}\left(G_C(Q^2)+\eta G_M(Q^2)+\frac{\eta}{3}G_Q(Q^2)\right)  \,, \\
	G_{00}^+(Q^2)&=&\frac{1}{1+\eta}\left((1-\eta)G_C(Q^2)+2\eta G_M(Q^2)-\frac{2\eta}{3}(1+2\eta)G_Q(Q^2)\right)\,, \\
	G_{0+}^+(Q^2)&=&-\frac{\sqrt{2\eta}}{1+\eta}\left(G_C(Q^2)-\frac{1}{2}(1-\eta)G_M(Q^2)+\frac{\eta}{3}G_Q(Q^2)\right)\,, \\
	G_{-+}^+(Q^2)&=&\frac{\eta}{1+\eta}\left(G_C(Q^2)-G_M(Q^2)-(1+\frac{2\eta}{3})G_Q(Q^2)\right).
\end{eqnarray}
Here, $+(-)$ denote the $+1(-1)$ spin projections. These matrix elements have been plotted in Fig. \ref{fig11} for both the mesons. Both the conserving matrix elements $G^+_{++}$ and $G^+_{00}$ shows positive behavior and are found to be unity at $Q^2=0$. A similar kind of results have also been seen in NJL model \cite{Zhang:2024nxl} and LFQM \cite{Kumar:2019eck}. The non-conserving ($G^+_{-+}(Q^2)$ and $G^+_{0+}(Q^2)$) helicity matrix elements show both positive and negative distributions. The $G_{0+}^+$ elements have negative distribution in between $0.1\le Q^2 \le 2.1$ GeV$^2$, while $G_{-+}^+$ shows negative distributions around $0 \le Q^2\le 1.7 $ GeV$^2$. In the case of LFQM \cite{Kumar:2019eck} and NJL model \cite{Zhang:2024nxl}, the non-conserving $G_{-+}^+$ matrix element shows only a negative distribution. Another important observation in this regard is that the non-conserving matrix elements have lower distributions compared to the conserving matrix elements.

\begin{figure*}
	(a)\includegraphics[width=7.5cm]{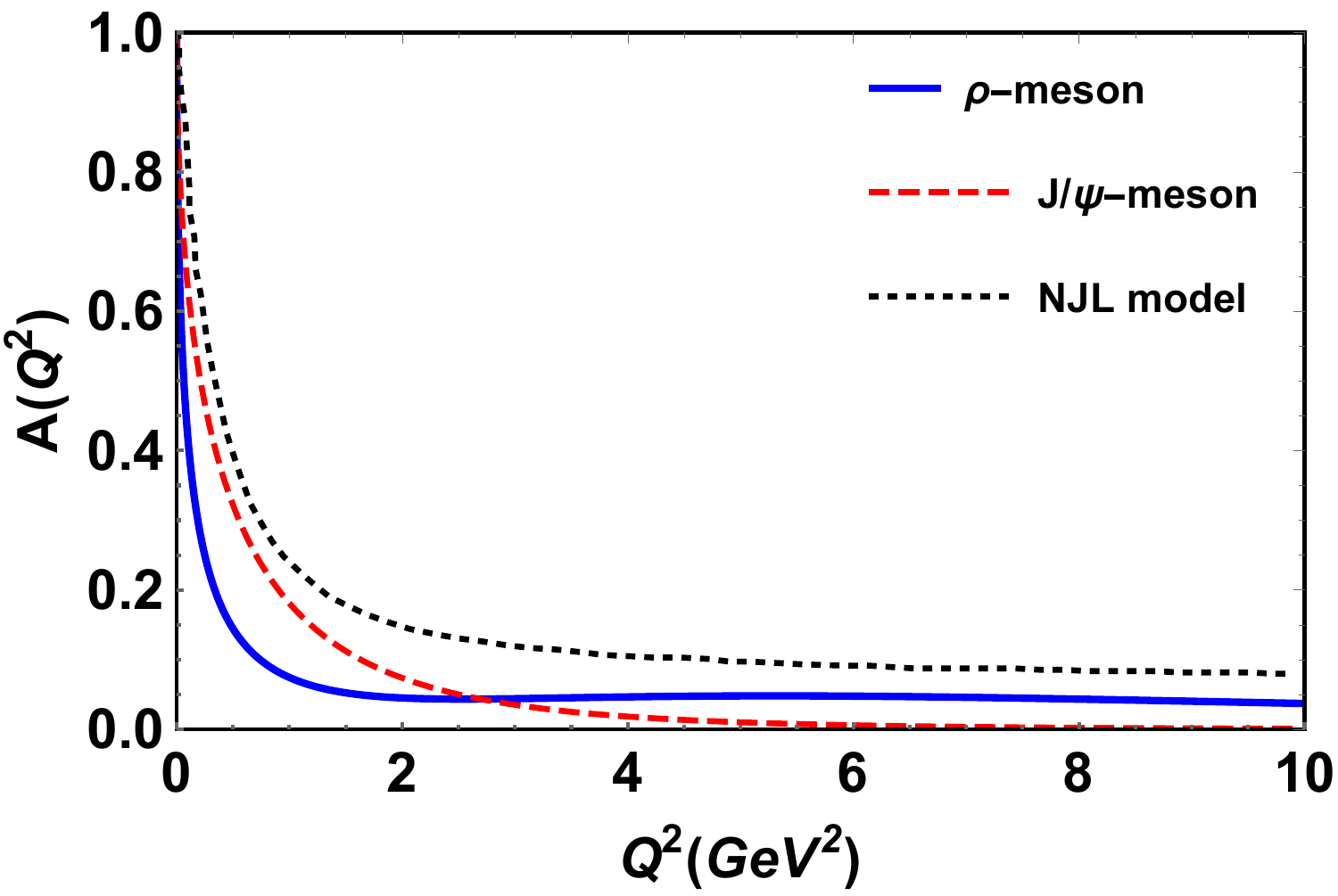}
	\hspace{0.03cm}	
	(b)\includegraphics[width=7.5cm]{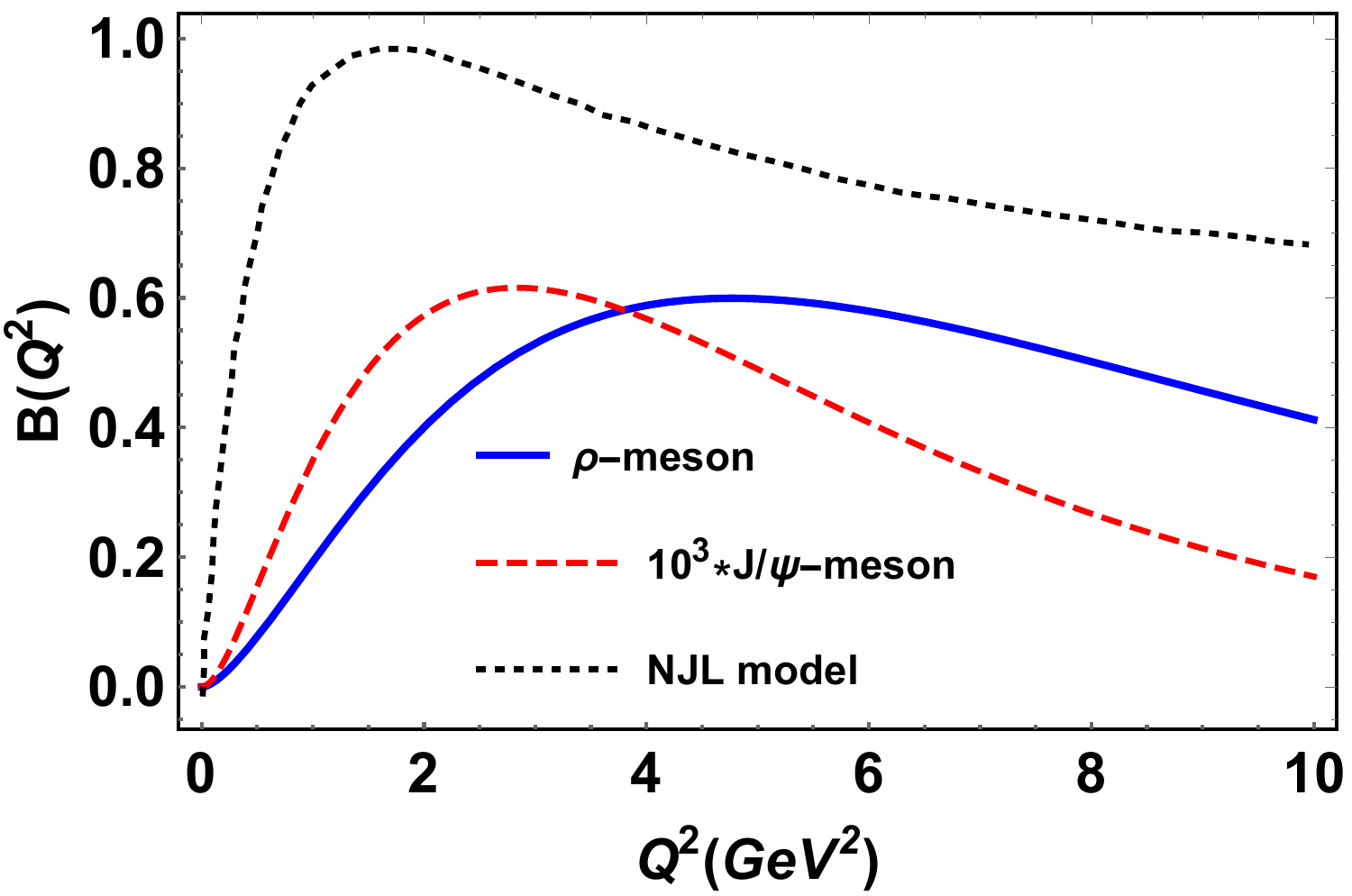} 
	\caption{\label{fig12} (Color online) The $A(Q^2)$ and $B(Q^2)$ structure function of light $\rho$ and heavy $J/\psi$ vector meson along with comparison with NJL model.}
\end{figure*}

An important reason to study these $G_C(Q^2)$, $G_M(Q^2)$, and $G_Q(Q^2)$ form factors is that they can be directly connected with the Rosenbluth cross section \cite{Hofstadter:1957wk} for elastic
electron scattering on a target of arbitrary spin in the
laboratory frame through $A(Q^2)$ and $B(Q^2)$ structure function along with tensor polarization $T_{20}(Q^2,\phi)$. The $A(Q^2)$, $B(Q^2)$, and $T_{20}(Q^2,\phi)$ can be calculated from $G_C(Q^2)$, $G_M(Q^2)$, and $G_Q(Q^2)$ as \cite{Zhang:2024nxl,Haftel:1980zz}
\begin{eqnarray}
	A(Q^2)&=&G_C^2(Q^2)+2/3\eta G_M^2(Q^2)+2(2/3)^2\eta^2G_Q^2(Q^2)\,,\\
	B(Q^2)&=&\frac{4}{3}\eta(1+\eta) G_M^2\,,\\
	T_{20}(Q^2,\phi)&=&-\eta\frac{\sqrt{2}}{3}
	\frac{\frac{4}{3}\eta G_Q^2(Q^2)+4G_Q(Q^2)G_C(Q^2)+(1/2+(1+\eta\tan^2\frac{\phi}{2}))G_M^2(Q^2) }{A(Q^2)+B(Q^2)\tan^2\frac{\phi}{2}}.
\end{eqnarray}
Using these, the cross-section of can be calculated as \cite{Hofstadter:1957wk,Zhang:2024nxl}
\begin{eqnarray}
	\frac{\mathrm{d}\sigma}{\mathrm{d}(-Q^2)}=\frac{4\pi \alpha^2}{(-Q^2)^2}\left[\left(1+\frac{(-Q^2)s}{(s-M_{\rho(J/\psi)}^2)^2}\right)A(Q^2)-\frac{M_{\rho(J/\psi)}^2(-Q^2) }{(s-M_{\rho(J/\psi)}^2)^2}B(Q^2) \right].
\end{eqnarray}
As there is no experimental data available for $A(Q^2)$ and $B(Q^2)$ structure functions,  our results can be compared  with the NJL model calculations \cite{Zhang:2024nxl} which exhibits a similar behavior as shown in Fig. \ref{fig12}. The $A(Q^2)$ structure function saturates after $Q^2=1.5$ GeV$^2$, as in the case of NJL model and LFQM \cite{Choi:2004ww}. Analogous to the form factor results, the structure functions also have a similar kind of distribution with slighter magnitudes. For the case of heavy $J/\psi$-vector meson, the $B(Q^2)$ structure function is very low compared to the $\rho$-meson. However, both the structure and function obey the sum rule
\begin{eqnarray}
	A(Q^2=0)=1,  \ \ \ \ B(Q^2=0)=0.
\end{eqnarray}

\section{Spin-1 parton distribution functions}\label{pdfs}
\begin{figure*}
	\centering
	\begin{minipage}[c]{0.98\textwidth}
		(a)\includegraphics[width=7.5cm]{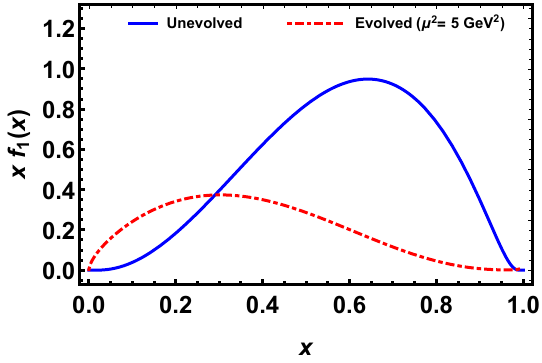}
		\hspace{0.03cm}	
		(b)\includegraphics[width=7.5cm]{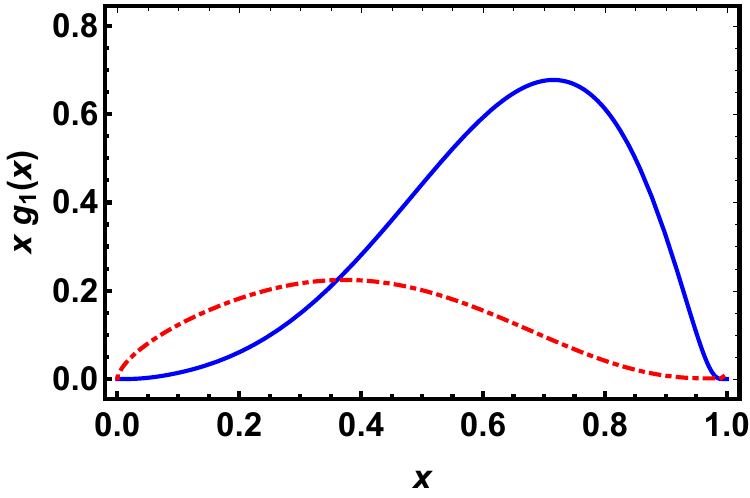} 
		\hspace{0.03cm}
	\end{minipage}
	\centering
	\begin{minipage}[c]{0.98\textwidth}
		(c)\includegraphics[width=7.5cm]{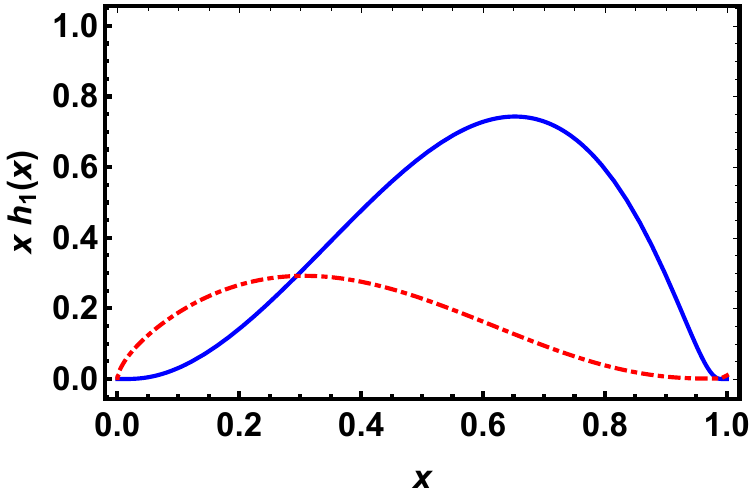}
		\hspace{0.03cm}	
	\end{minipage}
	\caption{\label{fig13} (Color online) The unpolarized $xf_1(x)$, helicity $xg_1(x)$, and transversity polarized $xh_1(x)$ quark PDFs have been plotted with respect to longitudinal momentum fraction $x$ for light $\rho$-meson at model scale $\mu^2_0=0.19$ GeV$^2$ (blue solid line) and at scale $\mu^2=5$ GeV$^2$ (red dot dashed line) in (a), (b) and (c), respectively.}
\end{figure*}

\begin{figure*}
	\centering
	\begin{minipage}[c]{0.98\textwidth}
		(a)\includegraphics[width=7.5cm]{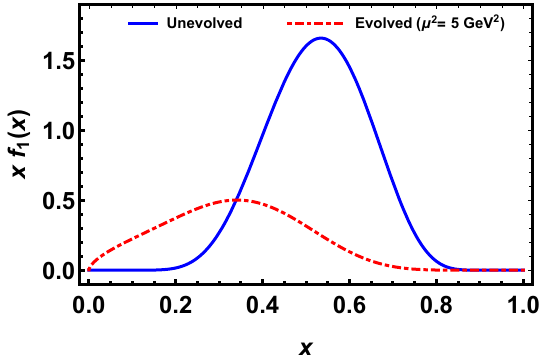}
		\hspace{0.03cm}	
		(b)\includegraphics[width=7.5cm]{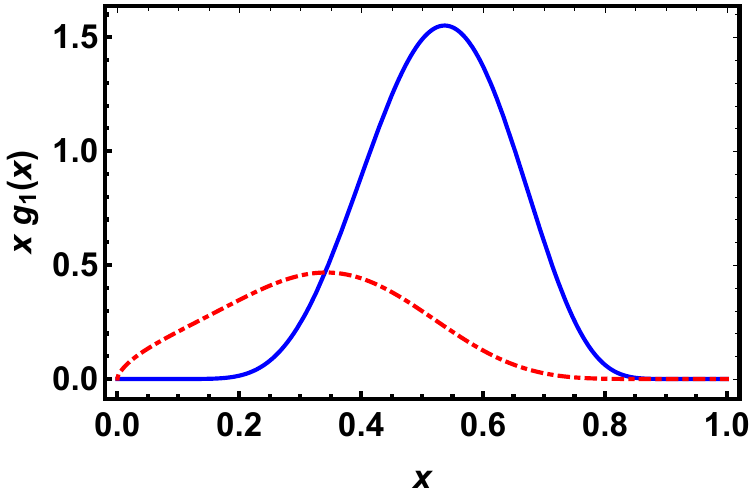} 
		\hspace{0.03cm}
	\end{minipage}
	\centering
	\begin{minipage}[c]{0.98\textwidth}
		(c)\includegraphics[width=7.5cm]{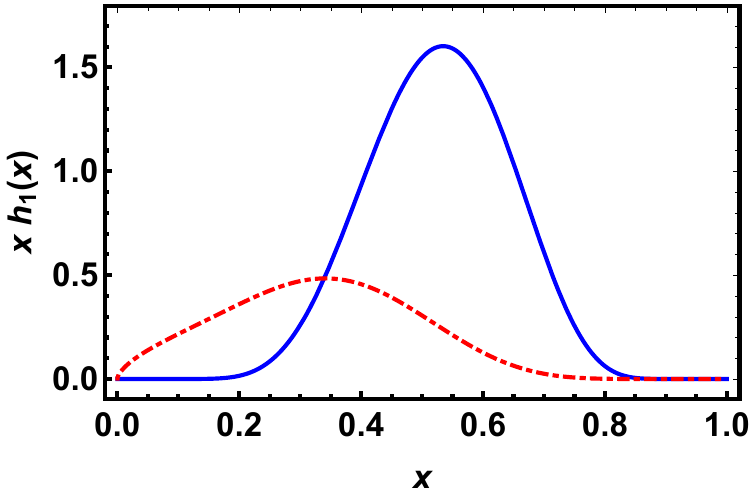}
		\hspace{0.03cm}	
	\end{minipage}
	\caption{\label{fig14} (Color online) The unpolarized $xf_1(x)$, helicity $xg_1(x)$, and transversity polarized $xh_1(x)$ quark PDFs have been plotted with respect to longitudinal momentum fraction $x$ for heavy $J/\psi$-meson at model scale $\mu^2_0=0.19$ GeV$^2$ (blue solid line) and at scale $\mu^2=5$ GeV$^2$ (red dot dashed line) in (a), (b) and (c), respectively.}
\end{figure*}

Without knowing the parton's inherent transverse momentum, the PDFs encode the longitudinal momentum distribution and the polarization carried by the partons at $P=P^{\prime}$ and $\Delta_\perp=0$. For the case of spin-1 vector mesons, there are a total of four quark PDFs present at the leading twist, which are unpolarized $f_1(x)$, transversity polarized $h_1(x)$, helicity $g_1(x)$, and tensor polarized $f_{1ll}(x)$ quark PDFs \cite{Kumano:2021fem}. From the unpolarized GPDs, we can only derive quark $f_1(x)$ and $f_{1LL}(x)$ PDFs. For the polarized quark $g_1(x)$ and $h_1(x)$ PDFs, we have considered the possible LFWFs. These quark PDFs can be calculated using the overlap form of LFWFs and from GPDs limit as \cite{Bacchetta:2000jk,Hino:1999qi,Puhan:2023hio,Zhang:2022zim}
\begin{eqnarray}
	f_1(x)&=&H_1(x,0,0),\nonumber\\
	&=&\int \frac{d^2\bfk}{2(2\pi)^3}\frac{1}{3} \sum_{\lambda_1,\lambda_2}\Bigg( |\Psi_{S_z=0}(x,\bfk,\lambda_1,\lambda_2)|^2+|\Psi_{S_z=1}(x,\bfk,\lambda_1,\lambda_2)|^2+|\Psi_{S_z=-1}(x,\bfk,\lambda_1,\lambda_2)|^2\Bigg),\\
	f_{1LL}(x)&=&H_5(x,0,0)\nonumber\\
	&=&b_1(x)\nonumber\\
	&=&\int \frac{d^2\bfk}{2(2\pi)^3} \sum_{\lambda_1,\lambda_2}\Bigg( |\Psi_{S_z=0}(x,\bfk,\lambda_1,\lambda_2)|^2-\Big(|\Psi_{S_z=1}(x,\bfk,\lambda_1,\lambda_2)|^2+|\Psi_{S_z=-1}(x,\bfk,\lambda_1,\lambda_2)|^2\Big)\Bigg),\\
	h_1(x)&=&\int \frac{d^2\bfk}{2(2\pi)^3}\frac{1}{2\sqrt{2}} \sum_{\lambda_1,\lambda_2}\Bigg( \Psi^{*}_{S_z=1}(x,\bfk,\uparrow,\lambda_2)\Psi_{S_z=0}(x,\bfk,\downarrow,\lambda_2)+\Psi^{*}_{S_z=0}(x,\bfk,\downarrow,\lambda_2)\Psi_{S_z=1}(x,\bfk,\uparrow,\lambda_2)\nonumber\\
	&+&\Psi^{*}_{S_z=-1}(x,\bfk,\downarrow,\lambda_2)\Psi_{S_z=0}(x,\bfk,\uparrow,\lambda_2)+\Psi^{*}_{S_z=0}(x,\bfk,\uparrow,\lambda_2)\Psi_{S_z=-1}(x,\bfk,\downarrow,\lambda_2)\Bigg),\\
	g_1(x)&=&\int \frac{d^2\bfk}{2(2\pi)^3}\frac{1}{2}\sum_{\lambda_1,\lambda_2} \Bigg(|\Psi_{S_z=1}(x,\bfk,\uparrow,\lambda_2)|^2-\Psi_{S_z=1}(x,\bfk,\downarrow,\lambda_2)|^2\nonumber\\
	&+&|\Psi_{S_z=-1}(x,\bfk,\downarrow,\lambda_2)|^2-\Psi_{S_z=-1}(x,\bfk,\uparrow,\lambda_2)|^2\Bigg).
\end{eqnarray}
The explicit form of these quark PDFs have been given in the appendix (Sec. \ref{app}). The tensor polarized $f_{1LL}$ carries information about the $b_1$ structure function of the vector mesons \cite{Cosyn:2017fbo}. However, in this work, the $f_{1LL}(x)$ is found to be zero.
Further, these PDFs obey the sum rule 
\begin{eqnarray}
	\int f_1(x) dx=1, \ \ \ \ \int f_{1LL} (x) dx=b_1(x) dx=0.
\end{eqnarray}
These quark PDFs  have been plotted with respect to the longitudinal momentum fraction $x$ in Figs. \ref{fig13} and \ref{fig14} for both light $\rho$ and heavy $J/\psi$ vector mesons, respectively. The PDfs have been plotted at the model scale ($\mu^2=0.19$ GeV$^2$) as well as at $\mu^2=5$ GeV$^2$ for which the evolution has been carried out using next to next to leading order Dokshitzer–Gribov–Lipatov–Altarelli–Parisi (DGLAP) equations \cite{Miyama:1995bd,Hirai:1997gb,Hirai:1997mm}. All the quark PDFs are found to have positive distributions and a maxima after $x=0.5$. From the $\rho$-meson quark PDFs, we observe that unpolarized $f_1(x)$ have higher distributions compared to $g_1(x)$ and $h_1(x)$ quark PDFs. Similar trend can also be seen at $\mu^2=5$ GeV$^2$. The $\rho$-meson quark PDFs have been explored in NJL model \cite{Ninomiya:2017ggn,Zhang:2024plq}, DS-BSE approach \cite{Shi:2023oll}, LFHM \cite{Kaur:2020emh} etc. Our results are found to have a similar kind of positive distributions as in these models. The tensor $f_{1LL}$ quark PDF is found to be zero for both the particles. Due to heavy quark mass of the $J/\psi$-meson, the quark PDFs show a sharp distribution around $x=0.5$ and a symmetry $x\leftrightarrow(1-x)$. We observe that there is no distribution of quark PDFs below $x\le 0.2$ and above $x\ge 0.8$ for $J/\psi$-meson. In LFHM \cite{Puhan:2023hio}, DS-BSE approach \cite{Shi:2023oll}, BLFQ \cite{Lan:2019img}, and in Ref. \cite{Li:2021cwv},  similar results and behavior have been reported. The positivity constraints on the PDFs mentioned in Refs. \cite{Puhan:2023hio,Kaur:2020emh,Ninomiya:2017ggn} of spin-$1$ mesons are also verified in our calculations. The average momentum fraction of order $n$ (Mellin moments) carried by the quark PDFs can be calculated as 
\begin{eqnarray}
	\langle x^n \rangle = \frac{\int dx x^n PDFs}{\int dx PDFs}.
\end{eqnarray}
The $ \langle x^n \rangle$ of both particles has been presented in Table \ref{tab2}. It is clear from the results that the unpolarized $f_1(x)$ carry the highest average momenta as compared to the other PDFs.



\begin{table}
	\centering
	\begin{tabular}{|c| c |c|c|c|c|}
		\hline
		& &$\sqrt{\langle r^2_c \rangle}$ fm & $ \mu_p$   & $ Q_p $   \\\hline
		$\rho(u\bar d)$ &Our result&0.95 & 2.19 & -0.023   \\
		& Ref. \cite{Bhagwat:2006pu} & 0.73 & 2.01 & -0.026 \\
		& Ref. \cite{Owen:2015gva} & 0.82& 2.07 & -0.045    \\
		&Ref.  \cite{Shultz:2015pfa}&0.82 & 2.48 & -0.070   \\
		&Ref. \cite{Hernandez-Pinto:2024kwg}& 0.56& 0.75 & -   \\
		&Ref. \cite{Luan:2015goa}&1.12 & 2.54 & -    \\
		
		$J/\psi(c\bar c)$&Our result& 0.69& 2.05 & -0.006   \\
		&Ref. \cite{Hernandez-Pinto:2024kwg}& 0.350& 2.03 & -   \\
		&Ref. \cite{Bhagwat:2006pu}& 0.24& 2.12 & -   \\
		&Ref. \cite{Dudek:2006ej}&0.066& 2.10 & -    \\\hline
	\end{tabular}
	\caption{The charge radii $\sqrt{\langle r^2_c \rangle}$ (in fm), magnetic moment $\mu_p$ (in units of Bohr magneton), and quadrupole moment $Q_P$ (in fm$^2$) of light $\rho$ and heavy $J/\psi$ vector mesons have been compared with available data \cite{Bhagwat:2006pu,Dudek:2006ej,Hernandez-Pinto:2024kwg,Luan:2015goa,Shultz:2015pfa,Owen:2015gva}.}
	\label{table1}
\end{table}

\begin{table}[]
	\centering
	\begin{tabular}{|c|c|c|c|c|}
		\hline
		\raisebox{-2.5ex}{\shortstack{PDFs}} & \multicolumn{2}{c|}{\(\langle x^{0}\rangle\)}& \multicolumn{2}{c|}{\(\langle x^{1}\rangle\)} \\
		\cline{2-5}
		& \(\rho\) &\(J/\psi\) & \(\rho\) &\(J/\psi\) \\
		\hline
		\(f_{1} (x)\)&1 &1&0.5 &0.5\\
		\hline
		\(g_{1} (x)\)&0.559&0.932&0.314&0.467\\
		\hline
		\(h_{1} (x)\)&0.779&0.966&0.392&0.483\\
		\hline       
	\end{tabular}
	
	\caption{The average longitudinal momentum fractions \(\langle x^{0(1)}\rangle\) for both \(\rho\) and \(J/\psi\) vector mesons. }
	\label{tab2}
\end{table}

\section{Conclusion}\label{con}
In this work, we have evaluated the unpolarized quark generalized parton distributions (GPDs) of spin-1 vector \(\rho\) and \(J/\psi \) mesons at zero skewness, using the light-cone quark model (LCQM). We have employed the Brodsky-Huang-Lepage (BHL) prescription as the momentum space wave function, and the spin part of the wave function is obtained from the Melosh-Wigner rotation. 
\par For spin-1 vector mesons, the unpolarized quark GPDs have been calculated using the correlator \(V_{S'_{z}, S_{z}}\), which is an overlap of the initial and final state of the light-front wave function. Out of five unpolarized GPDs, $H_4(x,0,-\Delta_\perp^2)$ is comes out to be zero. We have observed the three-dimensional behavior of the quark GPDs \(H_{1,2,3} (x,0,- \Delta_{\perp}^{2})\) with transverse momentum transferred and longitudinal momentum fraction. The GPDs of \(J/\psi\) mesons exhibit a sharper peak in \(x\) as compared to that in \(\rho\) mesons which can be attributed to the active quark carrying more momentum at lower \(x\) for the case of \(\rho\) mesons. Further, a symmetric distribution of \(H_{1} (x,0,- \Delta_{\perp}^{2})\), as a function of \(x\) and \(\Delta_{\perp}^{2}\) along with a contribution of superpositioned S, P and D-waves, has been observed for both the vector mesons which is due to identical quark content. On the other hand, the distribution \(H_{3} (x,0,- \Delta_{\perp}^{2})\) has a contribution of D-wave and \(H_{2} (x,0,- \Delta_{\perp}^{2})\) has a contribution from S and P-wave. The behavior of these unpolarized quark GPDs has also been studied at finite values of \(x\) and \(\Delta_{\perp}^{2}\). Additionally, we have examined the unpolarized quark GPDs in impact parameter space which is accomplished by taking the GPD's Fourier transform in momentum space.

In light of the recent progress, the electromagnetic form factors have also been extracted from the respective GPDs and have been further used to calculate the charge \(G_{C}(Q^{2})\), magnetic \(G_{M}(Q^{2})\) and quadrupole \(G_{Q}(Q^{2})\) form factors for both \(\rho\) and \(J/\psi\) mesons. Our results show a similar trend when compared with the lattice stimulation and other theoretical model results. Subsequently, physical observables such as the charge radii, magnetic moment, and quadrupole moment for both the vector mesons have also been computed and compared with the available data. We have also predicted the $A(Q^2)$ and $B(Q^2)$ structure functions for the Rosenbluth cross section in terms of charge, magnetic, and quadrupole form factors. Furthermore, we have also evaluated the four parton distribution functions (PDFs), namely, unpolarized \(f_{1}(x)\), helicity \(g_{1}(x)\), transversity \(h_{1}(x)\) and tensor \(f_{1LL}(x)\) PDFs of \(\rho\) and \(J/\psi\) mesons in LCQM at leading twist. Additionally, we have also evolved our PDFs from our model scale to higher scale \(\mu^{2}\) for both the vector mesons. This study may be significant to understand the $\rho$-meson lepton production \cite{HERMES:2017qwt,Cano:2003ju,Mankiewicz:1997uy,CLAS:2008rpm}, $\gamma \gamma^{*}\to \rho \rho$ \cite{Anikin:2003fr} process as well as the results from the future Electron-Ion 
Collider (EIC) experiments \cite{Boer:2011fh,AbdulKhalek:2021gbh}.

\section{Acknowledgement}
H.D. would like to thank  the Science and Engineering Research Board, Anusandhan-National Research Foundation, Government of India under the scheme SERB-POWER Fellowship (Ref No. SPF/2023/000116) for financial support.

\section{Appendix}\label{app}
The explicit form of quark PDFs (by considering $m_q=m_{\bar q}=m$) can be expressed as
\begin{eqnarray}
	f_1(x)&=& \int \frac{\mathrm{d}^2\mathbf{k}_{\perp}}{3 (16\pi^3)} \bigg[ ( (\mathbf{k}_{\perp} (1 - 2 x) M_{q \bar q})^2 + (2 \mathbf{k}_{\perp}^2+ (M_{q \bar q} + 2 m) m )^2) +
	\nonumber \\
	&+& 2 ( ( \mathbf{k}_{\perp}^2 (M_{q \bar q}
	+ 2 m) m ) + \mathbf{k}_{\perp}^2 (x M_{q \bar q} + m) + ( -\mathbf{k}_{\perp} (1 - x) M_{q \bar q} + m)^2 + \mathbf{k}_{\perp}^4 ) \bigg] \frac{|\psi_{q \bar q}(x, \mathbf{k}_{\perp}|^2}{\omega_{q \bar q}^2},\\
	g_{1} (x) &=& \int \frac{\mathrm{d}^2\mathbf{k}_{\perp}}{16 \pi^3} \bigg[ (\mathbf{k}_{\perp}^2 + (M_{q \bar q} + 2 m ) m)^2+ (\mathbf{k}_{\perp} (x M_{q \bar q} + m))^2 \nonumber\\
	&-& \mathbf{k}_{\perp}^4 - (\mathbf{k}_{\perp} ((1 - x) M_{q \bar q} + m))^2 \bigg] \frac{|\psi_{q \bar q}(x, \mathbf{k}_{\perp}|^2}{\omega_{q \bar q}^2}, \\
	h_1 (x) &=& \int \frac{\mathrm{d}^2\mathbf{k}_{\perp}}{16 \pi^3}  \bigg[ ( (2 \mathbf{k}_{\perp}^2 + (M_{q \bar q} + 2m)m) ( \mathbf{k}_{\perp}^2 +(M_{q \bar q} +2m )m) )\nonumber\\
	&+& (\frac{1}{2} \mathbf{k}_{\perp}^2 (1 - 2 x) (1 - x) M_{q \bar q}^2) - (\frac{1}{2} (x M_{q \bar q} + m) (1 - 2 x) M_{q \bar q} )\bigg]  \frac{|\psi_{q \bar q}(x, \mathbf{k}_{\perp}|^2}{\omega_{q \bar q}^2},\\
	f_{1ll}&=&0.
\end{eqnarray}

\section{Reference}

\bibliography{ref}

\begin{thebibliography}{95}%
\makeatletter
\providecommand \@ifxundefined [1]{%
 \@ifx{#1\undefined}
}%
\providecommand \@ifnum [1]{%
 \ifnum #1\expandafter \@firstoftwo
 \else \expandafter \@secondoftwo
 \fi
}%
\providecommand \@ifx [1]{%
 \ifx #1\expandafter \@firstoftwo
 \else \expandafter \@secondoftwo
 \fi
}%
\providecommand \natexlab [1]{#1}%
\providecommand \enquote  [1]{``#1''}%
\providecommand \bibnamefont  [1]{#1}%
\providecommand \bibfnamefont [1]{#1}%
\providecommand \citenamefont [1]{#1}%
\providecommand \href@noop [0]{\@secondoftwo}%
\providecommand \href [0]{\begingroup \@sanitize@url \@href}%
\providecommand \@href[1]{\@@startlink{#1}\@@href}%
\providecommand \@@href[1]{\endgroup#1\@@endlink}%
\providecommand \@sanitize@url [0]{\catcode `\\12\catcode `\$12\catcode
  `\&12\catcode `\#12\catcode `\^12\catcode `\_12\catcode `\%12\relax}%
\providecommand \@@startlink[1]{}%
\providecommand \@@endlink[0]{}%
\providecommand \url  [0]{\begingroup\@sanitize@url \@url }%
\providecommand \@url [1]{\endgroup\@href {#1}{\urlprefix }}%
\providecommand \urlprefix  [0]{URL }%
\providecommand \Eprint [0]{\href }%
\providecommand \doibase [0]{https://doi.org/}%
\providecommand \selectlanguage [0]{\@gobble}%
\providecommand \bibinfo  [0]{\@secondoftwo}%
\providecommand \bibfield  [0]{\@secondoftwo}%
\providecommand \translation [1]{[#1]}%
\providecommand \BibitemOpen [0]{}%
\providecommand \bibitemStop [0]{}%
\providecommand \bibitemNoStop [0]{.\EOS\space}%
\providecommand \EOS [0]{\spacefactor3000\relax}%
\providecommand \BibitemShut  [1]{\csname bibitem#1\endcsname}%
\let\auto@bib@innerbib\@empty
\bibitem [{\citenamefont {Volkov}\ and\ \citenamefont
  {Radzhabov}(2006)}]{Volkov:2005kw}%
  \BibitemOpen
  \bibfield  {author} {\bibinfo {author} {\bibfnamefont {M.~K.}\ \bibnamefont
  {Volkov}}\ and\ \bibinfo {author} {\bibfnamefont {A.~E.}\ \bibnamefont
  {Radzhabov}},\ }\bibfield  {title} {\bibinfo {title} {{The Nambu-Jona-Lasinio
  model and its development}},\ }\href
  {https://doi.org/10.1070/PU2006v049n06ABEH005905} {\bibfield  {journal}
  {\bibinfo  {journal} {Phys. Usp.}\ }\textbf {\bibinfo {volume} {49}},\
  \bibinfo {pages} {551} (\bibinfo {year} {2006})},\ \Eprint
  {https://arxiv.org/abs/hep-ph/0508263} {arXiv:hep-ph/0508263} \BibitemShut
  {NoStop}%
\bibitem [{\citenamefont {Arbuzov}\ \emph {et~al.}(2006)\citenamefont
  {Arbuzov}, \citenamefont {Volkov},\ and\ \citenamefont
  {Zaitsev}}]{Arbuzov:2006ia}%
  \BibitemOpen
  \bibfield  {author} {\bibinfo {author} {\bibfnamefont {B.~A.}\ \bibnamefont
  {Arbuzov}}, \bibinfo {author} {\bibfnamefont {M.~K.}\ \bibnamefont
  {Volkov}},\ and\ \bibinfo {author} {\bibfnamefont {I.~V.}\ \bibnamefont
  {Zaitsev}},\ }\bibfield  {title} {\bibinfo {title} {{NJL model derived from
  QCD}},\ }\href {https://doi.org/10.1142/S0217751X06033830} {\bibfield
  {journal} {\bibinfo  {journal} {Int. J. Mod. Phys. A}\ }\textbf {\bibinfo
  {volume} {21}},\ \bibinfo {pages} {5721} (\bibinfo {year} {2006})},\ \Eprint
  {https://arxiv.org/abs/hep-ph/0604051} {arXiv:hep-ph/0604051} \BibitemShut
  {NoStop}%
\bibitem [{\citenamefont {Ninomiya}\ \emph {et~al.}(2015)\citenamefont
  {Ninomiya}, \citenamefont {Bentz},\ and\ \citenamefont
  {Clo\"et}}]{Ninomiya:2014kja}%
  \BibitemOpen
  \bibfield  {author} {\bibinfo {author} {\bibfnamefont {Y.}~\bibnamefont
  {Ninomiya}}, \bibinfo {author} {\bibfnamefont {W.}~\bibnamefont {Bentz}},\
  and\ \bibinfo {author} {\bibfnamefont {I.~C.}\ \bibnamefont {Clo\"et}},\
  }\bibfield  {title} {\bibinfo {title} {{Dressed Quark Mass Dependence of Pion
  and Kaon Form Factors}},\ }\href {https://doi.org/10.1103/PhysRevC.91.025202}
  {\bibfield  {journal} {\bibinfo  {journal} {Phys. Rev. C}\ }\textbf {\bibinfo
  {volume} {91}},\ \bibinfo {pages} {025202} (\bibinfo {year} {2015})},\
  \Eprint {https://arxiv.org/abs/1406.7212} {arXiv:1406.7212 [nucl-th]}
  \BibitemShut {NoStop}%
\bibitem [{\citenamefont {Belyaev}\ and\ \citenamefont
  {Johnson}(1998)}]{Belyaev:1997iu}%
  \BibitemOpen
  \bibfield  {author} {\bibinfo {author} {\bibfnamefont {V.~M.}\ \bibnamefont
  {Belyaev}}\ and\ \bibinfo {author} {\bibfnamefont {M.~B.}\ \bibnamefont
  {Johnson}},\ }\bibfield  {title} {\bibinfo {title} {{Pion light cone wave
  functions and light front quark model}},\ }\href
  {https://doi.org/10.1016/S0370-2693(98)00152-X} {\bibfield  {journal}
  {\bibinfo  {journal} {Phys. Lett. B}\ }\textbf {\bibinfo {volume} {423}},\
  \bibinfo {pages} {379} (\bibinfo {year} {1998})},\ \Eprint
  {https://arxiv.org/abs/hep-ph/9707329} {arXiv:hep-ph/9707329} \BibitemShut
  {NoStop}%
\bibitem [{\citenamefont {Choi}\ \emph {et~al.}(2017)\citenamefont {Choi},
  \citenamefont {Ryu},\ and\ \citenamefont {Ji}}]{Choi:2017zxn}%
  \BibitemOpen
  \bibfield  {author} {\bibinfo {author} {\bibfnamefont {H.-M.}\ \bibnamefont
  {Choi}}, \bibinfo {author} {\bibfnamefont {H.-Y.}\ \bibnamefont {Ryu}},\ and\
  \bibinfo {author} {\bibfnamefont {C.-R.}\ \bibnamefont {Ji}},\ }\bibfield
  {title} {\bibinfo {title} {{Spacelike and timelike form factors for the
  $(\pi^0,\eta,\eta')\to\gamma^*\gamma$ transitions in the light-front quark
  model}},\ }\href {https://doi.org/10.1103/PhysRevD.96.056008} {\bibfield
  {journal} {\bibinfo  {journal} {Phys. Rev. D}\ }\textbf {\bibinfo {volume}
  {96}},\ \bibinfo {pages} {056008} (\bibinfo {year} {2017})},\ \Eprint
  {https://arxiv.org/abs/1708.00736} {arXiv:1708.00736 [hep-ph]} \BibitemShut
  {NoStop}%
\bibitem [{\citenamefont {Yadav}\ \emph {et~al.}(2025)\citenamefont {Yadav},
  \citenamefont {Puhan},\ and\ \citenamefont {Dahiya}}]{Yadav:2025txk}%
  \BibitemOpen
  \bibfield  {author} {\bibinfo {author} {\bibfnamefont {A.}~\bibnamefont
  {Yadav}}, \bibinfo {author} {\bibfnamefont {S.}~\bibnamefont {Puhan}},\ and\
  \bibinfo {author} {\bibfnamefont {H.}~\bibnamefont {Dahiya}},\ }\bibfield
  {title} {\bibinfo {title} {{Radiative Transitions for the Ground and Excited
  Charmonia States}},\ }\href@noop {} {\  (\bibinfo {year} {2025})},\ \Eprint
  {https://arxiv.org/abs/2504.14864} {arXiv:2504.14864 [hep-ph]} \BibitemShut
  {NoStop}%
\bibitem [{\citenamefont {Brodsky}\ \emph {et~al.}(2015)\citenamefont
  {Brodsky}, \citenamefont {de~Teramond}, \citenamefont {Dosch},\ and\
  \citenamefont {Erlich}}]{Brodsky:2014yha}%
  \BibitemOpen
  \bibfield  {author} {\bibinfo {author} {\bibfnamefont {S.~J.}\ \bibnamefont
  {Brodsky}}, \bibinfo {author} {\bibfnamefont {G.~F.}\ \bibnamefont
  {de~Teramond}}, \bibinfo {author} {\bibfnamefont {H.~G.}\ \bibnamefont
  {Dosch}},\ and\ \bibinfo {author} {\bibfnamefont {J.}~\bibnamefont
  {Erlich}},\ }\bibfield  {title} {\bibinfo {title} {{Light-Front Holographic
  QCD and Emerging Confinement}},\ }\href
  {https://doi.org/10.1016/j.physrep.2015.05.001} {\bibfield  {journal}
  {\bibinfo  {journal} {Phys. Rept.}\ }\textbf {\bibinfo {volume} {584}},\
  \bibinfo {pages} {1} (\bibinfo {year} {2015})},\ \Eprint
  {https://arxiv.org/abs/1407.8131} {arXiv:1407.8131 [hep-ph]} \BibitemShut
  {NoStop}%
\bibitem [{\citenamefont {de~Teramond}\ and\ \citenamefont
  {Brodsky}(2009)}]{deTeramond:2008ht}%
  \BibitemOpen
  \bibfield  {author} {\bibinfo {author} {\bibfnamefont {G.~F.}\ \bibnamefont
  {de~Teramond}}\ and\ \bibinfo {author} {\bibfnamefont {S.~J.}\ \bibnamefont
  {Brodsky}},\ }\bibfield  {title} {\bibinfo {title} {{Light-Front Holography:
  A First Approximation to QCD}},\ }\href
  {https://doi.org/10.1103/PhysRevLett.102.081601} {\bibfield  {journal}
  {\bibinfo  {journal} {Phys. Rev. Lett.}\ }\textbf {\bibinfo {volume} {102}},\
  \bibinfo {pages} {081601} (\bibinfo {year} {2009})},\ \Eprint
  {https://arxiv.org/abs/0809.4899} {arXiv:0809.4899 [hep-ph]} \BibitemShut
  {NoStop}%
\bibitem [{\citenamefont {Bacchetta}\ \emph {et~al.}(2022)\citenamefont
  {Bacchetta}, \citenamefont {Celiberto}, \citenamefont {Radici},\ and\
  \citenamefont {Taels}}]{Bacchetta:2021oht}%
  \BibitemOpen
  \bibfield  {author} {\bibinfo {author} {\bibfnamefont {A.}~\bibnamefont
  {Bacchetta}}, \bibinfo {author} {\bibfnamefont {F.~G.}\ \bibnamefont
  {Celiberto}}, \bibinfo {author} {\bibfnamefont {M.}~\bibnamefont {Radici}},\
  and\ \bibinfo {author} {\bibfnamefont {P.}~\bibnamefont {Taels}},\ }\bibfield
   {title} {\bibinfo {title} {{A spectator-model way to
  transverse-momentum-dependent gluon distribution functions}},\ }\href
  {https://doi.org/10.21468/SciPostPhysProc.8.040} {\bibfield  {journal}
  {\bibinfo  {journal} {SciPost Phys. Proc.}\ }\textbf {\bibinfo {volume}
  {8}},\ \bibinfo {pages} {040} (\bibinfo {year} {2022})},\ \Eprint
  {https://arxiv.org/abs/2107.13446} {arXiv:2107.13446 [hep-ph]} \BibitemShut
  {NoStop}%
\bibitem [{\citenamefont {Lorc\'e}\ \emph {et~al.}(2016)\citenamefont
  {Lorc\'e}, \citenamefont {Pasquini},\ and\ \citenamefont
  {Schweitzer}}]{Lorce:2016ugb}%
  \BibitemOpen
  \bibfield  {author} {\bibinfo {author} {\bibfnamefont {C.}~\bibnamefont
  {Lorc\'e}}, \bibinfo {author} {\bibfnamefont {B.}~\bibnamefont {Pasquini}},\
  and\ \bibinfo {author} {\bibfnamefont {P.}~\bibnamefont {Schweitzer}},\
  }\bibfield  {title} {\bibinfo {title} {{Transverse pion structure beyond
  leading twist in constituent models}},\ }\href
  {https://doi.org/10.1140/epjc/s10052-016-4257-8} {\bibfield  {journal}
  {\bibinfo  {journal} {Eur. Phys. J. C}\ }\textbf {\bibinfo {volume} {76}},\
  \bibinfo {pages} {415} (\bibinfo {year} {2016})},\ \Eprint
  {https://arxiv.org/abs/1605.00815} {arXiv:1605.00815 [hep-ph]} \BibitemShut
  {NoStop}%
\bibitem [{\citenamefont {Strodthoff}\ \emph {et~al.}(2012)\citenamefont
  {Strodthoff}, \citenamefont {Schaefer},\ and\ \citenamefont {von
  Smekal}}]{Strodthoff:2011tz}%
  \BibitemOpen
  \bibfield  {author} {\bibinfo {author} {\bibfnamefont {N.}~\bibnamefont
  {Strodthoff}}, \bibinfo {author} {\bibfnamefont {B.-J.}\ \bibnamefont
  {Schaefer}},\ and\ \bibinfo {author} {\bibfnamefont {L.}~\bibnamefont {von
  Smekal}},\ }\bibfield  {title} {\bibinfo {title} {{Quark-meson-diquark model
  for two-color QCD}},\ }\href {https://doi.org/10.1103/PhysRevD.85.074007}
  {\bibfield  {journal} {\bibinfo  {journal} {Phys. Rev. D}\ }\textbf {\bibinfo
  {volume} {85}},\ \bibinfo {pages} {074007} (\bibinfo {year} {2012})},\
  \Eprint {https://arxiv.org/abs/1112.5401} {arXiv:1112.5401 [hep-ph]}
  \BibitemShut {NoStop}%
\bibitem [{\citenamefont {Maji}\ and\ \citenamefont
  {Chakrabarti}(2016)}]{Maji:2016yqo}%
  \BibitemOpen
  \bibfield  {author} {\bibinfo {author} {\bibfnamefont {T.}~\bibnamefont
  {Maji}}\ and\ \bibinfo {author} {\bibfnamefont {D.}~\bibnamefont
  {Chakrabarti}},\ }\bibfield  {title} {\bibinfo {title} {{Light front
  quark-diquark model for the nucleons}},\ }\href
  {https://doi.org/10.1103/PhysRevD.94.094020} {\bibfield  {journal} {\bibinfo
  {journal} {Phys. Rev. D}\ }\textbf {\bibinfo {volume} {94}},\ \bibinfo
  {pages} {094020} (\bibinfo {year} {2016})},\ \Eprint
  {https://arxiv.org/abs/1608.07776} {arXiv:1608.07776 [hep-ph]} \BibitemShut
  {NoStop}%
\bibitem [{\citenamefont {Signal}\ and\ \citenamefont
  {Cao}(2022)}]{Signal:2021aum}%
  \BibitemOpen
  \bibfield  {author} {\bibinfo {author} {\bibfnamefont {A.~I.}\ \bibnamefont
  {Signal}}\ and\ \bibinfo {author} {\bibfnamefont {F.~G.}\ \bibnamefont
  {Cao}},\ }\bibfield  {title} {\bibinfo {title} {{Transverse momentum and
  transverse momentum distributions in the MIT bag model}},\ }\href
  {https://doi.org/10.1016/j.physletb.2022.136898} {\bibfield  {journal}
  {\bibinfo  {journal} {Phys. Lett. B}\ }\textbf {\bibinfo {volume} {826}},\
  \bibinfo {pages} {136898} (\bibinfo {year} {2022})},\ \Eprint
  {https://arxiv.org/abs/2108.12116} {arXiv:2108.12116 [hep-ph]} \BibitemShut
  {NoStop}%
\bibitem [{\citenamefont {Johnson}(1975)}]{Johnson:1975zp}%
  \BibitemOpen
  \bibfield  {author} {\bibinfo {author} {\bibfnamefont {K.}~\bibnamefont
  {Johnson}},\ }\bibfield  {title} {\bibinfo {title} {{The M.I.T. Bag Model}},\
  }\href@noop {} {\bibfield  {journal} {\bibinfo  {journal} {Acta Phys. Polon.
  B}\ }\textbf {\bibinfo {volume} {6}},\ \bibinfo {pages} {865} (\bibinfo
  {year} {1975})}\BibitemShut {NoStop}%
\bibitem [{\citenamefont {Zhang}(2024)}]{Zhang:2024adr}%
  \BibitemOpen
  \bibfield  {author} {\bibinfo {author} {\bibfnamefont {J.-L.}\ \bibnamefont
  {Zhang}},\ }\bibfield  {title} {\bibinfo {title} {{Kaon GTMDs in the
  Dyson-Schwinger equations using contact interaction}},\ }\href@noop {} {\
  (\bibinfo {year} {2024})},\ \Eprint {https://arxiv.org/abs/2409.04105}
  {arXiv:2409.04105 [hep-ph]} \BibitemShut {NoStop}%
\bibitem [{\citenamefont {Zhang}\ \emph {et~al.}(2021)\citenamefont {Zhang},
  \citenamefont {Cui}, \citenamefont {Ping},\ and\ \citenamefont
  {Roberts}}]{Zhang:2020ecj}%
  \BibitemOpen
  \bibfield  {author} {\bibinfo {author} {\bibfnamefont {J.-L.}\ \bibnamefont
  {Zhang}}, \bibinfo {author} {\bibfnamefont {Z.-F.}\ \bibnamefont {Cui}},
  \bibinfo {author} {\bibfnamefont {J.}~\bibnamefont {Ping}},\ and\ \bibinfo
  {author} {\bibfnamefont {C.~D.}\ \bibnamefont {Roberts}},\ }\bibfield
  {title} {\bibinfo {title} {{Contact interaction analysis of pion GTMDs}},\
  }\href {https://doi.org/10.1140/epjc/s10052-020-08791-1} {\bibfield
  {journal} {\bibinfo  {journal} {Eur. Phys. J. C}\ }\textbf {\bibinfo {volume}
  {81}},\ \bibinfo {pages} {6} (\bibinfo {year} {2021})},\ \Eprint
  {https://arxiv.org/abs/2009.11384} {arXiv:2009.11384 [hep-ph]} \BibitemShut
  {NoStop}%
\bibitem [{\citenamefont {Lin}\ \emph {et~al.}(2018)\citenamefont {Lin} \emph
  {et~al.}}]{Lin:2017snn}%
  \BibitemOpen
  \bibfield  {author} {\bibinfo {author} {\bibfnamefont {H.-W.}\ \bibnamefont
  {Lin}} \emph {et~al.},\ }\bibfield  {title} {\bibinfo {title} {{Parton
  distributions and lattice QCD calculations: a community white paper}},\
  }\href {https://doi.org/10.1016/j.ppnp.2018.01.007} {\bibfield  {journal}
  {\bibinfo  {journal} {Prog. Part. Nucl. Phys.}\ }\textbf {\bibinfo {volume}
  {100}},\ \bibinfo {pages} {107} (\bibinfo {year} {2018})},\ \Eprint
  {https://arxiv.org/abs/1711.07916} {arXiv:1711.07916 [hep-ph]} \BibitemShut
  {NoStop}%
\bibitem [{\citenamefont {Savage}(2012)}]{Savage:2011xk}%
  \BibitemOpen
  \bibfield  {author} {\bibinfo {author} {\bibfnamefont {M.~J.}\ \bibnamefont
  {Savage}},\ }\bibfield  {title} {\bibinfo {title} {{Nuclear Physics from
  Lattice QCD}},\ }\href {https://doi.org/10.1016/j.ppnp.2011.12.008}
  {\bibfield  {journal} {\bibinfo  {journal} {Prog. Part. Nucl. Phys.}\
  }\textbf {\bibinfo {volume} {67}},\ \bibinfo {pages} {140} (\bibinfo {year}
  {2012})},\ \Eprint {https://arxiv.org/abs/1110.5943} {arXiv:1110.5943
  [nucl-th]} \BibitemShut {NoStop}%
\bibitem [{\citenamefont {Glozman}\ \emph {et~al.}(2009)\citenamefont
  {Glozman}, \citenamefont {Lang},\ and\ \citenamefont
  {Limmer}}]{Glozman:2009rn}%
  \BibitemOpen
  \bibfield  {author} {\bibinfo {author} {\bibfnamefont {L.~Y.}\ \bibnamefont
  {Glozman}}, \bibinfo {author} {\bibfnamefont {C.~B.}\ \bibnamefont {Lang}},\
  and\ \bibinfo {author} {\bibfnamefont {M.}~\bibnamefont {Limmer}},\
  }\bibfield  {title} {\bibinfo {title} {{Angular momentum content of the
  rho-meson in lattice QCD}},\ }\href
  {https://doi.org/10.1103/PhysRevLett.103.121601} {\bibfield  {journal}
  {\bibinfo  {journal} {Phys. Rev. Lett.}\ }\textbf {\bibinfo {volume} {103}},\
  \bibinfo {pages} {121601} (\bibinfo {year} {2009})},\ \Eprint
  {https://arxiv.org/abs/0905.0811} {arXiv:0905.0811 [hep-lat]} \BibitemShut
  {NoStop}%
\bibitem [{\citenamefont {Puhan}\ \emph
  {et~al.}(2025{\natexlab{a}})\citenamefont {Puhan}, \citenamefont {Sharma},
  \citenamefont {Kumar},\ and\ \citenamefont {Dahiya}}]{Puhan:2025kzz}%
  \BibitemOpen
  \bibfield  {author} {\bibinfo {author} {\bibfnamefont {S.}~\bibnamefont
  {Puhan}}, \bibinfo {author} {\bibfnamefont {S.}~\bibnamefont {Sharma}},
  \bibinfo {author} {\bibfnamefont {N.}~\bibnamefont {Kumar}},\ and\ \bibinfo
  {author} {\bibfnamefont {H.}~\bibnamefont {Dahiya}},\ }\bibfield  {title}
  {\bibinfo {title} {{Understanding the Valence Quark Structure of the Pion
  through GTMDs}},\ }\href@noop {} {\  (\bibinfo {year}
  {2025}{\natexlab{a}})},\ \Eprint {https://arxiv.org/abs/2504.14982}
  {arXiv:2504.14982 [hep-ph]} \BibitemShut {NoStop}%
\bibitem [{\citenamefont {Echevarria}\ \emph {et~al.}(2016)\citenamefont
  {Echevarria}, \citenamefont {Idilbi}, \citenamefont {Kanazawa}, \citenamefont
  {Lorc\'e}, \citenamefont {Metz}, \citenamefont {Pasquini},\ and\
  \citenamefont {Schlegel}}]{Echevarria:2016mrc}%
  \BibitemOpen
  \bibfield  {author} {\bibinfo {author} {\bibfnamefont {M.~G.}\ \bibnamefont
  {Echevarria}}, \bibinfo {author} {\bibfnamefont {A.}~\bibnamefont {Idilbi}},
  \bibinfo {author} {\bibfnamefont {K.}~\bibnamefont {Kanazawa}}, \bibinfo
  {author} {\bibfnamefont {C.}~\bibnamefont {Lorc\'e}}, \bibinfo {author}
  {\bibfnamefont {A.}~\bibnamefont {Metz}}, \bibinfo {author} {\bibfnamefont
  {B.}~\bibnamefont {Pasquini}},\ and\ \bibinfo {author} {\bibfnamefont
  {M.}~\bibnamefont {Schlegel}},\ }\bibfield  {title} {\bibinfo {title}
  {{Proper definition and evolution of generalized transverse momentum
  dependent distributions}},\ }\href
  {https://doi.org/10.1016/j.physletb.2016.05.086} {\bibfield  {journal}
  {\bibinfo  {journal} {Phys. Lett. B}\ }\textbf {\bibinfo {volume} {759}},\
  \bibinfo {pages} {336} (\bibinfo {year} {2016})},\ \Eprint
  {https://arxiv.org/abs/1602.06953} {arXiv:1602.06953 [hep-ph]} \BibitemShut
  {NoStop}%
\bibitem [{\citenamefont {Meissner}\ \emph {et~al.}(2009)\citenamefont
  {Meissner}, \citenamefont {Metz},\ and\ \citenamefont
  {Schlegel}}]{Meissner:2009ww}%
  \BibitemOpen
  \bibfield  {author} {\bibinfo {author} {\bibfnamefont {S.}~\bibnamefont
  {Meissner}}, \bibinfo {author} {\bibfnamefont {A.}~\bibnamefont {Metz}},\
  and\ \bibinfo {author} {\bibfnamefont {M.}~\bibnamefont {Schlegel}},\
  }\bibfield  {title} {\bibinfo {title} {{Generalized parton correlation
  functions for a spin-1/2 hadron}},\ }\href
  {https://doi.org/10.1088/1126-6708/2009/08/056} {\bibfield  {journal}
  {\bibinfo  {journal} {JHEP}\ }\textbf {\bibinfo {volume} {08}},\ \bibinfo
  {pages} {056}},\ \Eprint {https://arxiv.org/abs/0906.5323} {arXiv:0906.5323
  [hep-ph]} \BibitemShut {NoStop}%
\bibitem [{\citenamefont {Meissner}\ \emph {et~al.}(2008)\citenamefont
  {Meissner}, \citenamefont {Metz}, \citenamefont {Schlegel},\ and\
  \citenamefont {Goeke}}]{Meissner:2008ay}%
  \BibitemOpen
  \bibfield  {author} {\bibinfo {author} {\bibfnamefont {S.}~\bibnamefont
  {Meissner}}, \bibinfo {author} {\bibfnamefont {A.}~\bibnamefont {Metz}},
  \bibinfo {author} {\bibfnamefont {M.}~\bibnamefont {Schlegel}},\ and\
  \bibinfo {author} {\bibfnamefont {K.}~\bibnamefont {Goeke}},\ }\bibfield
  {title} {\bibinfo {title} {{Generalized parton correlation functions for a
  spin-0 hadron}},\ }\href {https://doi.org/10.1088/1126-6708/2008/08/038}
  {\bibfield  {journal} {\bibinfo  {journal} {JHEP}\ }\textbf {\bibinfo
  {volume} {08}},\ \bibinfo {pages} {038}},\ \Eprint
  {https://arxiv.org/abs/0805.3165} {arXiv:0805.3165 [hep-ph]} \BibitemShut
  {NoStop}%
\bibitem [{\citenamefont {Diehl}(2003)}]{Diehl:2003ny}%
  \BibitemOpen
  \bibfield  {author} {\bibinfo {author} {\bibfnamefont {M.}~\bibnamefont
  {Diehl}},\ }\bibfield  {title} {\bibinfo {title} {{Generalized parton
  distributions}},\ }\href {https://doi.org/10.1016/j.physrep.2003.08.002}
  {\bibfield  {journal} {\bibinfo  {journal} {Phys. Rept.}\ }\textbf {\bibinfo
  {volume} {388}},\ \bibinfo {pages} {41} (\bibinfo {year} {2003})},\ \Eprint
  {https://arxiv.org/abs/hep-ph/0307382} {arXiv:hep-ph/0307382} \BibitemShut
  {NoStop}%
\bibitem [{\citenamefont {Chavez}\ \emph {et~al.}(2022)\citenamefont {Chavez},
  \citenamefont {Bertone}, \citenamefont {De~Soto~Borrero}, \citenamefont
  {Defurne}, \citenamefont {Mezrag}, \citenamefont {Moutarde}, \citenamefont
  {Rodr\'\i{}guez-Quintero},\ and\ \citenamefont {Segovia}}]{Chavez:2021llq}%
  \BibitemOpen
  \bibfield  {author} {\bibinfo {author} {\bibfnamefont {J.~M.~M.}\
  \bibnamefont {Chavez}}, \bibinfo {author} {\bibfnamefont {V.}~\bibnamefont
  {Bertone}}, \bibinfo {author} {\bibfnamefont {F.}~\bibnamefont
  {De~Soto~Borrero}}, \bibinfo {author} {\bibfnamefont {M.}~\bibnamefont
  {Defurne}}, \bibinfo {author} {\bibfnamefont {C.}~\bibnamefont {Mezrag}},
  \bibinfo {author} {\bibfnamefont {H.}~\bibnamefont {Moutarde}}, \bibinfo
  {author} {\bibfnamefont {J.}~\bibnamefont {Rodr\'\i{}guez-Quintero}},\ and\
  \bibinfo {author} {\bibfnamefont {J.}~\bibnamefont {Segovia}},\ }\bibfield
  {title} {\bibinfo {title} {{Pion generalized parton distributions: A path
  toward phenomenology}},\ }\href {https://doi.org/10.1103/PhysRevD.105.094012}
  {\bibfield  {journal} {\bibinfo  {journal} {Phys. Rev. D}\ }\textbf {\bibinfo
  {volume} {105}},\ \bibinfo {pages} {094012} (\bibinfo {year} {2022})},\
  \Eprint {https://arxiv.org/abs/2110.06052} {arXiv:2110.06052 [hep-ph]}
  \BibitemShut {NoStop}%
\bibitem [{\citenamefont {Broniowski}\ \emph {et~al.}(2023)\citenamefont
  {Broniowski}, \citenamefont {Shastry},\ and\ \citenamefont
  {Ruiz~Arriola}}]{Broniowski:2022iip}%
  \BibitemOpen
  \bibfield  {author} {\bibinfo {author} {\bibfnamefont {W.}~\bibnamefont
  {Broniowski}}, \bibinfo {author} {\bibfnamefont {V.}~\bibnamefont
  {Shastry}},\ and\ \bibinfo {author} {\bibfnamefont {E.}~\bibnamefont
  {Ruiz~Arriola}},\ }\bibfield  {title} {\bibinfo {title} {{Off-shell
  generalized parton distributions and form factors of the pion}},\ }\href
  {https://doi.org/10.1016/j.physletb.2023.137872} {\bibfield  {journal}
  {\bibinfo  {journal} {Phys. Lett. B}\ }\textbf {\bibinfo {volume} {840}},\
  \bibinfo {pages} {137872} (\bibinfo {year} {2023})},\ \Eprint
  {https://arxiv.org/abs/2211.11067} {arXiv:2211.11067 [hep-ph]} \BibitemShut
  {NoStop}%
\bibitem [{\citenamefont {Guidal}\ \emph {et~al.}(2005)\citenamefont {Guidal},
  \citenamefont {Polyakov}, \citenamefont {Radyushkin},\ and\ \citenamefont
  {Vanderhaeghen}}]{Guidal:2004nd}%
  \BibitemOpen
  \bibfield  {author} {\bibinfo {author} {\bibfnamefont {M.}~\bibnamefont
  {Guidal}}, \bibinfo {author} {\bibfnamefont {M.~V.}\ \bibnamefont
  {Polyakov}}, \bibinfo {author} {\bibfnamefont {A.~V.}\ \bibnamefont
  {Radyushkin}},\ and\ \bibinfo {author} {\bibfnamefont {M.}~\bibnamefont
  {Vanderhaeghen}},\ }\bibfield  {title} {\bibinfo {title} {{Nucleon
  form-factors from generalized parton distributions}},\ }\href
  {https://doi.org/10.1103/PhysRevD.72.054013} {\bibfield  {journal} {\bibinfo
  {journal} {Phys. Rev. D}\ }\textbf {\bibinfo {volume} {72}},\ \bibinfo
  {pages} {054013} (\bibinfo {year} {2005})},\ \Eprint
  {https://arxiv.org/abs/hep-ph/0410251} {arXiv:hep-ph/0410251} \BibitemShut
  {NoStop}%
\bibitem [{\citenamefont {Diehl}(2016)}]{Diehl:2015uka}%
  \BibitemOpen
  \bibfield  {author} {\bibinfo {author} {\bibfnamefont {M.}~\bibnamefont
  {Diehl}},\ }\bibfield  {title} {\bibinfo {title} {{Introduction to GPDs and
  TMDs}},\ }\href {https://doi.org/10.1140/epja/i2016-16149-3} {\bibfield
  {journal} {\bibinfo  {journal} {Eur. Phys. J. A}\ }\textbf {\bibinfo {volume}
  {52}},\ \bibinfo {pages} {149} (\bibinfo {year} {2016})},\ \Eprint
  {https://arxiv.org/abs/1512.01328} {arXiv:1512.01328 [hep-ph]} \BibitemShut
  {NoStop}%
\bibitem [{\citenamefont {Angeles-Martinez}\ \emph {et~al.}(2015)\citenamefont
  {Angeles-Martinez} \emph {et~al.}}]{Angeles-Martinez:2015sea}%
  \BibitemOpen
  \bibfield  {author} {\bibinfo {author} {\bibfnamefont {R.}~\bibnamefont
  {Angeles-Martinez}} \emph {et~al.},\ }\bibfield  {title} {\bibinfo {title}
  {{Transverse Momentum Dependent (TMD) parton distribution functions: status
  and prospects}},\ }\href {https://doi.org/10.5506/APhysPolB.46.2501}
  {\bibfield  {journal} {\bibinfo  {journal} {Acta Phys. Polon. B}\ }\textbf
  {\bibinfo {volume} {46}},\ \bibinfo {pages} {2501} (\bibinfo {year}
  {2015})},\ \Eprint {https://arxiv.org/abs/1507.05267} {arXiv:1507.05267
  [hep-ph]} \BibitemShut {NoStop}%
\bibitem [{\citenamefont {Pasquini}\ \emph {et~al.}(2008)\citenamefont
  {Pasquini}, \citenamefont {Cazzaniga},\ and\ \citenamefont
  {Boffi}}]{Pasquini:2008ax}%
  \BibitemOpen
  \bibfield  {author} {\bibinfo {author} {\bibfnamefont {B.}~\bibnamefont
  {Pasquini}}, \bibinfo {author} {\bibfnamefont {S.}~\bibnamefont
  {Cazzaniga}},\ and\ \bibinfo {author} {\bibfnamefont {S.}~\bibnamefont
  {Boffi}},\ }\bibfield  {title} {\bibinfo {title} {{Transverse momentum
  dependent parton distributions in a light-cone quark model}},\ }\href
  {https://doi.org/10.1103/PhysRevD.78.034025} {\bibfield  {journal} {\bibinfo
  {journal} {Phys. Rev. D}\ }\textbf {\bibinfo {volume} {78}},\ \bibinfo
  {pages} {034025} (\bibinfo {year} {2008})},\ \Eprint
  {https://arxiv.org/abs/0806.2298} {arXiv:0806.2298 [hep-ph]} \BibitemShut
  {NoStop}%
\bibitem [{\citenamefont {Puhan}\ and\ \citenamefont
  {Dahiya}(2024)}]{Puhan:2023hio}%
  \BibitemOpen
  \bibfield  {author} {\bibinfo {author} {\bibfnamefont {S.}~\bibnamefont
  {Puhan}}\ and\ \bibinfo {author} {\bibfnamefont {H.}~\bibnamefont {Dahiya}},\
  }\bibfield  {title} {\bibinfo {title} {{Leading twist T-even TMDs for the
  spin-1 heavy vector mesons}},\ }\href
  {https://doi.org/10.1103/PhysRevD.109.034005} {\bibfield  {journal} {\bibinfo
   {journal} {Phys. Rev. D}\ }\textbf {\bibinfo {volume} {109}},\ \bibinfo
  {pages} {034005} (\bibinfo {year} {2024})},\ \Eprint
  {https://arxiv.org/abs/2310.03465} {arXiv:2310.03465 [hep-ph]} \BibitemShut
  {NoStop}%
\bibitem [{\citenamefont {Collins}\ and\ \citenamefont
  {Soper}(1982)}]{Collins:1981uw}%
  \BibitemOpen
  \bibfield  {author} {\bibinfo {author} {\bibfnamefont {J.~C.}\ \bibnamefont
  {Collins}}\ and\ \bibinfo {author} {\bibfnamefont {D.~E.}\ \bibnamefont
  {Soper}},\ }\bibfield  {title} {\bibinfo {title} {{Parton Distribution and
  Decay Functions}},\ }\href {https://doi.org/10.1016/0550-3213(82)90021-9}
  {\bibfield  {journal} {\bibinfo  {journal} {Nucl. Phys. B}\ }\textbf
  {\bibinfo {volume} {194}},\ \bibinfo {pages} {445} (\bibinfo {year}
  {1982})}\BibitemShut {NoStop}%
\bibitem [{\citenamefont {Martin}\ \emph {et~al.}(1998)\citenamefont {Martin},
  \citenamefont {Roberts}, \citenamefont {Stirling},\ and\ \citenamefont
  {Thorne}}]{Martin:1998sq}%
  \BibitemOpen
  \bibfield  {author} {\bibinfo {author} {\bibfnamefont {A.~D.}\ \bibnamefont
  {Martin}}, \bibinfo {author} {\bibfnamefont {R.~G.}\ \bibnamefont {Roberts}},
  \bibinfo {author} {\bibfnamefont {W.~J.}\ \bibnamefont {Stirling}},\ and\
  \bibinfo {author} {\bibfnamefont {R.~S.}\ \bibnamefont {Thorne}},\ }\bibfield
   {title} {\bibinfo {title} {{Parton distributions: A New global analysis}},\
  }\href {https://doi.org/10.1007/s100520050220} {\bibfield  {journal}
  {\bibinfo  {journal} {Eur. Phys. J. C}\ }\textbf {\bibinfo {volume} {4}},\
  \bibinfo {pages} {463} (\bibinfo {year} {1998})},\ \Eprint
  {https://arxiv.org/abs/hep-ph/9803445} {arXiv:hep-ph/9803445} \BibitemShut
  {NoStop}%
\bibitem [{\citenamefont {Gluck}\ \emph {et~al.}(1995)\citenamefont {Gluck},
  \citenamefont {Reya},\ and\ \citenamefont {Vogt}}]{Gluck:1994uf}%
  \BibitemOpen
  \bibfield  {author} {\bibinfo {author} {\bibfnamefont {M.}~\bibnamefont
  {Gluck}}, \bibinfo {author} {\bibfnamefont {E.}~\bibnamefont {Reya}},\ and\
  \bibinfo {author} {\bibfnamefont {A.}~\bibnamefont {Vogt}},\ }\bibfield
  {title} {\bibinfo {title} {{Dynamical parton distributions of the proton and
  small x physics}},\ }\href {https://doi.org/10.1007/BF01624586} {\bibfield
  {journal} {\bibinfo  {journal} {Z. Phys. C}\ }\textbf {\bibinfo {volume}
  {67}},\ \bibinfo {pages} {433} (\bibinfo {year} {1995})}\BibitemShut
  {NoStop}%
\bibitem [{\citenamefont {Freese}\ and\ \citenamefont
  {Clo\"et}(2021)}]{Freese:2020mcx}%
  \BibitemOpen
  \bibfield  {author} {\bibinfo {author} {\bibfnamefont {A.}~\bibnamefont
  {Freese}}\ and\ \bibinfo {author} {\bibfnamefont {I.~C.}\ \bibnamefont
  {Clo\"et}},\ }\bibfield  {title} {\bibinfo {title} {{Quark spin and orbital
  angular momentum from proton generalized parton distributions}},\ }\href
  {https://doi.org/10.1103/PhysRevC.103.045204} {\bibfield  {journal} {\bibinfo
   {journal} {Phys. Rev. C}\ }\textbf {\bibinfo {volume} {103}},\ \bibinfo
  {pages} {045204} (\bibinfo {year} {2021})},\ \Eprint
  {https://arxiv.org/abs/2005.10286} {arXiv:2005.10286 [nucl-th]} \BibitemShut
  {NoStop}%
\bibitem [{\citenamefont {Guidal}\ \emph {et~al.}(2013)\citenamefont {Guidal},
  \citenamefont {Moutarde},\ and\ \citenamefont
  {Vanderhaeghen}}]{Guidal:2013rya}%
  \BibitemOpen
  \bibfield  {author} {\bibinfo {author} {\bibfnamefont {M.}~\bibnamefont
  {Guidal}}, \bibinfo {author} {\bibfnamefont {H.}~\bibnamefont {Moutarde}},\
  and\ \bibinfo {author} {\bibfnamefont {M.}~\bibnamefont {Vanderhaeghen}},\
  }\bibfield  {title} {\bibinfo {title} {{Generalized Parton Distributions in
  the valence region from Deeply Virtual Compton Scattering}},\ }\href
  {https://doi.org/10.1088/0034-4885/76/6/066202} {\bibfield  {journal}
  {\bibinfo  {journal} {Rept. Prog. Phys.}\ }\textbf {\bibinfo {volume} {76}},\
  \bibinfo {pages} {066202} (\bibinfo {year} {2013})},\ \Eprint
  {https://arxiv.org/abs/1303.6600} {arXiv:1303.6600 [hep-ph]} \BibitemShut
  {NoStop}%
\bibitem [{\citenamefont {Ji}(1997)}]{Ji:1996nm}%
  \BibitemOpen
  \bibfield  {author} {\bibinfo {author} {\bibfnamefont {X.-D.}\ \bibnamefont
  {Ji}},\ }\bibfield  {title} {\bibinfo {title} {{Deeply virtual Compton
  scattering}},\ }\href {https://doi.org/10.1103/PhysRevD.55.7114} {\bibfield
  {journal} {\bibinfo  {journal} {Phys. Rev. D}\ }\textbf {\bibinfo {volume}
  {55}},\ \bibinfo {pages} {7114} (\bibinfo {year} {1997})},\ \Eprint
  {https://arxiv.org/abs/hep-ph/9609381} {arXiv:hep-ph/9609381} \BibitemShut
  {NoStop}%
\bibitem [{\citenamefont {Xie}\ \emph {et~al.}(2023)\citenamefont {Xie},
  \citenamefont {Kou}, \citenamefont {Fu}, \citenamefont {Ye},\ and\
  \citenamefont {Chen}}]{Xie:2023xkz}%
  \BibitemOpen
  \bibfield  {author} {\bibinfo {author} {\bibfnamefont {G.}~\bibnamefont
  {Xie}}, \bibinfo {author} {\bibfnamefont {W.}~\bibnamefont {Kou}}, \bibinfo
  {author} {\bibfnamefont {Q.}~\bibnamefont {Fu}}, \bibinfo {author}
  {\bibfnamefont {Z.}~\bibnamefont {Ye}},\ and\ \bibinfo {author}
  {\bibfnamefont {X.}~\bibnamefont {Chen}},\ }\bibfield  {title} {\bibinfo
  {title} {{Deeply virtual compton scattering at future electron-ion
  colliders}},\ }\href {https://doi.org/10.1140/epjc/s10052-023-12065-x}
  {\bibfield  {journal} {\bibinfo  {journal} {Eur. Phys. J. C}\ }\textbf
  {\bibinfo {volume} {83}},\ \bibinfo {pages} {900} (\bibinfo {year} {2023})},\
  \Eprint {https://arxiv.org/abs/2306.02357} {arXiv:2306.02357 [hep-ph]}
  \BibitemShut {NoStop}%
\bibitem [{\citenamefont {Favart}\ \emph {et~al.}(2016)\citenamefont {Favart},
  \citenamefont {Guidal}, \citenamefont {Horn},\ and\ \citenamefont
  {Kroll}}]{Favart:2015umi}%
  \BibitemOpen
  \bibfield  {author} {\bibinfo {author} {\bibfnamefont {L.}~\bibnamefont
  {Favart}}, \bibinfo {author} {\bibfnamefont {M.}~\bibnamefont {Guidal}},
  \bibinfo {author} {\bibfnamefont {T.}~\bibnamefont {Horn}},\ and\ \bibinfo
  {author} {\bibfnamefont {P.}~\bibnamefont {Kroll}},\ }\bibfield  {title}
  {\bibinfo {title} {{Deeply Virtual Meson Production on the nucleon}},\ }\href
  {https://doi.org/10.1140/epja/i2016-16158-2} {\bibfield  {journal} {\bibinfo
  {journal} {Eur. Phys. J. A}\ }\textbf {\bibinfo {volume} {52}},\ \bibinfo
  {pages} {158} (\bibinfo {year} {2016})},\ \Eprint
  {https://arxiv.org/abs/1511.04535} {arXiv:1511.04535 [hep-ph]} \BibitemShut
  {NoStop}%
\bibitem [{\citenamefont {Alekhin}(2003)}]{Alekhin:2002fv}%
  \BibitemOpen
  \bibfield  {author} {\bibinfo {author} {\bibfnamefont {S.}~\bibnamefont
  {Alekhin}},\ }\bibfield  {title} {\bibinfo {title} {{Parton distributions
  from deep inelastic scattering data}},\ }\href
  {https://doi.org/10.1103/PhysRevD.68.014002} {\bibfield  {journal} {\bibinfo
  {journal} {Phys. Rev. D}\ }\textbf {\bibinfo {volume} {68}},\ \bibinfo
  {pages} {014002} (\bibinfo {year} {2003})},\ \Eprint
  {https://arxiv.org/abs/hep-ph/0211096} {arXiv:hep-ph/0211096} \BibitemShut
  {NoStop}%
\bibitem [{\citenamefont {Puhan}\ \emph {et~al.}(2024)\citenamefont {Puhan},
  \citenamefont {Sharma}, \citenamefont {Kaur}, \citenamefont {Kumar},\ and\
  \citenamefont {Dahiya}}]{Puhan2023}%
  \BibitemOpen
  \bibfield  {author} {\bibinfo {author} {\bibfnamefont {S.}~\bibnamefont
  {Puhan}}, \bibinfo {author} {\bibfnamefont {S.}~\bibnamefont {Sharma}},
  \bibinfo {author} {\bibfnamefont {N.}~\bibnamefont {Kaur}}, \bibinfo {author}
  {\bibfnamefont {N.}~\bibnamefont {Kumar}},\ and\ \bibinfo {author}
  {\bibfnamefont {H.}~\bibnamefont {Dahiya}},\ }\bibfield  {title} {\bibinfo
  {title} {{T-even TMDs for the spin-0 pseudo-scalar mesons upto twist-4 using
  light-front formalism}},\ }\href {https://doi.org/10.1007/JHEP02(2024)075}
  {\bibfield  {journal} {\bibinfo  {journal} {JHEP}\ }\textbf {\bibinfo
  {volume} {02}},\ \bibinfo {pages} {075}},\ \Eprint
  {https://arxiv.org/abs/2310.03464} {arXiv:2310.03464 [hep-ph]} \BibitemShut
  {NoStop}%
\bibitem [{\citenamefont {Qian}\ and\ \citenamefont {Ma}(2008)}]{Qian:2008px}%
  \BibitemOpen
  \bibfield  {author} {\bibinfo {author} {\bibfnamefont {W.}~\bibnamefont
  {Qian}}\ and\ \bibinfo {author} {\bibfnamefont {B.-Q.}\ \bibnamefont {Ma}},\
  }\bibfield  {title} {\bibinfo {title} {{Vector meson omega-phi mixing and
  their form factors in light-cone quark model}},\ }\href
  {https://doi.org/10.1103/PhysRevD.78.074002} {\bibfield  {journal} {\bibinfo
  {journal} {Phys. Rev. D}\ }\textbf {\bibinfo {volume} {78}},\ \bibinfo
  {pages} {074002} (\bibinfo {year} {2008})},\ \Eprint
  {https://arxiv.org/abs/0809.4411} {arXiv:0809.4411 [hep-ph]} \BibitemShut
  {NoStop}%
\bibitem [{\citenamefont {Brodsky}\ \emph {et~al.}(2001)\citenamefont
  {Brodsky}, \citenamefont {Diehl},\ and\ \citenamefont
  {Hwang}}]{Brodsky:2000xy}%
  \BibitemOpen
  \bibfield  {author} {\bibinfo {author} {\bibfnamefont {S.~J.}\ \bibnamefont
  {Brodsky}}, \bibinfo {author} {\bibfnamefont {M.}~\bibnamefont {Diehl}},\
  and\ \bibinfo {author} {\bibfnamefont {D.~S.}\ \bibnamefont {Hwang}},\
  }\bibfield  {title} {\bibinfo {title} {{Light cone wave function
  representation of deeply virtual Compton scattering}},\ }\href
  {https://doi.org/10.1016/S0550-3213(00)00695-7} {\bibfield  {journal}
  {\bibinfo  {journal} {Nucl. Phys. B}\ }\textbf {\bibinfo {volume} {596}},\
  \bibinfo {pages} {99} (\bibinfo {year} {2001})},\ \Eprint
  {https://arxiv.org/abs/hep-ph/0009254} {arXiv:hep-ph/0009254} \BibitemShut
  {NoStop}%
\bibitem [{\citenamefont {Cano}\ and\ \citenamefont
  {Pire}(2004)}]{Cano:2003ju}%
  \BibitemOpen
  \bibfield  {author} {\bibinfo {author} {\bibfnamefont {F.}~\bibnamefont
  {Cano}}\ and\ \bibinfo {author} {\bibfnamefont {B.}~\bibnamefont {Pire}},\
  }\bibfield  {title} {\bibinfo {title} {{Deep electroproduction of photons and
  mesons on the deuteron}},\ }\href
  {https://doi.org/10.1140/epja/i2003-10127-x} {\bibfield  {journal} {\bibinfo
  {journal} {Eur. Phys. J. A}\ }\textbf {\bibinfo {volume} {19}},\ \bibinfo
  {pages} {423} (\bibinfo {year} {2004})},\ \Eprint
  {https://arxiv.org/abs/hep-ph/0307231} {arXiv:hep-ph/0307231} \BibitemShut
  {NoStop}%
\bibitem [{\citenamefont {Zhang}\ \emph {et~al.}(2022)\citenamefont {Zhang},
  \citenamefont {Kang},\ and\ \citenamefont {Ping}}]{Zhang:2022zim}%
  \BibitemOpen
  \bibfield  {author} {\bibinfo {author} {\bibfnamefont {J.-L.}\ \bibnamefont
  {Zhang}}, \bibinfo {author} {\bibfnamefont {G.-Z.}\ \bibnamefont {Kang}},\
  and\ \bibinfo {author} {\bibfnamefont {J.-L.}\ \bibnamefont {Ping}},\
  }\bibfield  {title} {\bibinfo {title} {{\ensuremath{\rho} meson generalized
  parton distributions in the Nambu\textendash{}Jona-Lasinio model}},\ }\href
  {https://doi.org/10.1103/PhysRevD.105.094015} {\bibfield  {journal} {\bibinfo
   {journal} {Phys. Rev. D}\ }\textbf {\bibinfo {volume} {105}},\ \bibinfo
  {pages} {094015} (\bibinfo {year} {2022})},\ \Eprint
  {https://arxiv.org/abs/2204.14032} {arXiv:2204.14032 [hep-ph]} \BibitemShut
  {NoStop}%
\bibitem [{\citenamefont {Sun}\ and\ \citenamefont {Dong}(2019)}]{Sun:2018ldr}%
  \BibitemOpen
  \bibfield  {author} {\bibinfo {author} {\bibfnamefont {B.-D.}\ \bibnamefont
  {Sun}}\ and\ \bibinfo {author} {\bibfnamefont {Y.-B.}\ \bibnamefont {Dong}},\
  }\bibfield  {title} {\bibinfo {title} {{Polarized generalized parton
  distributions and structure functions of the $\rho$ meson}},\ }\href
  {https://doi.org/10.1103/PhysRevD.99.016023} {\bibfield  {journal} {\bibinfo
  {journal} {Phys. Rev. D}\ }\textbf {\bibinfo {volume} {99}},\ \bibinfo
  {pages} {016023} (\bibinfo {year} {2019})},\ \Eprint
  {https://arxiv.org/abs/1811.00666} {arXiv:1811.00666 [hep-ph]} \BibitemShut
  {NoStop}%
\bibitem [{\citenamefont {Kumar}(2019)}]{Kumar:2019eck}%
  \BibitemOpen
  \bibfield  {author} {\bibinfo {author} {\bibfnamefont {N.}~\bibnamefont
  {Kumar}},\ }\bibfield  {title} {\bibinfo {title} {{Transverse densities and
  generalized parton distributions of the $\rho$ meson in the light front quark
  model}},\ }\href {https://doi.org/10.1103/PhysRevD.99.014039} {\bibfield
  {journal} {\bibinfo  {journal} {Phys. Rev. D}\ }\textbf {\bibinfo {volume}
  {99}},\ \bibinfo {pages} {014039} (\bibinfo {year} {2019})},\ \Eprint
  {https://arxiv.org/abs/1901.02836} {arXiv:1901.02836 [hep-ph]} \BibitemShut
  {NoStop}%
\bibitem [{\citenamefont {Adhikari}\ \emph {et~al.}(2019)\citenamefont
  {Adhikari}, \citenamefont {Li}, \citenamefont {Li},\ and\ \citenamefont
  {Vary}}]{Adhikari:2018umb}%
  \BibitemOpen
  \bibfield  {author} {\bibinfo {author} {\bibfnamefont {L.}~\bibnamefont
  {Adhikari}}, \bibinfo {author} {\bibfnamefont {Y.}~\bibnamefont {Li}},
  \bibinfo {author} {\bibfnamefont {M.}~\bibnamefont {Li}},\ and\ \bibinfo
  {author} {\bibfnamefont {J.~P.}\ \bibnamefont {Vary}},\ }\bibfield  {title}
  {\bibinfo {title} {{Form factors and generalized parton distributions of
  heavy quarkonia in basis light front quantization}},\ }\href
  {https://doi.org/10.1103/PhysRevC.99.035208} {\bibfield  {journal} {\bibinfo
  {journal} {Phys. Rev. C}\ }\textbf {\bibinfo {volume} {99}},\ \bibinfo
  {pages} {035208} (\bibinfo {year} {2019})},\ \Eprint
  {https://arxiv.org/abs/1809.06475} {arXiv:1809.06475 [hep-ph]} \BibitemShut
  {NoStop}%
\bibitem [{\citenamefont {Shi}\ \emph {et~al.}(2023)\citenamefont {Shi},
  \citenamefont {Li}, \citenamefont {Yin},\ and\ \citenamefont
  {Jia}}]{Shi:2023oll}%
  \BibitemOpen
  \bibfield  {author} {\bibinfo {author} {\bibfnamefont {C.}~\bibnamefont
  {Shi}}, \bibinfo {author} {\bibfnamefont {J.}~\bibnamefont {Li}}, \bibinfo
  {author} {\bibfnamefont {P.-L.}\ \bibnamefont {Yin}},\ and\ \bibinfo {author}
  {\bibfnamefont {W.}~\bibnamefont {Jia}},\ }\bibfield  {title} {\bibinfo
  {title} {{Unpolarized generalized parton distributions of light and heavy
  vector mesons}},\ }\href {https://doi.org/10.1103/PhysRevD.107.074009}
  {\bibfield  {journal} {\bibinfo  {journal} {Phys. Rev. D}\ }\textbf {\bibinfo
  {volume} {107}},\ \bibinfo {pages} {074009} (\bibinfo {year} {2023})},\
  \Eprint {https://arxiv.org/abs/2302.02388} {arXiv:2302.02388 [hep-ph]}
  \BibitemShut {NoStop}%
\bibitem [{\citenamefont {Sun}\ and\ \citenamefont {Dong}(2017)}]{Sun:2017gtz}%
  \BibitemOpen
  \bibfield  {author} {\bibinfo {author} {\bibfnamefont {B.-D.}\ \bibnamefont
  {Sun}}\ and\ \bibinfo {author} {\bibfnamefont {Y.-B.}\ \bibnamefont {Dong}},\
  }\bibfield  {title} {\bibinfo {title} {{$\rho$ meson unpolarized generalized
  parton distributions with a light-front constituent quark model}},\ }\href
  {https://doi.org/10.1103/PhysRevD.96.036019} {\bibfield  {journal} {\bibinfo
  {journal} {Phys. Rev. D}\ }\textbf {\bibinfo {volume} {96}},\ \bibinfo
  {pages} {036019} (\bibinfo {year} {2017})},\ \Eprint
  {https://arxiv.org/abs/1707.03972} {arXiv:1707.03972 [hep-ph]} \BibitemShut
  {NoStop}%
\bibitem [{\citenamefont {Puhan}\ \emph
  {et~al.}(2025{\natexlab{b}})\citenamefont {Puhan}, \citenamefont {Kumar},\
  and\ \citenamefont {Dahiya}}]{Puhan:2025uat}%
  \BibitemOpen
  \bibfield  {author} {\bibinfo {author} {\bibfnamefont {S.}~\bibnamefont
  {Puhan}}, \bibinfo {author} {\bibfnamefont {N.}~\bibnamefont {Kumar}},\ and\
  \bibinfo {author} {\bibfnamefont {H.}~\bibnamefont {Dahiya}},\ }\bibfield
  {title} {\bibinfo {title} {{Spin-1 unpolarized GPDs in light front
  formalism}},\ }\href@noop {} {\bibfield  {journal} {\bibinfo  {journal} {DAE
  Symp. Nucl. Phys.}\ }\textbf {\bibinfo {volume} {68}},\ \bibinfo {pages}
  {825} (\bibinfo {year} {2025}{\natexlab{b}})}\BibitemShut {NoStop}%
\bibitem [{\citenamefont {Choi}\ and\ \citenamefont {Ji}(2004)}]{Choi:2004ww}%
  \BibitemOpen
  \bibfield  {author} {\bibinfo {author} {\bibfnamefont {H.-M.}\ \bibnamefont
  {Choi}}\ and\ \bibinfo {author} {\bibfnamefont {C.-R.}\ \bibnamefont {Ji}},\
  }\bibfield  {title} {\bibinfo {title} {{Electromagnetic structure of the rho
  meson in the light front quark model}},\ }\href
  {https://doi.org/10.1103/PhysRevD.70.053015} {\bibfield  {journal} {\bibinfo
  {journal} {Phys. Rev. D}\ }\textbf {\bibinfo {volume} {70}},\ \bibinfo
  {pages} {053015} (\bibinfo {year} {2004})},\ \Eprint
  {https://arxiv.org/abs/hep-ph/0402114} {arXiv:hep-ph/0402114} \BibitemShut
  {NoStop}%
\bibitem [{\citenamefont {Hofstadter}(1957)}]{Hofstadter:1957wk}%
  \BibitemOpen
  \bibfield  {author} {\bibinfo {author} {\bibfnamefont {R.}~\bibnamefont
  {Hofstadter}},\ }\bibfield  {title} {\bibinfo {title} {{Nuclear and nucleon
  scattering of high-energy electrons}},\ }\href
  {https://doi.org/10.1146/annurev.ns.07.120157.001311} {\bibfield  {journal}
  {\bibinfo  {journal} {Ann. Rev. Nucl. Part. Sci.}\ }\textbf {\bibinfo
  {volume} {7}},\ \bibinfo {pages} {231} (\bibinfo {year} {1957})}\BibitemShut
  {NoStop}%
\bibitem [{\citenamefont {Karmanov}(1996)}]{Karmanov:1996qc}%
  \BibitemOpen
  \bibfield  {author} {\bibinfo {author} {\bibfnamefont {V.~A.}\ \bibnamefont
  {Karmanov}},\ }\bibfield  {title} {\bibinfo {title} {{On ambiguities of the
  spin-1 electromagnetic form-factors in light front dynamics}},\ }\href
  {https://doi.org/10.1016/0375-9474(96)00260-6} {\bibfield  {journal}
  {\bibinfo  {journal} {Nucl. Phys. A}\ }\textbf {\bibinfo {volume} {608}},\
  \bibinfo {pages} {316} (\bibinfo {year} {1996})}\BibitemShut {NoStop}%
\bibitem [{\citenamefont {Chung}\ \emph {et~al.}(1988)\citenamefont {Chung},
  \citenamefont {Coester},\ and\ \citenamefont {Polyzou}}]{Chung:1988mu}%
  \BibitemOpen
  \bibfield  {author} {\bibinfo {author} {\bibfnamefont {P.~L.}\ \bibnamefont
  {Chung}}, \bibinfo {author} {\bibfnamefont {F.}~\bibnamefont {Coester}},\
  and\ \bibinfo {author} {\bibfnamefont {W.~N.}\ \bibnamefont {Polyzou}},\
  }\bibfield  {title} {\bibinfo {title} {{Charge Form-Factors of Quark Model
  Pions}},\ }\href {https://doi.org/10.1016/0370-2693(88)90995-1} {\bibfield
  {journal} {\bibinfo  {journal} {Phys. Lett. B}\ }\textbf {\bibinfo {volume}
  {205}},\ \bibinfo {pages} {545} (\bibinfo {year} {1988})}\BibitemShut
  {NoStop}%
\bibitem [{\citenamefont {Brodsky}\ and\ \citenamefont
  {Hiller}(1992)}]{Brodsky:1992px}%
  \BibitemOpen
  \bibfield  {author} {\bibinfo {author} {\bibfnamefont {S.~J.}\ \bibnamefont
  {Brodsky}}\ and\ \bibinfo {author} {\bibfnamefont {J.~R.}\ \bibnamefont
  {Hiller}},\ }\bibfield  {title} {\bibinfo {title} {{Universal properties of
  the electromagnetic interactions of spin one systems}},\ }\href
  {https://doi.org/10.1103/PhysRevD.46.2141} {\bibfield  {journal} {\bibinfo
  {journal} {Phys. Rev. D}\ }\textbf {\bibinfo {volume} {46}},\ \bibinfo
  {pages} {2141} (\bibinfo {year} {1992})}\BibitemShut {NoStop}%
\bibitem [{\citenamefont {Frankfurt}\ \emph {et~al.}(1993)\citenamefont
  {Frankfurt}, \citenamefont {Strikman},\ and\ \citenamefont
  {Frederico}}]{Frankfurt:1993ut}%
  \BibitemOpen
  \bibfield  {author} {\bibinfo {author} {\bibfnamefont {L.~L.}\ \bibnamefont
  {Frankfurt}}, \bibinfo {author} {\bibfnamefont {M.}~\bibnamefont
  {Strikman}},\ and\ \bibinfo {author} {\bibfnamefont {T.}~\bibnamefont
  {Frederico}},\ }\bibfield  {title} {\bibinfo {title} {{Deuteron form-factors
  in the light cone quantum mechanics 'good' component approach}},\ }\href
  {https://doi.org/10.1103/PhysRevC.48.2182} {\bibfield  {journal} {\bibinfo
  {journal} {Phys. Rev. C}\ }\textbf {\bibinfo {volume} {48}},\ \bibinfo
  {pages} {2182} (\bibinfo {year} {1993})}\BibitemShut {NoStop}%
\bibitem [{\citenamefont {De~Melo}(2019)}]{DeMelo:2018bim}%
  \BibitemOpen
  \bibfield  {author} {\bibinfo {author} {\bibfnamefont {J.~P. B.~C.}\
  \bibnamefont {De~Melo}},\ }\bibfield  {title} {\bibinfo {title} {{Unambiguous
  Extraction of the Electromagnetic Form Factors for Spin-1 Particles on the
  Light-Front}},\ }\href {https://doi.org/10.1016/j.physletb.2018.11.003}
  {\bibfield  {journal} {\bibinfo  {journal} {Phys. Lett. B}\ }\textbf
  {\bibinfo {volume} {788}},\ \bibinfo {pages} {152} (\bibinfo {year}
  {2019})},\ \Eprint {https://arxiv.org/abs/1810.11478} {arXiv:1810.11478
  [hep-ph]} \BibitemShut {NoStop}%
\bibitem [{\citenamefont {Allahverdiyeva}\ and\ \citenamefont
  {Mamedov}(2023)}]{Allahverdiyeva:2023fhn}%
  \BibitemOpen
  \bibfield  {author} {\bibinfo {author} {\bibfnamefont {M.}~\bibnamefont
  {Allahverdiyeva}}\ and\ \bibinfo {author} {\bibfnamefont {S.}~\bibnamefont
  {Mamedov}},\ }\bibfield  {title} {\bibinfo {title} {{Vector meson
  gravitational form factors and generalized parton distributions at finite
  temperature within the soft-wall AdS/QCD model}},\ }\href
  {https://doi.org/10.1140/epjc/s10052-023-11607-7} {\bibfield  {journal}
  {\bibinfo  {journal} {Eur. Phys. J. C}\ }\textbf {\bibinfo {volume} {83}},\
  \bibinfo {pages} {447} (\bibinfo {year} {2023})},\ \Eprint
  {https://arxiv.org/abs/2302.03383} {arXiv:2302.03383 [hep-ph]} \BibitemShut
  {NoStop}%
\bibitem [{\citenamefont {Gurtler}\ \emph {et~al.}(2008)\citenamefont {Gurtler}
  \emph {et~al.}}]{QCDSF:2008tjq}%
  \BibitemOpen
  \bibfield  {author} {\bibinfo {author} {\bibfnamefont {M.}~\bibnamefont
  {Gurtler}} \emph {et~al.} (\bibinfo {collaboration} {QCDSF}),\ }\bibfield
  {title} {\bibinfo {title} {{Vector meson electromagnetic form factors}},\
  }\href {https://doi.org/10.22323/1.066.0051} {\bibfield  {journal} {\bibinfo
  {journal} {PoS}\ }\textbf {\bibinfo {volume} {LATTICE2008}},\ \bibinfo
  {pages} {051} (\bibinfo {year} {2008})}\BibitemShut {NoStop}%
\bibitem [{\citenamefont {Shultz}\ \emph {et~al.}(2015)\citenamefont {Shultz},
  \citenamefont {Dudek},\ and\ \citenamefont {Edwards}}]{Shultz:2015pfa}%
  \BibitemOpen
  \bibfield  {author} {\bibinfo {author} {\bibfnamefont {C.~J.}\ \bibnamefont
  {Shultz}}, \bibinfo {author} {\bibfnamefont {J.~J.}\ \bibnamefont {Dudek}},\
  and\ \bibinfo {author} {\bibfnamefont {R.~G.}\ \bibnamefont {Edwards}},\
  }\bibfield  {title} {\bibinfo {title} {{Excited meson radiative transitions
  from lattice QCD using variationally optimized operators}},\ }\href
  {https://doi.org/10.1103/PhysRevD.91.114501} {\bibfield  {journal} {\bibinfo
  {journal} {Phys. Rev. D}\ }\textbf {\bibinfo {volume} {91}},\ \bibinfo
  {pages} {114501} (\bibinfo {year} {2015})},\ \Eprint
  {https://arxiv.org/abs/1501.07457} {arXiv:1501.07457 [hep-lat]} \BibitemShut
  {NoStop}%
\bibitem [{\citenamefont {Bacchetta}\ and\ \citenamefont
  {Mulders}(2000)}]{Bacchetta:2000jk}%
  \BibitemOpen
  \bibfield  {author} {\bibinfo {author} {\bibfnamefont {A.}~\bibnamefont
  {Bacchetta}}\ and\ \bibinfo {author} {\bibfnamefont {P.~J.}\ \bibnamefont
  {Mulders}},\ }\bibfield  {title} {\bibinfo {title} {{Deep inelastic
  leptoproduction of spin-one hadrons}},\ }\href
  {https://doi.org/10.1103/PhysRevD.62.114004} {\bibfield  {journal} {\bibinfo
  {journal} {Phys. Rev. D}\ }\textbf {\bibinfo {volume} {62}},\ \bibinfo
  {pages} {114004} (\bibinfo {year} {2000})},\ \Eprint
  {https://arxiv.org/abs/hep-ph/0007120} {arXiv:hep-ph/0007120} \BibitemShut
  {NoStop}%
\bibitem [{\citenamefont {Lepage}\ and\ \citenamefont
  {Brodsky}(1980)}]{Lepage:1980fj}%
  \BibitemOpen
  \bibfield  {author} {\bibinfo {author} {\bibfnamefont {G.~P.}\ \bibnamefont
  {Lepage}}\ and\ \bibinfo {author} {\bibfnamefont {S.~J.}\ \bibnamefont
  {Brodsky}},\ }\bibfield  {title} {\bibinfo {title} {{Exclusive Processes in
  Perturbative Quantum Chromodynamics}},\ }\href
  {https://doi.org/10.1103/PhysRevD.22.2157} {\bibfield  {journal} {\bibinfo
  {journal} {Phys. Rev. D}\ }\textbf {\bibinfo {volume} {22}},\ \bibinfo
  {pages} {2157} (\bibinfo {year} {1980})}\BibitemShut {NoStop}%
\bibitem [{\citenamefont {Choi}\ and\ \citenamefont {Ji}(1997)}]{Choi:1996mq}%
  \BibitemOpen
  \bibfield  {author} {\bibinfo {author} {\bibfnamefont {H.~M.}\ \bibnamefont
  {Choi}}\ and\ \bibinfo {author} {\bibfnamefont {C.-R.}\ \bibnamefont {Ji}},\
  }\bibfield  {title} {\bibinfo {title} {{Light cone quark model predictions
  for radiative meson decays}},\ }\href
  {https://doi.org/10.1016/S0375-9474(97)00052-3} {\bibfield  {journal}
  {\bibinfo  {journal} {Nucl. Phys. A}\ }\textbf {\bibinfo {volume} {618}},\
  \bibinfo {pages} {291} (\bibinfo {year} {1997})}\BibitemShut {NoStop}%
\bibitem [{\citenamefont {Brodsky}\ \emph {et~al.}(1998)\citenamefont
  {Brodsky}, \citenamefont {Pauli},\ and\ \citenamefont
  {Pinsky}}]{Brodsky:1997de}%
  \BibitemOpen
  \bibfield  {author} {\bibinfo {author} {\bibfnamefont {S.~J.}\ \bibnamefont
  {Brodsky}}, \bibinfo {author} {\bibfnamefont {H.-C.}\ \bibnamefont {Pauli}},\
  and\ \bibinfo {author} {\bibfnamefont {S.~S.}\ \bibnamefont {Pinsky}},\
  }\bibfield  {title} {\bibinfo {title} {{Quantum chromodynamics and other
  field theories on the light cone}},\ }\href
  {https://doi.org/10.1016/S0370-1573(97)00089-6} {\bibfield  {journal}
  {\bibinfo  {journal} {Phys. Rept.}\ }\textbf {\bibinfo {volume} {301}},\
  \bibinfo {pages} {299} (\bibinfo {year} {1998})},\ \Eprint
  {https://arxiv.org/abs/hep-ph/9705477} {arXiv:hep-ph/9705477} \BibitemShut
  {NoStop}%
\bibitem [{\citenamefont {Pasquini}\ \emph {et~al.}(2023)\citenamefont
  {Pasquini}, \citenamefont {Rodini},\ and\ \citenamefont
  {Venturini}}]{Pasquini:2023aaf}%
  \BibitemOpen
  \bibfield  {author} {\bibinfo {author} {\bibfnamefont {B.}~\bibnamefont
  {Pasquini}}, \bibinfo {author} {\bibfnamefont {S.}~\bibnamefont {Rodini}},\
  and\ \bibinfo {author} {\bibfnamefont {S.}~\bibnamefont {Venturini}}
  (\bibinfo {collaboration} {MAP (Multi-dimensional Analyses of Partonic
  distributions)}),\ }\bibfield  {title} {\bibinfo {title} {{Valence quark,
  sea, and gluon content of the pion from the parton distribution functions and
  the electromagnetic form factor}},\ }\href
  {https://doi.org/10.1103/PhysRevD.107.114023} {\bibfield  {journal} {\bibinfo
   {journal} {Phys. Rev. D}\ }\textbf {\bibinfo {volume} {107}},\ \bibinfo
  {pages} {114023} (\bibinfo {year} {2023})},\ \Eprint
  {https://arxiv.org/abs/2303.01789} {arXiv:2303.01789 [hep-ph]} \BibitemShut
  {NoStop}%
\bibitem [{\citenamefont {Shi}\ \emph {et~al.}(2022)\citenamefont {Shi},
  \citenamefont {Li}, \citenamefont {Li}, \citenamefont {Chen},\ and\
  \citenamefont {Jia}}]{Shi:2022erw}%
  \BibitemOpen
  \bibfield  {author} {\bibinfo {author} {\bibfnamefont {C.}~\bibnamefont
  {Shi}}, \bibinfo {author} {\bibfnamefont {J.}~\bibnamefont {Li}}, \bibinfo
  {author} {\bibfnamefont {M.}~\bibnamefont {Li}}, \bibinfo {author}
  {\bibfnamefont {X.}~\bibnamefont {Chen}},\ and\ \bibinfo {author}
  {\bibfnamefont {W.}~\bibnamefont {Jia}},\ }\bibfield  {title} {\bibinfo
  {title} {{Transverse momentum distributions of valence quarks in light and
  heavy vector mesons}},\ }\href {https://doi.org/10.1103/PhysRevD.106.014026}
  {\bibfield  {journal} {\bibinfo  {journal} {Phys. Rev. D}\ }\textbf {\bibinfo
  {volume} {106}},\ \bibinfo {pages} {014026} (\bibinfo {year} {2022})},\
  \Eprint {https://arxiv.org/abs/2205.02757} {arXiv:2205.02757 [hep-ph]}
  \BibitemShut {NoStop}%
\bibitem [{\citenamefont {Xiao}\ \emph {et~al.}(2002)\citenamefont {Xiao},
  \citenamefont {Qian},\ and\ \citenamefont {Ma}}]{Xiao:2002iv}%
  \BibitemOpen
  \bibfield  {author} {\bibinfo {author} {\bibfnamefont {B.-W.}\ \bibnamefont
  {Xiao}}, \bibinfo {author} {\bibfnamefont {X.}~\bibnamefont {Qian}},\ and\
  \bibinfo {author} {\bibfnamefont {B.-Q.}\ \bibnamefont {Ma}},\ }\bibfield
  {title} {\bibinfo {title} {{The Kaon form-factor in the light cone quark
  model}},\ }\href {https://doi.org/10.1140/epja/i2002-10059-y} {\bibfield
  {journal} {\bibinfo  {journal} {Eur. Phys. J. A}\ }\textbf {\bibinfo {volume}
  {15}},\ \bibinfo {pages} {523} (\bibinfo {year} {2002})},\ \Eprint
  {https://arxiv.org/abs/hep-ph/0209138} {arXiv:hep-ph/0209138} \BibitemShut
  {NoStop}%
\bibitem [{\citenamefont {Kaur}\ \emph {et~al.}(2021)\citenamefont {Kaur},
  \citenamefont {Mondal},\ and\ \citenamefont {Dahiya}}]{Kaur:2020emh}%
  \BibitemOpen
  \bibfield  {author} {\bibinfo {author} {\bibfnamefont {S.}~\bibnamefont
  {Kaur}}, \bibinfo {author} {\bibfnamefont {C.}~\bibnamefont {Mondal}},\ and\
  \bibinfo {author} {\bibfnamefont {H.}~\bibnamefont {Dahiya}},\ }\bibfield
  {title} {\bibinfo {title} {{Light-front holographic $\rho$-meson
  distributions in the momentum space}},\ }\href
  {https://doi.org/10.1007/JHEP01(2021)136} {\bibfield  {journal} {\bibinfo
  {journal} {JHEP}\ }\textbf {\bibinfo {volume} {01}},\ \bibinfo {pages}
  {136}},\ \Eprint {https://arxiv.org/abs/2009.04288} {arXiv:2009.04288
  [hep-ph]} \BibitemShut {NoStop}%
\bibitem [{\citenamefont {Kaur}\ \emph {et~al.}(2020)\citenamefont {Kaur},
  \citenamefont {Kumar}, \citenamefont {Lan}, \citenamefont {Mondal},\ and\
  \citenamefont {Dahiya}}]{Kaur:2020vkq}%
  \BibitemOpen
  \bibfield  {author} {\bibinfo {author} {\bibfnamefont {S.}~\bibnamefont
  {Kaur}}, \bibinfo {author} {\bibfnamefont {N.}~\bibnamefont {Kumar}},
  \bibinfo {author} {\bibfnamefont {J.}~\bibnamefont {Lan}}, \bibinfo {author}
  {\bibfnamefont {C.}~\bibnamefont {Mondal}},\ and\ \bibinfo {author}
  {\bibfnamefont {H.}~\bibnamefont {Dahiya}},\ }\bibfield  {title} {\bibinfo
  {title} {{Tomography of light mesons in the light-cone quark model}},\ }\href
  {https://doi.org/10.1103/PhysRevD.102.014021} {\bibfield  {journal} {\bibinfo
   {journal} {Phys. Rev. D}\ }\textbf {\bibinfo {volume} {102}},\ \bibinfo
  {pages} {014021} (\bibinfo {year} {2020})},\ \Eprint
  {https://arxiv.org/abs/2002.01199} {arXiv:2002.01199 [hep-ph]} \BibitemShut
  {NoStop}%
\bibitem [{\citenamefont {Berger}\ \emph {et~al.}(2001)\citenamefont {Berger},
  \citenamefont {Cano}, \citenamefont {Diehl},\ and\ \citenamefont
  {Pire}}]{Berger:2001zb}%
  \BibitemOpen
  \bibfield  {author} {\bibinfo {author} {\bibfnamefont {E.~R.}\ \bibnamefont
  {Berger}}, \bibinfo {author} {\bibfnamefont {F.}~\bibnamefont {Cano}},
  \bibinfo {author} {\bibfnamefont {M.}~\bibnamefont {Diehl}},\ and\ \bibinfo
  {author} {\bibfnamefont {B.}~\bibnamefont {Pire}},\ }\bibfield  {title}
  {\bibinfo {title} {{Generalized parton distributions in the deuteron}},\
  }\href {https://doi.org/10.1103/PhysRevLett.87.142302} {\bibfield  {journal}
  {\bibinfo  {journal} {Phys. Rev. Lett.}\ }\textbf {\bibinfo {volume} {87}},\
  \bibinfo {pages} {142302} (\bibinfo {year} {2001})},\ \Eprint
  {https://arxiv.org/abs/hep-ph/0106192} {arXiv:hep-ph/0106192} \BibitemShut
  {NoStop}%
\bibitem [{\citenamefont {Lasscock}\ \emph {et~al.}(2006)\citenamefont
  {Lasscock}, \citenamefont {Hedditch}, \citenamefont {Leinweber},\ and\
  \citenamefont {Williams}}]{Lasscock:2006nh}%
  \BibitemOpen
  \bibfield  {author} {\bibinfo {author} {\bibfnamefont {B.~G.}\ \bibnamefont
  {Lasscock}}, \bibinfo {author} {\bibfnamefont {J.}~\bibnamefont {Hedditch}},
  \bibinfo {author} {\bibfnamefont {D.~B.}\ \bibnamefont {Leinweber}},\ and\
  \bibinfo {author} {\bibfnamefont {A.~G.}\ \bibnamefont {Williams}},\
  }\bibfield  {title} {\bibinfo {title} {{Vector meson electromagnetic form
  factors}},\ }\href {https://doi.org/10.22323/1.032.0114} {\bibfield
  {journal} {\bibinfo  {journal} {PoS}\ }\textbf {\bibinfo {volume}
  {LAT2006}},\ \bibinfo {pages} {114} (\bibinfo {year} {2006})},\ \Eprint
  {https://arxiv.org/abs/hep-lat/0611029} {arXiv:hep-lat/0611029} \BibitemShut
  {NoStop}%
\bibitem [{\citenamefont {Zhang}(2025)}]{Zhang:2024nxl}%
  \BibitemOpen
  \bibfield  {author} {\bibinfo {author} {\bibfnamefont {J.-L.}\ \bibnamefont
  {Zhang}},\ }\bibfield  {title} {\bibinfo {title} {{\ensuremath{\rho} meson
  form factors and parton distribution functions in impact parameter space*}},\
  }\href {https://doi.org/10.1088/1674-1137/adab61} {\bibfield  {journal}
  {\bibinfo  {journal} {Chin. Phys. C}\ }\textbf {\bibinfo {volume} {49}},\
  \bibinfo {pages} {043104} (\bibinfo {year} {2025})},\ \Eprint
  {https://arxiv.org/abs/2409.19525} {arXiv:2409.19525 [hep-ph]} \BibitemShut
  {NoStop}%
\bibitem [{\citenamefont {Hern\'andez-Pinto}\ \emph {et~al.}(2024)\citenamefont
  {Hern\'andez-Pinto}, \citenamefont {Guti\'errez-Guerrero}, \citenamefont
  {Bedolla},\ and\ \citenamefont {Bashir}}]{Hernandez-Pinto:2024kwg}%
  \BibitemOpen
  \bibfield  {author} {\bibinfo {author} {\bibfnamefont {R.~J.}\ \bibnamefont
  {Hern\'andez-Pinto}}, \bibinfo {author} {\bibfnamefont {L.~X.}\ \bibnamefont
  {Guti\'errez-Guerrero}}, \bibinfo {author} {\bibfnamefont {M.~A.}\
  \bibnamefont {Bedolla}},\ and\ \bibinfo {author} {\bibfnamefont
  {A.}~\bibnamefont {Bashir}},\ }\bibfield  {title} {\bibinfo {title}
  {{Electric, magnetic, and quadrupole form factors and charge radii of vector
  mesons: From light to heavy sectors in a contact interaction}},\ }\href
  {https://doi.org/10.1103/PhysRevD.110.114015} {\bibfield  {journal} {\bibinfo
   {journal} {Phys. Rev. D}\ }\textbf {\bibinfo {volume} {110}},\ \bibinfo
  {pages} {114015} (\bibinfo {year} {2024})},\ \Eprint
  {https://arxiv.org/abs/2410.23813} {arXiv:2410.23813 [hep-ph]} \BibitemShut
  {NoStop}%
\bibitem [{\citenamefont {Bhagwat}\ and\ \citenamefont
  {Maris}(2008)}]{Bhagwat:2006pu}%
  \BibitemOpen
  \bibfield  {author} {\bibinfo {author} {\bibfnamefont {M.~S.}\ \bibnamefont
  {Bhagwat}}\ and\ \bibinfo {author} {\bibfnamefont {P.}~\bibnamefont
  {Maris}},\ }\bibfield  {title} {\bibinfo {title} {{Vector meson form factors
  and their quark-mass dependence}},\ }\href
  {https://doi.org/10.1103/PhysRevC.77.025203} {\bibfield  {journal} {\bibinfo
  {journal} {Phys. Rev. C}\ }\textbf {\bibinfo {volume} {77}},\ \bibinfo
  {pages} {025203} (\bibinfo {year} {2008})},\ \Eprint
  {https://arxiv.org/abs/nucl-th/0612069} {arXiv:nucl-th/0612069} \BibitemShut
  {NoStop}%
\bibitem [{\citenamefont {Dudek}\ \emph {et~al.}(2006)\citenamefont {Dudek},
  \citenamefont {Edwards},\ and\ \citenamefont {Richards}}]{Dudek:2006ej}%
  \BibitemOpen
  \bibfield  {author} {\bibinfo {author} {\bibfnamefont {J.~J.}\ \bibnamefont
  {Dudek}}, \bibinfo {author} {\bibfnamefont {R.~G.}\ \bibnamefont {Edwards}},\
  and\ \bibinfo {author} {\bibfnamefont {D.~G.}\ \bibnamefont {Richards}},\
  }\bibfield  {title} {\bibinfo {title} {{Radiative transitions in charmonium
  from lattice QCD}},\ }\href {https://doi.org/10.1103/PhysRevD.73.074507}
  {\bibfield  {journal} {\bibinfo  {journal} {Phys. Rev. D}\ }\textbf {\bibinfo
  {volume} {73}},\ \bibinfo {pages} {074507} (\bibinfo {year} {2006})},\
  \Eprint {https://arxiv.org/abs/hep-ph/0601137} {arXiv:hep-ph/0601137}
  \BibitemShut {NoStop}%
\bibitem [{\citenamefont {Luan}\ \emph {et~al.}(2015)\citenamefont {Luan},
  \citenamefont {Chen},\ and\ \citenamefont {Deng}}]{Luan:2015goa}%
  \BibitemOpen
  \bibfield  {author} {\bibinfo {author} {\bibfnamefont {Y.-L.}\ \bibnamefont
  {Luan}}, \bibinfo {author} {\bibfnamefont {X.-L.}\ \bibnamefont {Chen}},\
  and\ \bibinfo {author} {\bibfnamefont {W.-Z.}\ \bibnamefont {Deng}},\
  }\bibfield  {title} {\bibinfo {title} {{Meson electro-magnetic form factors
  in an extended Nambu\textendash{}Jona-Lasinio model including heavy quark
  flavors}},\ }\href {https://doi.org/10.1088/1674-1137/39/11/113103}
  {\bibfield  {journal} {\bibinfo  {journal} {Chin. Phys. C}\ }\textbf
  {\bibinfo {volume} {39}},\ \bibinfo {pages} {113103} (\bibinfo {year}
  {2015})},\ \Eprint {https://arxiv.org/abs/1504.03799} {arXiv:1504.03799
  [hep-ph]} \BibitemShut {NoStop}%
\bibitem [{\citenamefont {Owen}\ \emph {et~al.}(2015)\citenamefont {Owen},
  \citenamefont {Kamleh}, \citenamefont {Leinweber}, \citenamefont {Menadue},\
  and\ \citenamefont {Mahbub}}]{Owen:2015gva}%
  \BibitemOpen
  \bibfield  {author} {\bibinfo {author} {\bibfnamefont {B.}~\bibnamefont
  {Owen}}, \bibinfo {author} {\bibfnamefont {W.}~\bibnamefont {Kamleh}},
  \bibinfo {author} {\bibfnamefont {D.}~\bibnamefont {Leinweber}}, \bibinfo
  {author} {\bibfnamefont {B.}~\bibnamefont {Menadue}},\ and\ \bibinfo {author}
  {\bibfnamefont {S.}~\bibnamefont {Mahbub}},\ }\bibfield  {title} {\bibinfo
  {title} {{Light Meson Form Factors at near Physical Masses}},\ }\href
  {https://doi.org/10.1103/PhysRevD.91.074503} {\bibfield  {journal} {\bibinfo
  {journal} {Phys. Rev. D}\ }\textbf {\bibinfo {volume} {91}},\ \bibinfo
  {pages} {074503} (\bibinfo {year} {2015})},\ \Eprint
  {https://arxiv.org/abs/1501.02561} {arXiv:1501.02561 [hep-lat]} \BibitemShut
  {NoStop}%
\bibitem [{\citenamefont {Haftel}\ \emph {et~al.}(1980)\citenamefont {Haftel},
  \citenamefont {Mathelitsch},\ and\ \citenamefont {Zingl}}]{Haftel:1980zz}%
  \BibitemOpen
  \bibfield  {author} {\bibinfo {author} {\bibfnamefont {M.~I.}\ \bibnamefont
  {Haftel}}, \bibinfo {author} {\bibfnamefont {L.}~\bibnamefont
  {Mathelitsch}},\ and\ \bibinfo {author} {\bibfnamefont {H.~F.~K.}\
  \bibnamefont {Zingl}},\ }\bibfield  {title} {\bibinfo {title}
  {{Electron-deuteron tensor polarization and the two-nucleon force}},\ }\href
  {https://doi.org/10.1103/PhysRevC.22.1285} {\bibfield  {journal} {\bibinfo
  {journal} {Phys. Rev. C}\ }\textbf {\bibinfo {volume} {22}},\ \bibinfo
  {pages} {1285} (\bibinfo {year} {1980})}\BibitemShut {NoStop}%
\bibitem [{\citenamefont {Kumano}\ and\ \citenamefont
  {Song}(2021)}]{Kumano:2021fem}%
  \BibitemOpen
  \bibfield  {author} {\bibinfo {author} {\bibfnamefont {S.}~\bibnamefont
  {Kumano}}\ and\ \bibinfo {author} {\bibfnamefont {Q.-T.}\ \bibnamefont
  {Song}},\ }\bibfield  {title} {\bibinfo {title} {{Twist-2 relation and sum
  rule for tensor-polarized parton distribution functions of spin-1 hadrons}},\
  }\href {https://doi.org/10.1007/JHEP09(2021)141} {\bibfield  {journal}
  {\bibinfo  {journal} {JHEP}\ }\textbf {\bibinfo {volume} {09}},\ \bibinfo
  {pages} {141}},\ \Eprint {https://arxiv.org/abs/2106.15849} {arXiv:2106.15849
  [hep-ph]} \BibitemShut {NoStop}%
\bibitem [{\citenamefont {Hino}\ and\ \citenamefont
  {Kumano}(1999)}]{Hino:1999qi}%
  \BibitemOpen
  \bibfield  {author} {\bibinfo {author} {\bibfnamefont {S.}~\bibnamefont
  {Hino}}\ and\ \bibinfo {author} {\bibfnamefont {S.}~\bibnamefont {Kumano}},\
  }\bibfield  {title} {\bibinfo {title} {{Structure functions in the polarized
  Drell-Yan processes with spin 1/2 and spin 1 hadrons. 2. Parton model}},\
  }\href {https://doi.org/10.1103/PhysRevD.60.054018} {\bibfield  {journal}
  {\bibinfo  {journal} {Phys. Rev. D}\ }\textbf {\bibinfo {volume} {60}},\
  \bibinfo {pages} {054018} (\bibinfo {year} {1999})},\ \Eprint
  {https://arxiv.org/abs/hep-ph/9902258} {arXiv:hep-ph/9902258} \BibitemShut
  {NoStop}%
\bibitem [{\citenamefont {Cosyn}\ \emph {et~al.}(2017)\citenamefont {Cosyn},
  \citenamefont {Dong}, \citenamefont {Kumano},\ and\ \citenamefont
  {Sargsian}}]{Cosyn:2017fbo}%
  \BibitemOpen
  \bibfield  {author} {\bibinfo {author} {\bibfnamefont {W.}~\bibnamefont
  {Cosyn}}, \bibinfo {author} {\bibfnamefont {Y.-B.}\ \bibnamefont {Dong}},
  \bibinfo {author} {\bibfnamefont {S.}~\bibnamefont {Kumano}},\ and\ \bibinfo
  {author} {\bibfnamefont {M.}~\bibnamefont {Sargsian}},\ }\bibfield  {title}
  {\bibinfo {title} {{Tensor-polarized structure function $b_1$ in standard
  convolution description of deuteron}},\ }\href
  {https://doi.org/10.1103/PhysRevD.95.074036} {\bibfield  {journal} {\bibinfo
  {journal} {Phys. Rev. D}\ }\textbf {\bibinfo {volume} {95}},\ \bibinfo
  {pages} {074036} (\bibinfo {year} {2017})},\ \Eprint
  {https://arxiv.org/abs/1702.05337} {arXiv:1702.05337 [hep-ph]} \BibitemShut
  {NoStop}%
\bibitem [{\citenamefont {Miyama}\ and\ \citenamefont
  {Kumano}(1996)}]{Miyama:1995bd}%
  \BibitemOpen
  \bibfield  {author} {\bibinfo {author} {\bibfnamefont {M.}~\bibnamefont
  {Miyama}}\ and\ \bibinfo {author} {\bibfnamefont {S.}~\bibnamefont
  {Kumano}},\ }\bibfield  {title} {\bibinfo {title} {{Numerical solution of
  Q**2 evolution equations in a brute force method}},\ }\href
  {https://doi.org/10.1016/0010-4655(96)00013-6} {\bibfield  {journal}
  {\bibinfo  {journal} {Comput. Phys. Commun.}\ }\textbf {\bibinfo {volume}
  {94}},\ \bibinfo {pages} {185} (\bibinfo {year} {1996})},\ \Eprint
  {https://arxiv.org/abs/hep-ph/9508246} {arXiv:hep-ph/9508246} \BibitemShut
  {NoStop}%
\bibitem [{\citenamefont {Hirai}\ \emph
  {et~al.}(1998{\natexlab{a}})\citenamefont {Hirai}, \citenamefont {Kumano},\
  and\ \citenamefont {Miyama}}]{Hirai:1997gb}%
  \BibitemOpen
  \bibfield  {author} {\bibinfo {author} {\bibfnamefont {M.}~\bibnamefont
  {Hirai}}, \bibinfo {author} {\bibfnamefont {S.}~\bibnamefont {Kumano}},\ and\
  \bibinfo {author} {\bibfnamefont {M.}~\bibnamefont {Miyama}},\ }\bibfield
  {title} {\bibinfo {title} {{Numerical solution of Q**2 evolution equations
  for polarized structure functions}},\ }\href
  {https://doi.org/10.1016/S0010-4655(97)00129-X} {\bibfield  {journal}
  {\bibinfo  {journal} {Comput. Phys. Commun.}\ }\textbf {\bibinfo {volume}
  {108}},\ \bibinfo {pages} {38} (\bibinfo {year} {1998}{\natexlab{a}})},\
  \Eprint {https://arxiv.org/abs/hep-ph/9707220} {arXiv:hep-ph/9707220}
  \BibitemShut {NoStop}%
\bibitem [{\citenamefont {Hirai}\ \emph
  {et~al.}(1998{\natexlab{b}})\citenamefont {Hirai}, \citenamefont {Kumano},\
  and\ \citenamefont {Miyama}}]{Hirai:1997mm}%
  \BibitemOpen
  \bibfield  {author} {\bibinfo {author} {\bibfnamefont {M.}~\bibnamefont
  {Hirai}}, \bibinfo {author} {\bibfnamefont {S.}~\bibnamefont {Kumano}},\ and\
  \bibinfo {author} {\bibfnamefont {M.}~\bibnamefont {Miyama}},\ }\bibfield
  {title} {\bibinfo {title} {{Numerical solution of Q**2 evolution equation for
  the transversity distribution Delta(T)q}},\ }\href
  {https://doi.org/10.1016/S0010-4655(98)00028-9} {\bibfield  {journal}
  {\bibinfo  {journal} {Comput. Phys. Commun.}\ }\textbf {\bibinfo {volume}
  {111}},\ \bibinfo {pages} {150} (\bibinfo {year} {1998}{\natexlab{b}})},\
  \Eprint {https://arxiv.org/abs/hep-ph/9712410} {arXiv:hep-ph/9712410}
  \BibitemShut {NoStop}%
\bibitem [{\citenamefont {Ninomiya}\ \emph {et~al.}(2017)\citenamefont
  {Ninomiya}, \citenamefont {Bentz},\ and\ \citenamefont
  {Clo\"et}}]{Ninomiya:2017ggn}%
  \BibitemOpen
  \bibfield  {author} {\bibinfo {author} {\bibfnamefont {Y.}~\bibnamefont
  {Ninomiya}}, \bibinfo {author} {\bibfnamefont {W.}~\bibnamefont {Bentz}},\
  and\ \bibinfo {author} {\bibfnamefont {I.~C.}\ \bibnamefont {Clo\"et}},\
  }\bibfield  {title} {\bibinfo {title} {{Transverse-momentum-dependent quark
  distribution functions of spin-one targets: Formalism and covariant
  calculations}},\ }\href {https://doi.org/10.1103/PhysRevC.96.045206}
  {\bibfield  {journal} {\bibinfo  {journal} {Phys. Rev. C}\ }\textbf {\bibinfo
  {volume} {96}},\ \bibinfo {pages} {045206} (\bibinfo {year} {2017})},\
  \Eprint {https://arxiv.org/abs/1707.03787} {arXiv:1707.03787 [nucl-th]}
  \BibitemShut {NoStop}%
\bibitem [{\citenamefont {Zhang}\ and\ \citenamefont
  {Wu}(2025)}]{Zhang:2024plq}%
  \BibitemOpen
  \bibfield  {author} {\bibinfo {author} {\bibfnamefont {J.-L.}\ \bibnamefont
  {Zhang}}\ and\ \bibinfo {author} {\bibfnamefont {J.}~\bibnamefont {Wu}},\
  }\bibfield  {title} {\bibinfo {title} {{$\rho $ meson transverse
  momentum-dependent parton distributions}},\ }\href
  {https://doi.org/10.1140/epjc/s10052-024-13682-w} {\bibfield  {journal}
  {\bibinfo  {journal} {Eur. Phys. J. C}\ }\textbf {\bibinfo {volume} {85}},\
  \bibinfo {pages} {13} (\bibinfo {year} {2025})},\ \Eprint
  {https://arxiv.org/abs/2408.13569} {arXiv:2408.13569 [hep-ph]} \BibitemShut
  {NoStop}%
\bibitem [{\citenamefont {Lan}\ \emph {et~al.}(2020)\citenamefont {Lan},
  \citenamefont {Mondal}, \citenamefont {Li}, \citenamefont {Li}, \citenamefont
  {Tang}, \citenamefont {Zhao},\ and\ \citenamefont {Vary}}]{Lan:2019img}%
  \BibitemOpen
  \bibfield  {author} {\bibinfo {author} {\bibfnamefont {J.}~\bibnamefont
  {Lan}}, \bibinfo {author} {\bibfnamefont {C.}~\bibnamefont {Mondal}},
  \bibinfo {author} {\bibfnamefont {M.}~\bibnamefont {Li}}, \bibinfo {author}
  {\bibfnamefont {Y.}~\bibnamefont {Li}}, \bibinfo {author} {\bibfnamefont
  {S.}~\bibnamefont {Tang}}, \bibinfo {author} {\bibfnamefont {X.}~\bibnamefont
  {Zhao}},\ and\ \bibinfo {author} {\bibfnamefont {J.~P.}\ \bibnamefont
  {Vary}},\ }\bibfield  {title} {\bibinfo {title} {{Parton Distribution
  Functions of Heavy Mesons on the Light Front}},\ }\href
  {https://doi.org/10.1103/PhysRevD.102.014020} {\bibfield  {journal} {\bibinfo
   {journal} {Phys. Rev. D}\ }\textbf {\bibinfo {volume} {102}},\ \bibinfo
  {pages} {014020} (\bibinfo {year} {2020})},\ \Eprint
  {https://arxiv.org/abs/1911.11676} {arXiv:1911.11676 [nucl-th]} \BibitemShut
  {NoStop}%
\bibitem [{\citenamefont {Li}\ \emph {et~al.}(2022)\citenamefont {Li},
  \citenamefont {Li}, \citenamefont {Chen}, \citenamefont {Lappi},\ and\
  \citenamefont {Vary}}]{Li:2021cwv}%
  \BibitemOpen
  \bibfield  {author} {\bibinfo {author} {\bibfnamefont {M.}~\bibnamefont
  {Li}}, \bibinfo {author} {\bibfnamefont {Y.}~\bibnamefont {Li}}, \bibinfo
  {author} {\bibfnamefont {G.}~\bibnamefont {Chen}}, \bibinfo {author}
  {\bibfnamefont {T.}~\bibnamefont {Lappi}},\ and\ \bibinfo {author}
  {\bibfnamefont {J.~P.}\ \bibnamefont {Vary}},\ }\bibfield  {title} {\bibinfo
  {title} {{Light-front wavefunctions of mesons by design}},\ }\href
  {https://doi.org/10.1140/epjc/s10052-022-10988-5} {\bibfield  {journal}
  {\bibinfo  {journal} {Eur. Phys. J. C}\ }\textbf {\bibinfo {volume} {82}},\
  \bibinfo {pages} {1045} (\bibinfo {year} {2022})},\ \Eprint
  {https://arxiv.org/abs/2111.07087} {arXiv:2111.07087 [hep-ph]} \BibitemShut
  {NoStop}%
\bibitem [{\citenamefont {Airapetian}\ \emph {et~al.}(2017)\citenamefont
  {Airapetian} \emph {et~al.}}]{HERMES:2017qwt}%
  \BibitemOpen
  \bibfield  {author} {\bibinfo {author} {\bibfnamefont {A.}~\bibnamefont
  {Airapetian}} \emph {et~al.} (\bibinfo {collaboration} {HERMES}),\ }\bibfield
   {title} {\bibinfo {title} {{Ratios of helicity amplitudes for exclusive
  $\rho ^0$ electroproduction on transversely polarized protons}},\ }\href
  {https://doi.org/10.1140/epjc/s10052-017-4899-1} {\bibfield  {journal}
  {\bibinfo  {journal} {Eur. Phys. J. C}\ }\textbf {\bibinfo {volume} {77}},\
  \bibinfo {pages} {378} (\bibinfo {year} {2017})},\ \Eprint
  {https://arxiv.org/abs/1702.00345} {arXiv:1702.00345 [hep-ex]} \BibitemShut
  {NoStop}%
\bibitem [{\citenamefont {Mankiewicz}\ \emph {et~al.}(1998)\citenamefont
  {Mankiewicz}, \citenamefont {Piller},\ and\ \citenamefont
  {Weigl}}]{Mankiewicz:1997uy}%
  \BibitemOpen
  \bibfield  {author} {\bibinfo {author} {\bibfnamefont {L.}~\bibnamefont
  {Mankiewicz}}, \bibinfo {author} {\bibfnamefont {G.}~\bibnamefont {Piller}},\
  and\ \bibinfo {author} {\bibfnamefont {T.}~\bibnamefont {Weigl}},\ }\bibfield
   {title} {\bibinfo {title} {{Hard exclusive meson production and nonforward
  parton distributions}},\ }\href {https://doi.org/10.1007/s100520050253}
  {\bibfield  {journal} {\bibinfo  {journal} {Eur. Phys. J. C}\ }\textbf
  {\bibinfo {volume} {5}},\ \bibinfo {pages} {119} (\bibinfo {year} {1998})},\
  \Eprint {https://arxiv.org/abs/hep-ph/9711227} {arXiv:hep-ph/9711227}
  \BibitemShut {NoStop}%
\bibitem [{\citenamefont {Morrow}\ \emph {et~al.}(2009)\citenamefont {Morrow}
  \emph {et~al.}}]{CLAS:2008rpm}%
  \BibitemOpen
  \bibfield  {author} {\bibinfo {author} {\bibfnamefont {S.~A.}\ \bibnamefont
  {Morrow}} \emph {et~al.} (\bibinfo {collaboration} {CLAS}),\ }\bibfield
  {title} {\bibinfo {title} {{Exclusive rho0 electroproduction on the proton at
  CLAS}},\ }\href {https://doi.org/10.1140/epja/i2008-10683-5} {\bibfield
  {journal} {\bibinfo  {journal} {Eur. Phys. J. A}\ }\textbf {\bibinfo {volume}
  {39}},\ \bibinfo {pages} {5} (\bibinfo {year} {2009})},\ \Eprint
  {https://arxiv.org/abs/0807.3834} {arXiv:0807.3834 [hep-ex]} \BibitemShut
  {NoStop}%
\bibitem [{\citenamefont {Anikin}\ \emph {et~al.}(2004)\citenamefont {Anikin},
  \citenamefont {Pire},\ and\ \citenamefont {Teryaev}}]{Anikin:2003fr}%
  \BibitemOpen
  \bibfield  {author} {\bibinfo {author} {\bibfnamefont {I.~V.}\ \bibnamefont
  {Anikin}}, \bibinfo {author} {\bibfnamefont {B.}~\bibnamefont {Pire}},\ and\
  \bibinfo {author} {\bibfnamefont {O.~V.}\ \bibnamefont {Teryaev}},\
  }\bibfield  {title} {\bibinfo {title} {{On gamma gamma* production of two
  rho0 mesons}},\ }\href {https://doi.org/10.1103/PhysRevD.69.014018}
  {\bibfield  {journal} {\bibinfo  {journal} {Phys. Rev. D}\ }\textbf {\bibinfo
  {volume} {69}},\ \bibinfo {pages} {014018} (\bibinfo {year} {2004})},\
  \Eprint {https://arxiv.org/abs/hep-ph/0307059} {arXiv:hep-ph/0307059}
  \BibitemShut {NoStop}%
\bibitem [{\citenamefont {Boer}\ \emph {et~al.}(2011)\citenamefont {Boer} \emph
  {et~al.}}]{Boer:2011fh}%
  \BibitemOpen
  \bibfield  {author} {\bibinfo {author} {\bibfnamefont {D.}~\bibnamefont
  {Boer}} \emph {et~al.},\ }\bibfield  {title} {\bibinfo {title} {{Gluons and
  the quark sea at high energies: Distributions, polarization, tomography}},\
  }\href@noop {} {\  (\bibinfo {year} {2011})},\ \Eprint
  {https://arxiv.org/abs/1108.1713} {arXiv:1108.1713 [nucl-th]} \BibitemShut
  {NoStop}%
\bibitem [{\citenamefont {Abdul~Khalek}\ \emph {et~al.}(2022)\citenamefont
  {Abdul~Khalek} \emph {et~al.}}]{AbdulKhalek:2021gbh}%
  \BibitemOpen
  \bibfield  {author} {\bibinfo {author} {\bibfnamefont {R.}~\bibnamefont
  {Abdul~Khalek}} \emph {et~al.},\ }\bibfield  {title} {\bibinfo {title}
  {{Science Requirements and Detector Concepts for the Electron-Ion Collider}:
  {EIC Yellow Report}},\ }\href
  {https://doi.org/10.1016/j.nuclphysa.2022.122447} {\bibfield  {journal}
  {\bibinfo  {journal} {Nucl. Phys. A}\ }\textbf {\bibinfo {volume} {1026}},\
  \bibinfo {pages} {122447} (\bibinfo {year} {2022})},\ \Eprint
  {https://arxiv.org/abs/2103.05419} {arXiv:2103.05419 [physics.ins-det]}
  \BibitemShut {NoStop}%
\end{thebibliography}%

\end{document}